\documentclass[aps]{revtex4-1}
\usepackage[utf8]{inputenc}
\usepackage[english]{babel}
\usepackage{subcaption}
\usepackage{mhchem}
\usepackage{graphicx}

\begin{document}

\title{Explaining key properties of lithiation in \ce{TiO2}-anatase Li-ion battery electrodes using phase-field modelling}

\author{Niek J.J. de Klerk}
 \altaffiliation{Contributed equally to this work.}
\author{Alexandros Vasileiadis}%
 \altaffiliation{Contributed equally to this work.}
\affiliation{Department of Radiation Science and Technology, Delft University of Technology, Mekelweg 15, 2629JB Delft, The Netherlands}

\author{Raymond B. Smith}
\affiliation{Department of Chemical Engineering, Massachusetts Institute of Technology, Cambridge, Massachusetts 02139, USA}
\author{Martin Z. Bazant}
\affiliation{Department of Chemical Engineering, Massachusetts Institute of Technology, Cambridge, Massachusetts 02139, USA}
\affiliation{Department of Mathematics, Massachusetts Institute of Technology, Cambridge, Massachusetts 02139, USA}
\author{Marnix Wagemaker}
\email{m.wagemaker@tudelft.nl}
\affiliation{Department of Radiation Science and Technology, Delft University of Technology, Mekelweg 15, 2629JB Delft, The Netherlands}


\begin{abstract}
The improvement of Li-ion battery performance requires development of models that capture the essential physics and chemistry in Li-ion battery electrode materials. Phase-field modelling has recently been shown to have this ability, providing new opportunities to gain understanding of these complex systems. In this paper a novel electrochemical phase-field model is presented that captures the thermodynamic and kinetic properties of lithium-insertion in \ce{TiO2}-anatase, a well-known and intensively studied Li-ion battery electrode material. Using a linear combination of two regular solution models the two phase transitions during lithiation are described as lithiation of two separate lattices with different physical properties.
Previous elaborate experimental work on lithiated anatase \ce{TiO2} provides all parameters necessary for the phase-field simulations, giving the opportunity to gain fundamental insight in the lithiation of anatase and validate this phase-field model. 
The phase-field model captures the essential experimentally observed phenomena, rationalising the impact of C-rate, particle size, surface area, and the memory effect on the performance of anatase as a Li-ion battery electrode. Thereby a comprehensive physical picture of the lithiation of anatase \ce{TiO2} is provided.
The results of the simulations demonstrate that the performance of anatase is limited by the formation of the poor Li-ion diffusion in the \ce{Li1TiO2} phase at the surface of the particles. Unlike other electrode materials, the kinetic limitations of individual anatase particles limit the performance of full electrodes. Hence, rather than improving the ionic and electronic network in electrodes, improving the performance of anatase \ce{TiO2} electrodes requires preventing the formation of a blocking \ce{Li1TiO2} phase at the surface of particles. 
Additionally, the qualitative agreement of the phase-field model, containing only parameters from literature, with a broad spectrum of experiments demonstrates the capabilities of phase-field models for understanding Li-ion electrode materials, and its promise for guiding the design of electrodes through a thorough understanding of material properties and their interactions. 
\end{abstract}

\maketitle

\section{Introduction} 
High energy densities realised by Li-ion batteries have enabled mobile applications scaling from mobile phones, tablets, and laptops, up to electrical vehicles. The application of batteries in electric vehicles in particular has driven the demand for faster and more efficient electricity storage.
Different mechanisms may limit battery performance \cite{Deng_2015, Franco_2013, Zhang_2015a}: the electronic wiring in the electrodes, ionic transport through the electrolyte, the charge transfer reaction, and the solid state transport process. Which of these mechanisms is limiting depends on the applied current and the morphology of the electrodes \cite{Zhang_2015a}.
To understand the complex interplay of the processes in batteries and to enable improved battery design, various models have been developed \cite{Notten_2014, Vo_2015,  Sethuraman_2012,  Landstorfer_2011, Dargaville_2013, Bazant_2013, Salvadori_2015}.
Using these models it is possible to design better battery management systems \cite{Notten_2014}, decrease charging times \cite{Vo_2015}, estimate the effect of side-reactions on performance \cite{Sethuraman_2012}, and study what limits the performance of a battery \cite{Landstorfer_2011}. \\
The challenge for models describing batteries is taking into account microscopic processes, such as phase transitions and interfaces, in combination with macroscopic phenomena such as many particle effects \cite{Li_2014} and charge transport.
The non-equilibrium conditions in complete electrodes will lead to macroscopic gradients in diffusing species, and the associated potential gradients can change phase-transition kinetics, as has been demonstrated for \ce{LiFePO4} \cite{Li_2014}. Even when a model is obtained which reasonably describes the processes, it often involves a number of unknown physical parameters, which require fitting to experimental data. 
Although this may result in an appropriate model for conditions similar to those of the fitted experimental data, extrapolation to other operating conditions is uncertain \cite{Franco_2013}, making accurate model validation under different conditions vital. \\
The introduction of phase-field modelling to the battery field \cite{Singh_2008, Burch_2009, Ferguson_2012, Bazant_2013} has enabled accurate prediction of the phase transitions both in individual electrode particles and multi-particle systems \cite{Li_2014} representing entire electrodes. This is computationally feasible because the phase-interface is taken implicitly into account \cite{Bazant_2013}, making it unnecessary to evaluate the phase transition kinetics in every position in an electrode particle.
Using phase-field models for \ce{LiFePO4} the observed decreasing miscibility and spinodal gap in nano-particles \cite{Wagemaker_2011} has been explained \cite{Burch_2009, Welland_2015}, the observed transition from a first order phase transition to a solid solution reaction at high overpotentials \cite{Zhang_2014, Liu_2014} has been predicted, and the transition from particle-by-particle to a concurrent mechanism was predicted \cite{Li_2014} consistent with observations \cite{Zhang_2015b}. Recently a three-dimensional phase-field model has been presented for \ce{LiFePO4} \cite{Welland_2015}, and crack formation and the effects this causes have also been incorporated \cite{Oconnor_2016}.   
The phase-field method has also been used to describe the lithiation of graphite electrodes \cite{Ferguson_2014, Guo_2016}, requiring the introduction of two first-order phase-transformations, which is relatively straight forward in a phase-field model, resulting in good agreement with experiments \cite{Ferguson_2014}. \\
These results demonstrate the success of phase-field modelling of battery electrodes, and anatase \ce{TiO2} is another ideal candidate for applying phase-field modelling. It has been extensively studied for more than two decades, in which all parameters required for the phase-field model have been measured experimentally. 
This will allow comparison of a parameter free phase-field model towards a broad range of experimental results available in literature. 
Anatase \ce{TiO2} is an attractive Li-ion battery electrode material, based on its cheap and abundant elements, high theoretical capacity of 335 mAh/g, small volume expansion during lithiation \cite{Lafont_2010}, and good electronic conductivity \cite{Singh_JPC}. \\
\begin{figure}[htbp]
	\begin{center}
	\includegraphics[width=0.75\textwidth]{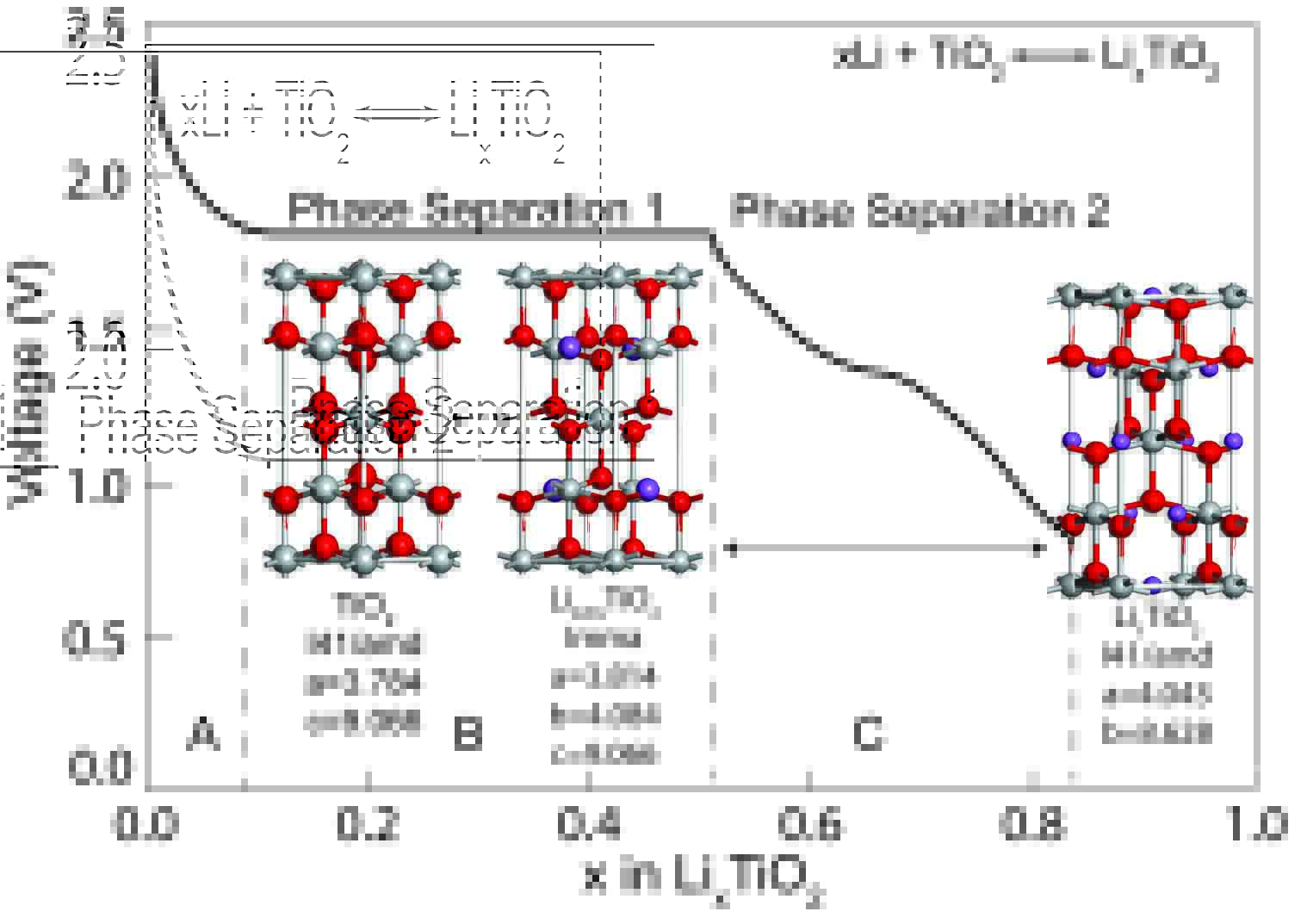}	
	\caption{The crystal structure \cite{Lafont_2010} and typical voltage profile of anatase during lithiation.}
	\label{fig:structure_and_voltage}
	\end{center}
\end{figure}
The TiO$_2$-anatase lattice consists of stacked one dimensional zigzag chains of TiO$_6$ octahedra sharing distorted edges, as shown in Figure \ref{fig:structure_and_voltage}. This stacking leads to empty zigzag channels with octahedral and tetrahedral interstitial sites that can accommodate lithium. 
A typical voltage profile for lithiation of anatase is shown in Figure \ref{fig:structure_and_voltage}. At low Li-concentrations a solid solution is formed (region A), the length of which depends on the particle size \cite{Shen_2014, Wagemaker_2007A, Sudant_2005}. Past the solid solution limit phase separation occurs, reflected by the plateau in region B, where half of the octahedral sites are filled to form the Li-titanate phase (Li$_{0.5}$TiO$_2$). This is followed by a pseudo-plateau (region C) during which the remaining octahedral sites are filled, forming Li$_{1}$TiO$_2$. Even though this phase transition usually does not show a voltage plateau, it is reported to occur via a phase separation mechanism \cite{Morgan_2011, Wagemaker_2007A}. \\
In bulk anatase roughly 0.6 lithium per formula unit is reversibly inserted in most experiments \cite{Sudant_2005, Wagemaker_2007A}, leading to capacities of approximately 200 mAh/g. However, many parameters have been shown to affect the capacity drastically, including the preparation method \cite{Sondergaard_2015}, removing water from the anatase crystals before assembling the battery \cite{Madej_2014A}, the atmosphere during annealing \cite{Wang_2011}, the morphology of the crystals \cite{Sun_2010}, and the cut-off potential used during cycling \cite{Madej_2014C}. However, the most decisive factor appears to be the particle size \cite{Gentili_2012, Sudant_2005, Rai_2013, Wagemaker_2007A}. By nano-sizing anatase particles the \ce{Li1TiO2}-phase can be obtained at room temperature, realising the theoretical capacity of 335 mAh/g \cite{Lafont_2010, Gentili_2012}. \\
For bulk samples complete lithiation via electrochemical experiments has been reported, but only when kinetic restrictions were removed, either by lithiating at 120$^{\circ}$C \cite{Zachau_1988, Macklin_1992}, or by allowing the anatase electrode to equilibrate during GITT measurements \cite{Shen_2014, Sudant_2005}. Computational results also indicate that full lithiation is energetically favourable \cite{Belak_2012, Morgan_2011}, and attribute the fact that experimentally only small particles can fully lithiate to the slow Li-diffusion in the lithium rich phase (Li$_x$TiO$_2$, x \textgreater 0.5) \cite{Belak_2012, Yildirim_2011}, which has also been measured by NMR spectroscopy \cite{Borghols_2009a}. It has been suggested that the slow Li-diffusion makes the \ce{Li1TiO2}-layer act as a blocking layer, preventing further lithiation \cite{Borghols_2009a}. \\
Despite the large amount of research regarding anatase, a comprehensive explanation for its complex behaviour during lithiation is absent. 
In the present study a phase-field model free of fitted parameters for the lithiation of anatase \ce{TiO2} is presented, based on microscopic parameters from the literature, describing both first-order phase-transitions. 
The phase-field model for anatase consistently explains the experimentally observed phenomena, improving the understanding of  \ce{TiO2}-anatase during Li-intercalation, and shedding light on the limitations and possibilities for anatase as an electrode material. 
Considering that this is achieved with a model that only contains parameters from the literature, this provides important validation for the physical foundation of phase-field modelling, especially considering the complex behaviour of anatase during lithiation. Furthermore, our work strengthens the background of simulating materials that undergo multiple phase transitions during lithiation, which poses a considerable challenge for conventional computational models.

\section{Phase-field model for anatase}
\begin{figure}[htbp]
	\begin{center}
	\includegraphics[width=0.75\textwidth]{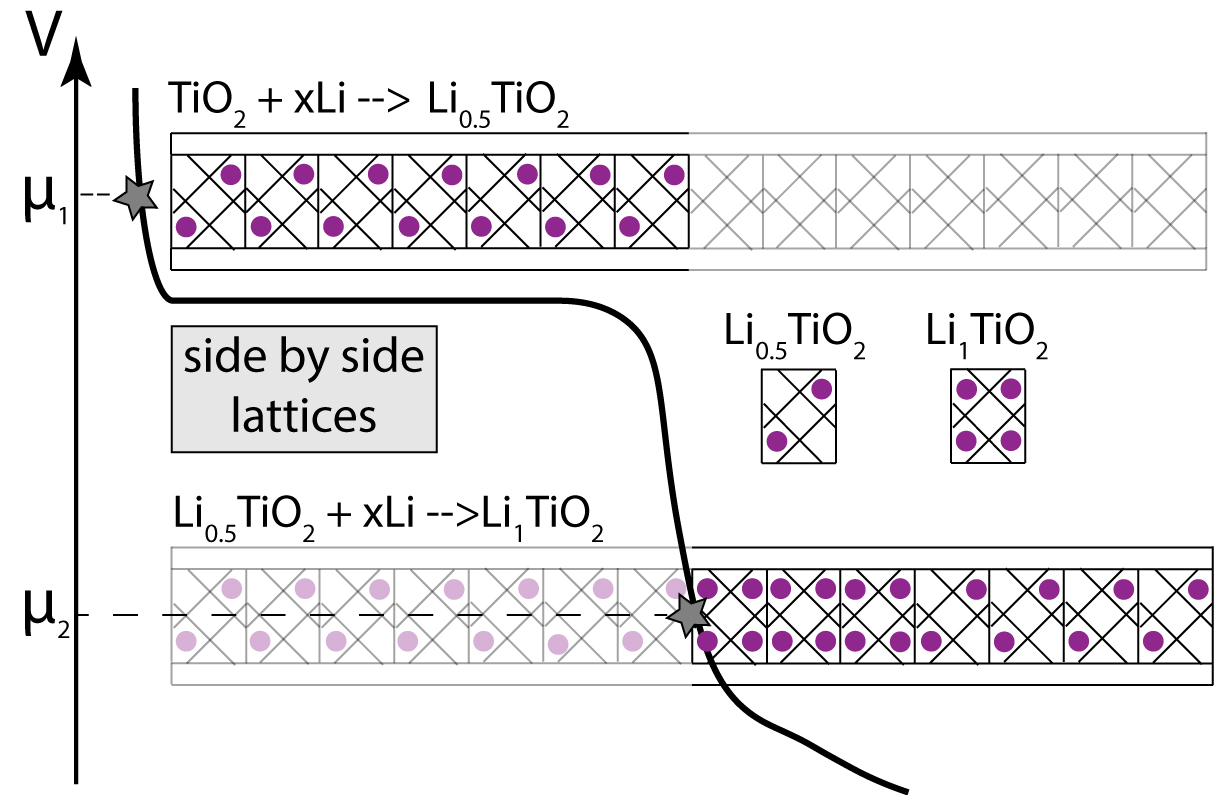}	
	\caption{Schematic representation of the two lattice phase-field model for lithiation in anatase \ce{TiO2}.}
	\label{fig:twolattice}
	\end{center}
\end{figure}

In this section the phase-field model for lithiation in anatase-\ce{TiO2} is presented. For a thorough background on phase-field modelling the reader is referred to several comprehensive publications \cite{Bazant_2013, Smith_MPET}.
The most important macroscopic output variable for phase-field modelling of batteries is the measured cell voltage ($V_\mathrm{cell}$) given by:
\begin{equation}
	V_\mathrm{cell} = - \Delta\mu/e + \eta_\mathrm{cell} 
	\label{eq:V_out}
\end{equation}
where $\Delta\mu$ is the change in chemical potential, $e$ the electron charge, and $\eta_\mathrm{cell}$ is the total cell overpotential. \\
The change in chemical potential is the difference in free energy of lithium at the solid-electrolyte interfaces of the anode and cathode material. For the simulated Li-metal/anatase system $\Delta\mu$ is the change in free energy for the reaction:  
\begin{align}
	\ce{x Li + TiO2 <-> Li_{x}TiO2}
\end{align}
Li-metal is defined as the reference electrode, and consequentially its chemical potential is defined as zero. Furthermore, the overpotential of the Li-metal electrode is assumed to be zero, which appears to be a good approximation given the small overpotentials experimentally observed for Li-metal electrodes \cite{Munich_1994}. These simplifications lead to a phase-field model in which only the lithiation of anatase needs to be taken into account to describe $V_\mathrm{cell}$. \\
The two first-order phase-transitions occurring upon lithiation of anatase can be considered as two independent chemical reactions, since (locally) the two reactions cannot occur simultaneously, and can therefore be described by two independent lattices having their own free energy functional. The first lattice represents the reaction \ce{TiO2 + 0.5 Li+ + e- <-> Li_{0.5}TiO2}, and the second lattice represents the reaction \ce{Li_{0.5}TiO2 + 0.5 Li+ + e- <-> Li1TiO2}.
Similar to the phase-field model for graphite \cite{Smith_2017} this requires the introduction of two parameters ($c_1$ and $c_2$) that describe the Li-concentration in the first and second lattice, respectively. 
In both lattices the Gibbs free energy ($g(\tilde{c_{i}})$) is described by a Cahn-Hilliard regular solution model \cite{Bazant_2013}: 
\begin{equation}
	g(\tilde{c_{i}}) = k_\mathrm{B} T \left( \tilde{c_{i}} \ln(\tilde{c_{i}}) + (1 - \tilde{c_{i}}) \ln(1 - \tilde{c_{i}}) \right) + \Omega_{i} \tilde{c_{i}} (1 - \tilde{c_{i}}) + \frac{1}{2} \frac{\kappa_i}{c_\mathrm{max}} |\nabla \tilde{c_{i}}|^2 + \tilde{c_{i}} \mu_{i}^{\Theta}
	\label{eq:g_lattice}
\end{equation}
where $k_\mathrm{B}$ is Boltzmann's constant, $T$ the temperature in Kelvin, $\tilde{c_{i}}$ the normalised concentration in lattice $i$ ($\tilde{c_{i}} =\frac{c_{i}}{c_\mathrm{max}}$), $\Omega_{i}$ the enthalpy of mixing, $\mu_{i}^{\Theta}$ the reference potential versus. Li/\ce{Li+}, and $\kappa_i$ the gradient penalty parameter. \\
The first term in Equation \ref{eq:g_lattice} describes the entropy change upon adding Li-ions ($\tilde{c_{i}}$) and removing Li-vacancies ($1 - \tilde{c_{i}}$). 
The enthalpy of mixing ($\Omega_{i}$) describes the interactions between Li-atoms in an intercalation material. Positive values for $\Omega_{i}$ correspond to attractive forces between Li-atoms, favouring phase separation into the end member phase (a Li-rich and a Li-poor phase).
The $\kappa_i$-term represents the energy penalty for the existence of concentration gradients when phase-separation occurs, with larger values for $\kappa_i$ leading to a wider interface region between Li-rich and Li-poor phases. 
Large entropy and $\kappa_i$-terms in Equation \ref{eq:g_lattice} promote solid solution behaviour, while a large $\Omega_{i}$-term will promote phase-separation. Which term dominates, and thus determines the phase-behaviour of a material, not only depends on the values of the parameters, but also on C-rate, temperature, and particle size \cite{Bai_2011, Borghols_2009b, Zhang_2014, Zhang_2015a}. \\
The diffusional chemical potential ($\mu_{i}$) of lithium in anatase is given by the variational derivative of the free energy with respect to concentration \cite{Smith_MPET}: 
\begin{equation}
	\mu_{i} = \frac{\partial g_{i}}{\partial \tilde{c_{i}}} - \nabla \cdot \frac{\partial g_{i}}{\partial \nabla  \tilde{c_{i}}}
\end{equation}
Using Equation \ref{eq:g_lattice} this gives:
\begin{equation}
	 \mu_{i} = k_\mathrm{B} T \ln(\frac{\tilde{c_{i}}}{1 - \tilde{c_{i}}}) + \Omega (1 - 2 \tilde{c_{i}}) - \frac{\kappa_i}{c_\mathrm{max}} \nabla^{2} \tilde{c_{i}} + \mu_{i}^{\Theta} 
	 \label{eq:mu}
\end{equation}
From the diffusional chemical potential the flux of lithium ($F_{i}$) through the particle can be determined based on the gradient of the diffusional chemical potential ($\nabla \mu_i$) \cite{Bazant_2013}: 
\begin{equation}
        F_i = -M_i c_i \nabla \mu_i = - \frac{D_{i} c_\mathrm{max} \tilde{c_i}}{k_\mathrm{B} T} \nabla \mu_i
	\label{eq:mobil_flux}
\end{equation}
where $M_i$ is the mobility and $D_{i}$ is the tracer diffusivity. \\
It is known that the lithium diffusion in \ce{TiO2} anatase is dependent on the lithium concentration, but the effect of the Li-concentration on Li-diffusion is unclear. Papers with calculations show contradicting results, with some reporting a large \cite{Belak_2012} or small \cite{Lunell_1997} increase in activation energy for Li-diffusion with increasing Li-concentration, while others show a large \cite{Tielens_2005} or small \cite{Yildirim_2011} decrease in activation energy at higher Li-contents. Experiments by Sussman et al. \cite{Sussman_2014} show a decrease in Li-diffusivity with increasing Li-content, although the magnitude of this effect strongly depends on the synthesis procedure.
The simplest approximation for the tracer diffusivity on a lattice is proportional to the vacancy concentration, $D_i \sim (1 - c_i)$, in order to account for site exclusion \cite{Bazant_2013, Ferguson_2012} and for thermodynamic consistency with binary species mixing \cite{Nauman_2001}, but we find that this model is not able to reproduce the general features of the experimental voltage profiles. On the other hand, ab initio calculations predict a much stronger concentration dependence, where the chemical diffusivity drops by many orders of magnitude between the \ce{TiO2}-, \ce{Li_{0.5}TiO2}- and \ce{Li1TiO2}-phases \cite{Belak_2012}, thus indicating stronger cooperative diffusion barriers. As a first approximation of such effects, we introduce a simple power-law correction: 
\begin{equation}
	D_{i} = D^*_i \frac{(1 - \tilde{c_i})}{\tilde{c_i}}
	\label{eq:tracer_def}
\end{equation}
where $D_{i}^{*}$ is the reference tracer diffusivity in lattice $i$ at $c_i = 0.5$. Despite the unphysical divergence at $c_i = 0$, the diffusivity effectively saturates at realistic values in our phase-field simulations, since the regular solution model only allows small, but finite, concentrations. Combining Equations \ref{eq:mobil_flux} and \ref{eq:tracer_def}, the flux of lithium is given by:
\begin{equation}
        F_{i} = - \frac{D_{i}^{*} c_\mathrm{max} (1 - \tilde{c_{i}})}{k_\mathrm{B} T} \nabla \mu_{i}
	\label{eq:Li_flux}
\end{equation} 
which is simply proportional to the vacancy concentration. The implied chemical diffusivity $D_{i}^\mathrm{chem} = D_{i}^{*} \left( \frac{(1-\tilde{c_{i}})}{\tilde{c_{i}}} - 2 \Omega_i (1-\tilde{c_{i}})^2 \right)$, is negative in the spinodal regions of thermodynamic instability, while capturing the strongly decreasing trend across the solid solution phases \cite{Belak_2012}, similar to the experiments of Sussman et al. \cite{Sussman_2014}. We find that this model is also capable of providing a good fit of the experimental voltage profiles. \\
Using Equation \ref{eq:mu} and \ref{eq:Li_flux} the behaviour of lithium inside anatase particles can be described, but to determine the battery voltage and influx of lithium the charge-transfer reaction at the electrode-electrolyte interface must also be described. This can be done using the Butler-Volmer equation \cite{Smith_MPET}:  
\begin{equation}
	I_i = \frac{k_{0} n e (a_{O} a_e^n)^{1 - \alpha} a_{R,i}^{\alpha}}{\gamma^{\ddag}_{i}} \left( \exp \left( - \frac{\alpha e \eta_{\mathrm{eff},i}}{k_\mathrm{B} T} \right) - \exp \left( \frac{(1-\alpha) e \eta_{\mathrm{eff},i}}{k_\mathrm{B} T} \right) \right)
	\label{eq:charge_transfer}
\end{equation} 
where  $I_i$ is the current density in lattice $i$, $k_{0}$ the reaction rate constant per surface area of the particle, $\alpha$ the reaction symmetry factor (assumed to be 0.5), $n$ the number of electrons participating in the reaction (one in this case), and $e$ the electronic charge. \\
The charge-transfer overpotential ($\eta_{\mathrm{eff},i}$) is defined as: $e \eta_{\mathrm{eff},i} = \mu_{R,i} - \mu_O$, where $\mu_{R,i}$ (the chemical potential of the reduced state of Li) is obtained from Equation \ref{eq:mu}, $\mu_O$ (the chemical potential of the oxidised state of Li) depends on the Li-concentration in the electrolyte ($c_\mathrm{lyte}$) and is approximated using a dilute electrolyte model as: $\mu_{O} = k_\mathrm{B} T ln(c_\mathrm{lyte})$. 
The activity of the oxidised state ($a_{O}$) is equal to $c_\mathrm{lyte}$, the activity of the electrons $a_e$ is taken to be unity. The activity of the reduced state ($a_{R,i}$) depends on the diffusional chemical potential ($\mu_{i}$) of lithium inside the particle: $a_{R,i} = \exp \left( \frac{\mu_{i} -\mu^{\Theta}}{k_\mathrm{B} T} \right)$, and the activity of the transition state ($\gamma^{\ddag}_{i}$) depends on the concentration of lithium-vacancies \cite{Bai_2011}: $\gamma^{\ddag}_{i} = \frac{1}{1 - \tilde{c_{i}}}$. During constant current simulations the applied current ($I_\mathrm{applied} = (I_1 + I_2)*\mathrm{Area}$) is known, thus $\eta_{\mathrm{eff},i}$ can be calculated. \\
The charge-transfer overpotential given by Equation \ref{eq:charge_transfer} describes the thermodynamic driving force for a lithium-ion to enter/leave the electrode particle. All the terms in Equation \ref{eq:charge_transfer} depend on the diffusional chemical potential or concentration of lithium, i.e. the size of the charge-transfer overpotential is determined by the Li-concentration of the electrode and electrolyte near the electrode-electrolyte interface. 
Using the equations given above, the diffusional chemical potential of Li in the particles (Equation \ref{eq:mu}), the Li-flow through the particle (Equation \ref{eq:Li_flux}), and the Li-flow into the anatase particles (Equation \ref{eq:charge_transfer}) can be described. Using the appropriate set of boundary conditions this set of equations can be solved \cite{Zeng_2014}, ultimately giving the cell voltage (Equation \ref{eq:V_out}), and the Li-concentration inside the anatase particles. \\
In contrast to the graphite phase-field model \cite{Ferguson_2014, Guo_2016}, which directly couples the two phase transitions, the anatase model consists of two independent lattices. The reason for this are the very different physical properties of the first and second phase transition in anatase, while in graphite the only difference between the two phase transitions (relevant to the phase-field model) is the voltage.
To describe the two phase transitions in anatase different parameters are necessary, which can be implemented by  introducing two independent lattices with different physical properties, schematically shown in Figure \ref{fig:twolattice}. At the start of the lithiation process the first phase transition will occur due to its higher intercalation potential, filling the first lattice with Li-ions. When the first lattice fills the charge-transfer overpotential will increase (lowering the voltage), because it gradually becomes harder to add more Li-ions. When the intercalation potential of the second lattice is reached lithiation of the second lattice becomes favourable, and the second phase transition will start.  
There are no interaction terms between the two lattices, since the effect the first lattice has on the second is already incorporated by the different parameters that are used. In Table \ref{tab:parameters} all parameters and their values from literature are listed, as well as what each one is based on. \\
\begin{table}[htbp]
	\begin{center}
	\begin{tabular}{| c | c | c | l |}
	\hline
	Parameter & Value & Units & Based on \\ \hline
	$\mu_{1}^{\Theta}$ & 1.82 & V & GITT experiments \cite{Shen_2014, Sudant_2005} \\ \hline	
	$\mu_{2}^{\Theta}$ & 1.56 & V & Electrochemical experiments at 120$^{\circ}$C \cite{Zachau_1988, Macklin_1992} \\ \hline	
	$D_1^*$ & $1*10^{-16}$ & cm$^2$/sec & Electrochemical experiments \cite{Wang_2007, Hwan_2015, Lindstrom_1997} \\ \hline	
	$D_2^*$ & $1*10^{-17}$ & cm$^2$/sec & Force-field molecular dynamics simulations \cite{Yildirim_2011} \\ \hline	
	$\Omega_1$ & $0.6*10^{-20}$ & J/Li & DFT calculations \cite{Belak_2012, Morgan_2011} \\ \hline
	$\Omega_2$ & $1.6*10^{-20}$ & J/Li & DFT calculations \cite{Belak_2012, Morgan_2011} \\ \hline
	$\kappa_{1}$ & $5.3*10^{-8}$ & J/m & Phase diagram (see text) \cite{Wagemaker_2007A} \\ \hline
	$\kappa_{2}$ & $0.8*10^{-8}$ & J/m & Phase diagram (see text) \cite{Wagemaker_2007A} \\ \hline
	$c_\mathrm{max}$ &	$1.419*10^{28}$	& Li/m$^3$ & Neutron diffraction \cite{Wagemaker_2003} \\ \hline 
	$k_0$ & $0.049$ & A/m$^2$ & NMR experiments \cite{Ganapathy_2009}, also see supporting info \\ \hline	
	\end{tabular}
	\caption{The values of the parameters and on what information they are based.}
	\label{tab:parameters}
	\end{center}
\end{table}
Normally, several physical parameters necessary for phase-field modelling are not available, either experimentally or computationally, and are therefore fitted by optimising the phase-field model towards experimental voltage profiles. 
Intensive research towards lithiation of anatase \ce{TiO2} during the last decades makes it possible to quantify all parameters necessary for the present phase-field model. Thus providing a unique opportunity to validate a phase-field model, using only parameters from the literature, with micro- and macroscopic observations. \\
The reference potentials $\mu_{1}^{\Theta}$ and $\mu_{2}^{\Theta}$ are based on literature data which are closest to equilibrium conditions at room temperature. $\mu_{1}^{\Theta}$ is based on GITT measurements \cite{Shen_2014, Sudant_2005} performed at room temperature, but for the second phase transition equilibrium is not even reached during the reported GITT measurements. Therefore the value for $\mu_{2}^{\Theta}$ is taken from experiments  performed at 120$^{\circ}$C, in which the second plateau indicates that equilibrium was reached \cite{Zachau_1988, Macklin_1992}. \\
For anatase electrodes electrochemical experiments have reported diffusivities in the first lattice between $5*10^{-10}$ and $4*10^{-20}$ cm$^2$/sec \cite{Lindstrom_1997, Wang_2007, Shin_2012, Hwan_2015, Kavan_2014, Zhiyuan_2014, Sussman_2014, Yang_2015}, and changes of 2 orders of magnitude during charging have been reported \cite{Sussman_2014}. 
Furthermore, calculations on Li-diffusion in anatase also show strongly differing results \cite{Belak_2012, Lunell_1997, Tielens_2005, Yildirim_2011}, and NMR experiments indicate that diffusion over the interface between the anatase and Li-titanate phases \cite{Wagemaker_2001, Wagemaker_2002} is the limiting step. For the second lattice no experimental value for the diffusivity has been reported, but NMR experiments \cite{Borghols_2009a} and calculations \cite{Belak_2012, Yildirim_2011} have shown that it is smaller than in the first lattice . \\
Given the large range of values in the literature for Li-diffusion in \ce{TiO2} anatase several values from the literature were used for testing, after which the simulated voltage profiles were compared to experimental ones. The best agreement with experiments was obtained using a value of $1*10^{-16}$ cm$^2$/sec for $D_1^*$, which has been reported by several experimental studies \cite{Wang_2007, Hwan_2015, Lindstrom_1997}. For $D_2^*$ a value of $1*10^{-17}$ cm$^2$/sec gave the best results, which is obtained from molecular dynamics simulations \cite{Yildirim_2011}. \\
Since it is impossible to experimentally measure the enthalpy of mixing the values for $\Omega_1$ and $\Omega_2$ are based on DFT calculations \cite{Belak_2012, Morgan_2011}, the values were determined by the difference between the convex hull and the configurational energies. 
Values of $\kappa_{1}$ and $\kappa_{2}$ are based on the particle size at which two phase coexistence inside a particle no longer occurs \cite{Wagemaker_2007A}. This means that for the radial 1D-model presented here the interface width ($\lambda_i$) corresponds to half of the particle size \cite{Wagemaker_2011}, which gives interface widths of 25 and 6 nm for lattice 1 and 2, respectively. Using these interface widths $\kappa_i$ can be calculated using \cite{Bazant_2013}: 
\begin{equation}
	\kappa_i = \lambda_i^2 c_\mathrm{max} \Omega_i 
\end{equation}
The maximum Li-concentration ($c_\mathrm{max}$) is calculated based on the four Li-sites per unit cell upon complete lithiation and the volume \cite{Wagemaker_2003} of the unit cell of \ce{Li_{0.5}TiO2} divided over the two lattices. There is a small volume change (3\%) upon lithiation from \ce{TiO2} to \ce{Li_{0.5}TiO2}, but given that the volume differs by just 0.1\% between \ce{Li_{0.5}TiO2} and \ce{Li1TiO2}, the volume of \ce{Li_{0.5}TiO2} is the best approximation over the range of possible Li-concentrations. \\
The equilibrium charge transfer constant ($k_0$) is typically not known because it is very hard to experimentally distinguish the Li reaction between electrolyte and electrode from other processes occurring simultaneously. However, using NMR this has been shown to be possible by Ganapathy et al. \cite{Ganapathy_2009}, reporting $k_0$ for the \ce{Li_{0.5}TiO2} phase, which is at present assumed to be representative for both lattices. 
Please note that physically the second phase transformation can only occur after the first phase transformation has happened (locally), and although this is not formally implemented in the model the 0.26 V lower insertion potential satisfies this condition during the simulations. 
To keep the model simple all properties of anatase were assumed to be isotropic, a reasonable assumption given the 3D-diffusion pathway \cite{Yildirim_2011} and small changes in lattice parameters upon lithiation \cite{Lafont_2010}. 
During lithiation of anatase \ce{TiO2} the interfaces with the Li-rich phase are predicted to occur along strain invariant planes \cite{Belak_2012}. For this reason it was assumed that the role of strain and of stress assisted diffusion can be neglected in the present 1D-simulations. \\
The simulations were performed using a modified version of the publicly available MPET code 
\cite{MPET_code}, in which the coupled differential equations are solved using the DAE tools package \cite{Nikolic_2016}. A 1D-model along the radial direction of the particles is used for the simulations, and unless stated otherwise simulations were performed on a single particle with a radius of 20 nm., a C-rate of 0.5C, a temperature of 298 K, and a cut-off voltage of 1 V vs. Li/Li$^+$. For the single particle simulations the Li-concentration in the electrolyte was assumed to be constant, in multi-particle simulations the dilute electrolyte model as implemented in the MPET code \cite{Smith_MPET} was used to describe the Li-concentration in the electrolyte.

\section{Results}
The results of the phase-field model for lithiation of anatase \ce{TiO2} are compared to a broad spectrum of experimental results available in literature. The aim here is qualitative validation of the phase-field model and understanding of the physical processes that determine the performance of anatase electrodes.
A qualitative validation rather than a quantitative validation is motivated by the many experimental parameters that affect the performance of anatase electrodes, resulting in a wide distribution of performances, even for equivalent electrochemical conditions \cite{Sondergaard_2015}.

\subsection{Impact of lithiation rate}
Similar to other electrode materials the (dis)charge rate, expressed in the C-rate (a 2C rate corresponds to (dis)charge of the full theoretical battery capacity in 1/2 hour, 1C in 1 hour, 0.5C in 2 hours, etc.) has a large impact on the voltage profile of anatase electrodes. Typically the capacity drops by approximately 25\% when going from cycling at 0.5C to 1C \cite{Yang_2015}, and at higher C-rates a significant drop of the plateau voltage is detected \cite{Kim_2010}. \\
\begin{figure}[htbp]
	\begin{center}
		\begin{subfigure}{0.45\textwidth}
			\includegraphics[width=\textwidth]{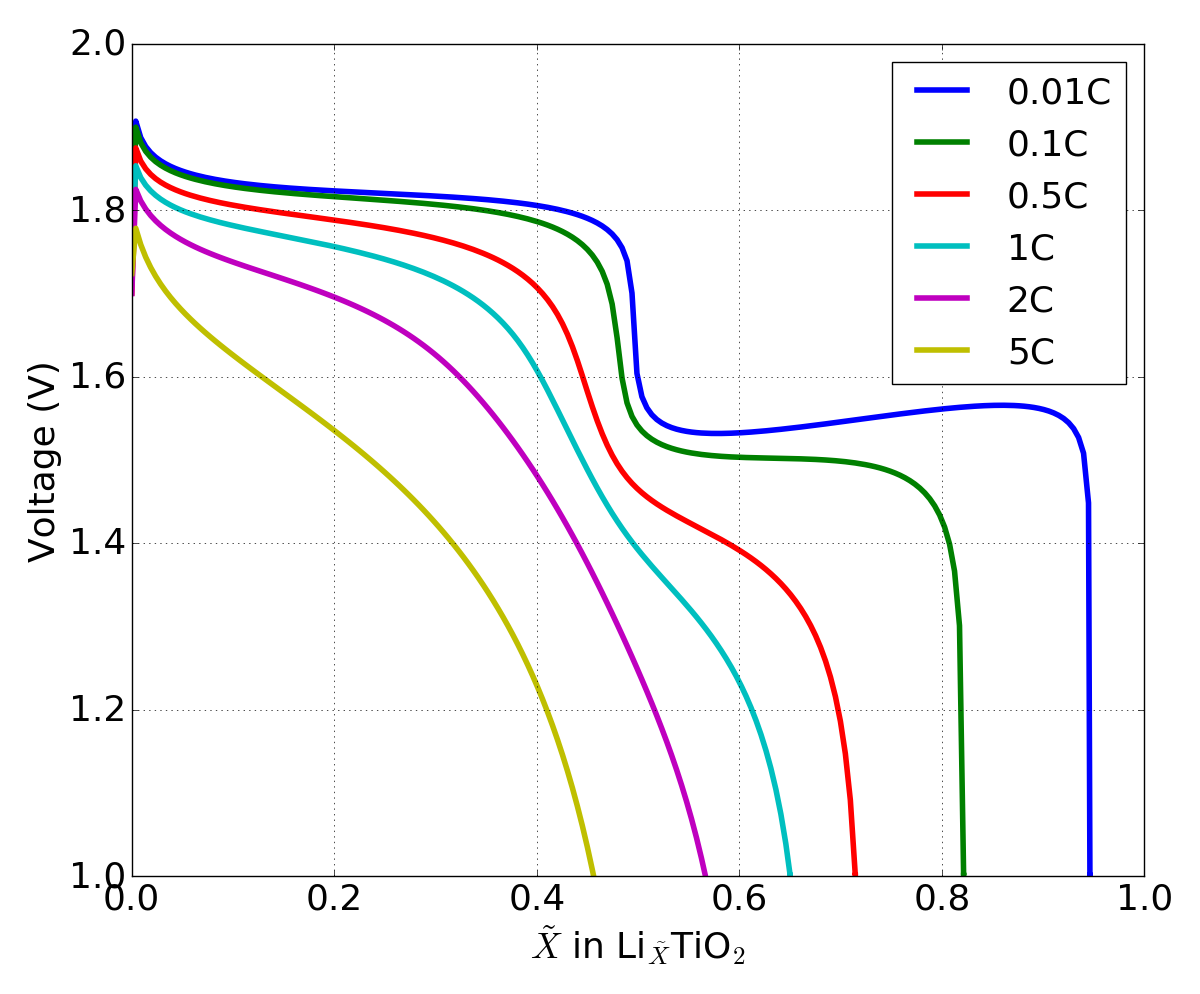}
			\caption{~} 
		\end{subfigure}
		\begin{subfigure}{0.45\textwidth}
			\includegraphics[width=0.8\textwidth]{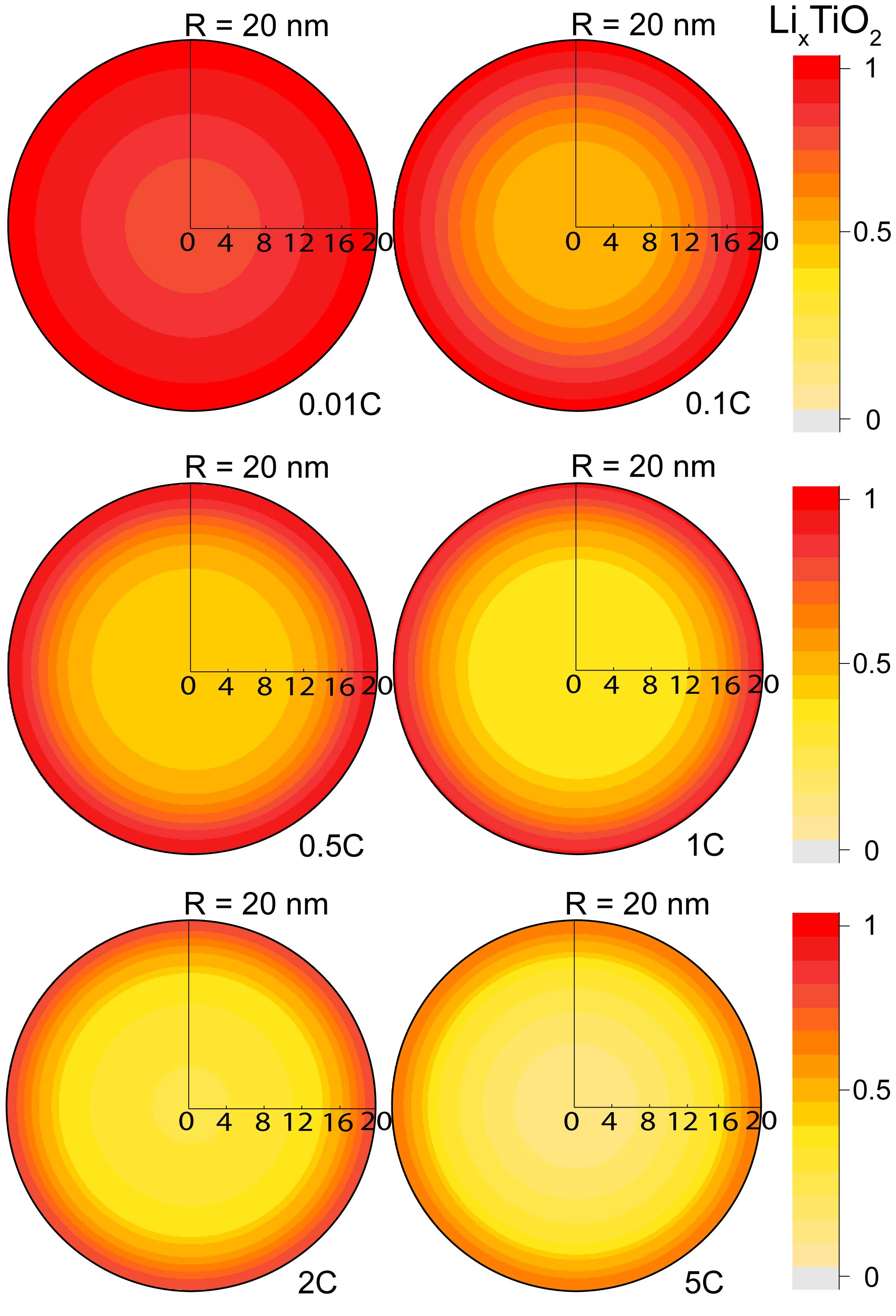} 
			\caption{~}
		\end{subfigure}
		\caption{(a) Voltages profiles versus average concentration ($\tilde{X}$) and (b) final concentration profiles in the particle at different C-rates.}
		\label{fig:c-rates}
	\end{center}
\end{figure}
The drop in capacity and voltage with increasing lithiation rate are both consistently reflected in the simulated voltage profiles for a single anatase particle with a radius of 20 nm shown in Figure \ref{fig:c-rates}a. 
At 5C the simulation leads to a maximum composition of \ce{Li_{0.45}TiO2}, increasing to \ce{Li_{0.7}TiO2} at 0.5C, and at 0.01C the anatase particle is almost completely lithiated. 
With increasing lithiation rate the increasing charge-transfer overpotential results in a voltage drop in Figure \ref{fig:c-rates}a, driven by limited Li transport away from the surface. 
The significant decrease in voltage upon increasing the current from 0.01C to 0.1C indicates poor Li-ion kinetics in anatase, in particular considering the small particle radius of 20 nm. Generally, 0.1C results in close to equilibrium conditions in most nano-structured electrode materials, whereas in anatase \ce{TiO2} Li-ion kinetics still restricts the capacity at this rate. \\
The large voltage drop at high C-rates for the second voltage plateau indicates that the formation of the \ce{Li1TiO2} phase limits the charge transport away from the surface, thus increasing the charge-transfer overpotential. An estimate for the time it takes a Li-ion to reach the center of the particle can be obtained by calculating the characteristic diffusion time \cite{Bruce_2008}, $t_\mathrm{D}$, defined as: 
\begin{equation}
        t_\mathrm{D} = \frac{R^2}{D}
	\label{eq:charac_time}
\end{equation}
where $R$ is the particle radius and $D$ the diffusion constant. 
For a particle with a radius of 20 nm, the characteristic time for diffusion in the \ce{Li_{0.5}TiO2} and \ce{Li1TiO2} are approximately $4*10^4$ and $4*10^5$ seconds, respectively. 
For the first voltage plateau, the phase transition towards \ce{Li_{0.5}TiO2}, this roughly corresponds to 0.1C. Therefore, at this rate the entire voltage plateau associated with the first phase transition should be observed, consistent with Figure \ref{fig:c-rates}a. \\
For the second voltage plateau, the phase transition towards \ce{Li1TiO2}, the characteristic time roughly corresponds to 0.01C, consistently reflected by the complete appearance of the second voltage plateau at this C-rate in Figure \ref{fig:c-rates}a. At time-scales shorter than $t_D$ the lithium ions are unable to reach the centre of the particle within the given time, i.e. kinetic limitations will restrict the capacity and decrease the cell voltage, as visible at higher C-rates in Figure \ref{fig:c-rates}a. 
This is confirmed by the Li-ion concentration profiles shown in Figure \ref{fig:c-rates}b, in which at 0.01C most of the particle is transformed to the \ce{Li1TiO2}-phase. At 0.5C only a thin layer at the surface approaches the maximum composition \ce{Li1TiO2}, and the inner 10 nm is only transformed to the \ce{Li_{0.5}TiO2}-phase. At 5C this effect is augmented, with a large part of the particle having a Li-concentration below $x = 0.5$, and only near the surface the Li-concentration exceeds $x = 0.5$. \\
For experimental electrochemical lithiation at room temperature only the onset of the second voltage plateau is observed, as consistently predicted by the simulation at 0.5C in Figure \ref{fig:c-rates}a. Raising the temperature to 120 $^{\circ}$C will significantly enhance Li-diffusion, largely lifting the diffusional limitations of the second lattice, resulting in a clear experimental observation of the second voltage plateau at 120 $^{\circ}$C \cite{Zachau_1988, Macklin_1992}.
The simulation at a very slow lithiation rate, 0.01C, predicts that particles with a 20 nm. radius can also be fully lithiated at room temperature. 
Although no experimental evidence showing this appears present (to the best of our knowledge), GITT measurements on particles with a diameter of 130 nm have been shown to reach full lithiation \cite{Shen_2014}. \\
The increasing voltage at 0.01C is caused by simulating only a single particle. For a single particle the voltage follows the spinodal potential, giving an upwards slope in the voltage profile \cite{Orvananos_2015}. When multiple particles are present interparticle phase-separation can occur, which smooths the voltage curve \cite{Ferguson_2014, Orvananos_2015}.
As demonstrated in the supporting information (Figure S1) the upward tilt disappears when the simulations are performed on multiple particles. \\
In order to gain understanding of the rate limiting kinetic mechanism in anatase \ce{TiO2} electrodes, multi-particle simulations were performed. A 50 $\mu$m porous electrode was separated into five volumes connected in series reflecting different depths inside the electrode, and each volume contained five particles with a 20($\pm$2) nm radius.
The chosen rate is 2C, since at this rate the performance of the material already results in a significant decrease in the capacity and voltage, as shown in Figure \ref{fig:c-rates}.
\begin{figure}[htbp]
	\begin{center}
	\includegraphics[width=0.75\textwidth]{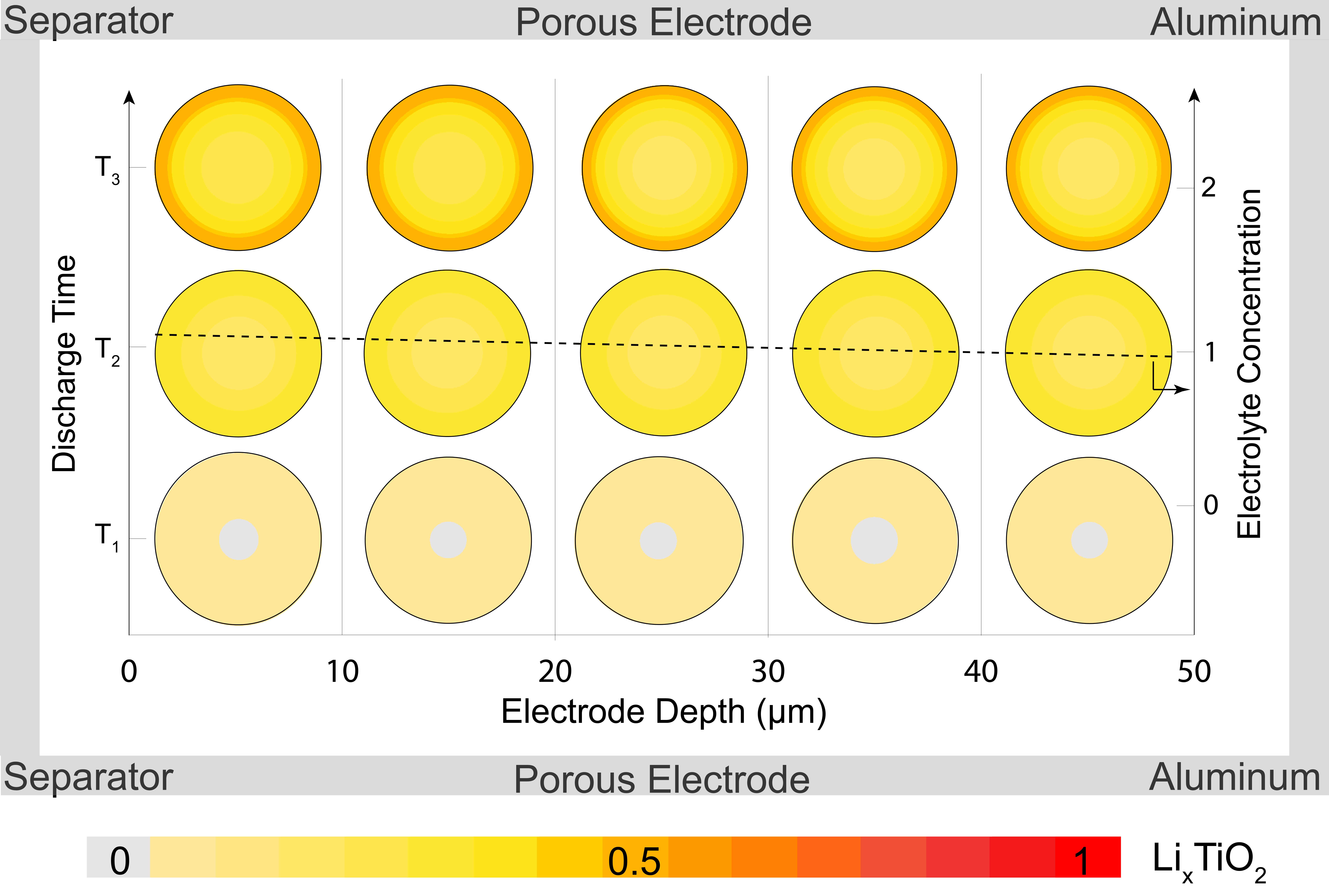}	
	\caption{Li-ion concentration profiles in the electrolyte and in electrode particles at different electrode depths at different times during 2C discharge. The filling fraction in the displayed particles is the average of that in the simulated particles in each volume.}
	\label{fig:multiparticle}
	\end{center}
\end{figure}
The results of the multi-particle simulations in Figure \ref{fig:multiparticle} demonstrate that the lithiation process proceeds concurrently at any given depth of the electrode. All particles are transforming simultaneously, which implies that the Li-ion diffusion in a single anatase grain is rate limiting, even when the particles are nano sized. 
In an actual electrode the consequence is that all grains are actively participating in delivering the current, thus electrode performance can be improved significantly by increasing the Li-diffusivity in the anatase lattice.
For comparison, in the simulation shown in Figure \ref{fig:multiparticle} the typical diffusion time ($t_\mathrm{D}$) through the electrolyte is 10 seconds (using an ambipolar diffusivity of $2.5*10^{-6}$ cm$^2$/sec), three orders of magnitude below the $t_\mathrm{D}$ inside the particles. \\
Furthermore, Singh et al. \cite{Singh_JPC} have shown that \ce{TiO2} anatase without electron conducting additives has excellent cycling behaviour. Thus electrode performance of \ce{TiO2} anatase can primarily be improved by increasing the Li-diffusivity in the anatase lattice, and only slightly by enhancing the ionic and electronic wiring.
This behaviour differs from other electrode materials, such as \ce{LiFePO4}, \ce{Li4Ti5O12}, and \ce{LiCoO2}, where it has been shown that for full electrodes the ionic and electronic wiring dominate the internal resistance from small to large (dis)charge rates \cite{Liu_2016, Strobridge_2015, Li_2015, Singh_2013, Kim_2013}. \\
Interestingly, in \ce{LiFePO4} the increasing overpotential when increasing the C-rate widens the interface between the coexisting phases, at some critical rate leading to a solid-solution reaction as predicted by phase-field modelling \cite{Bai_2011} and observed experimentally \cite{Zhang_2014, Zhang_2015b, Liu_2014}. In anatase \ce{TiO2} the phase interface also widens when increasing the C-rate; however, even at large C-rates the \ce{Li1TiO2} phase forms at the particle surface, because of the poor Li-ion diffusivity. Thus the model predicts that anatase \ce{TiO2} will undergo phase separation regardless of the imposed current.

\subsection{Impact of Li-diffusion coefficient}
The impact of the C-rate on the capacity and voltage for anatase \ce{TiO2} electrodes reveals that the Li-ion diffusivity in the anatase lattice is the key limiting factor.
Experimentally the Li-diffusivity has been increased by annealing in argon \cite{Wang_2011}, by hydrogen treatment \cite{Shin_2012}, and by \ce{Ti^{3+}} doping \cite{Ren_2014}, which all increase the amount of oxygen vacancies in \ce{TiO2}. Impedance measurements have shown that this can increase the Li-diffusivity by one order of magnitude \cite{Shin_2012}.  
\begin{figure}[htbp]
	\begin{center}
		\begin{subfigure}{0.45\textwidth}
			\includegraphics[width=\textwidth]{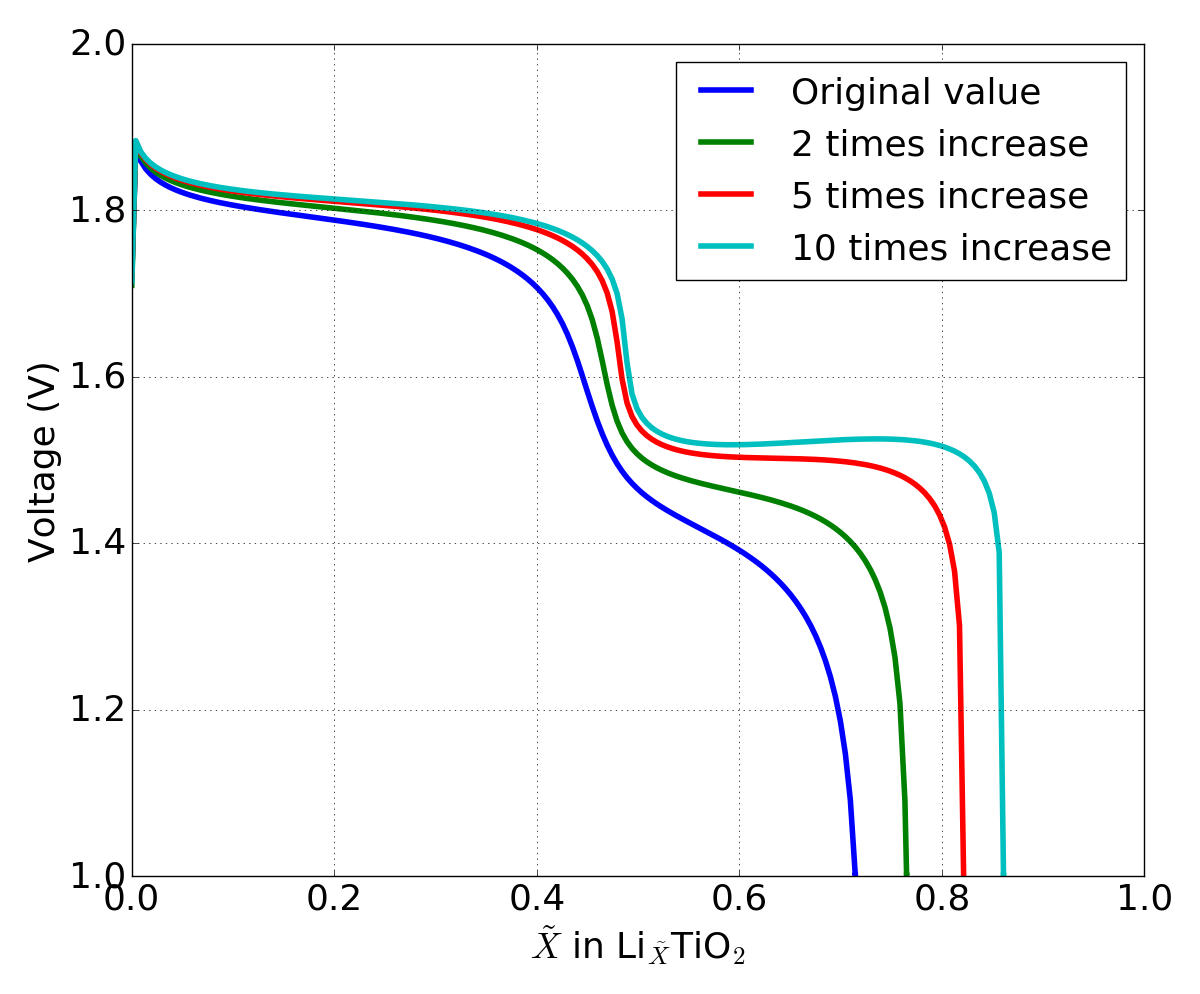} 
			\caption{~}
		\end{subfigure}		
		\begin{subfigure}{0.45\textwidth}
			\includegraphics[width=\textwidth]{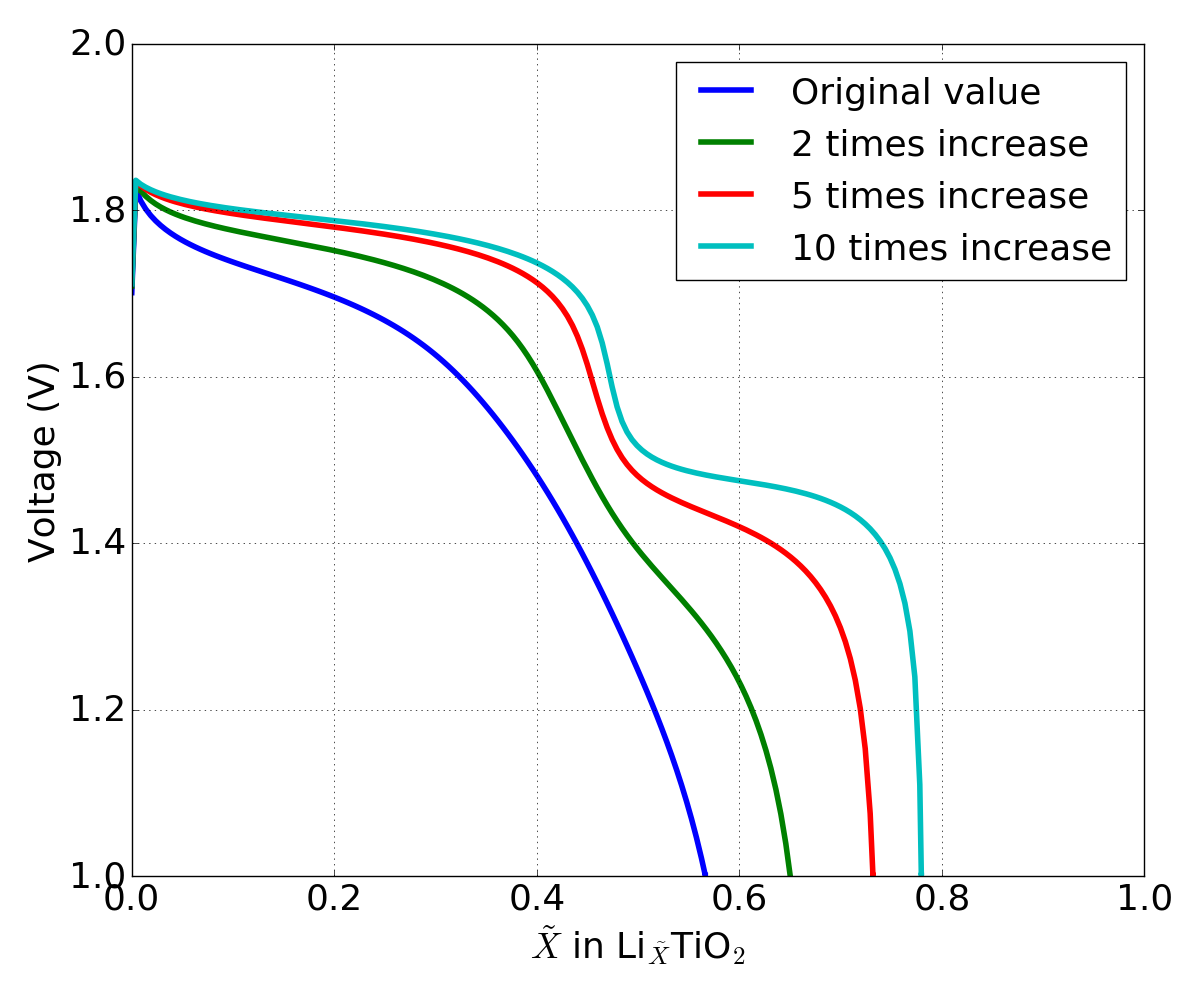} 
			\caption{~}
		\end{subfigure}
		\begin{subfigure}{0.45\textwidth}
			\includegraphics[width=\textwidth]{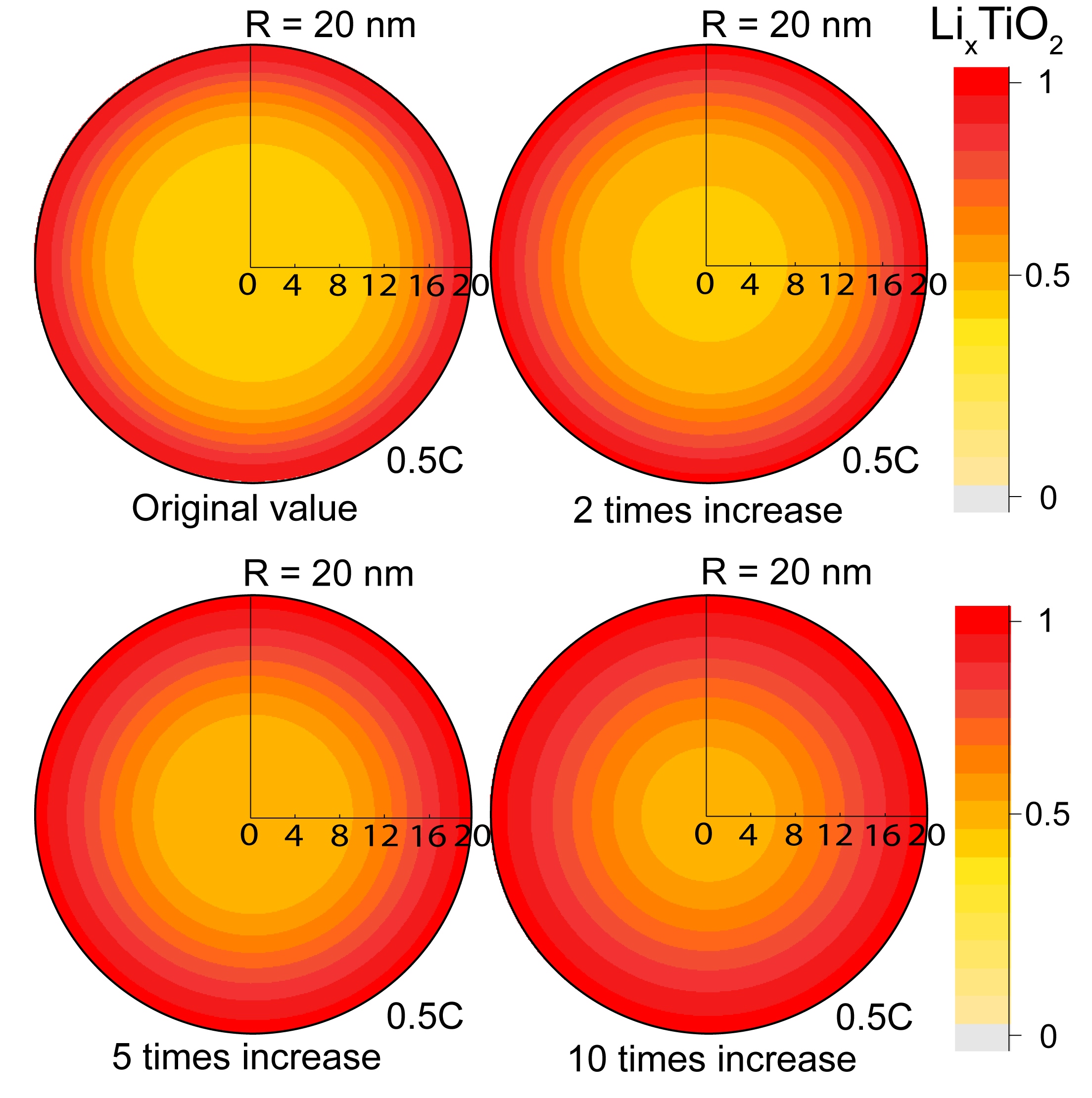} 
			\caption{~}
		\end{subfigure}		
		\begin{subfigure}{0.45\textwidth}
			\includegraphics[width=\textwidth]{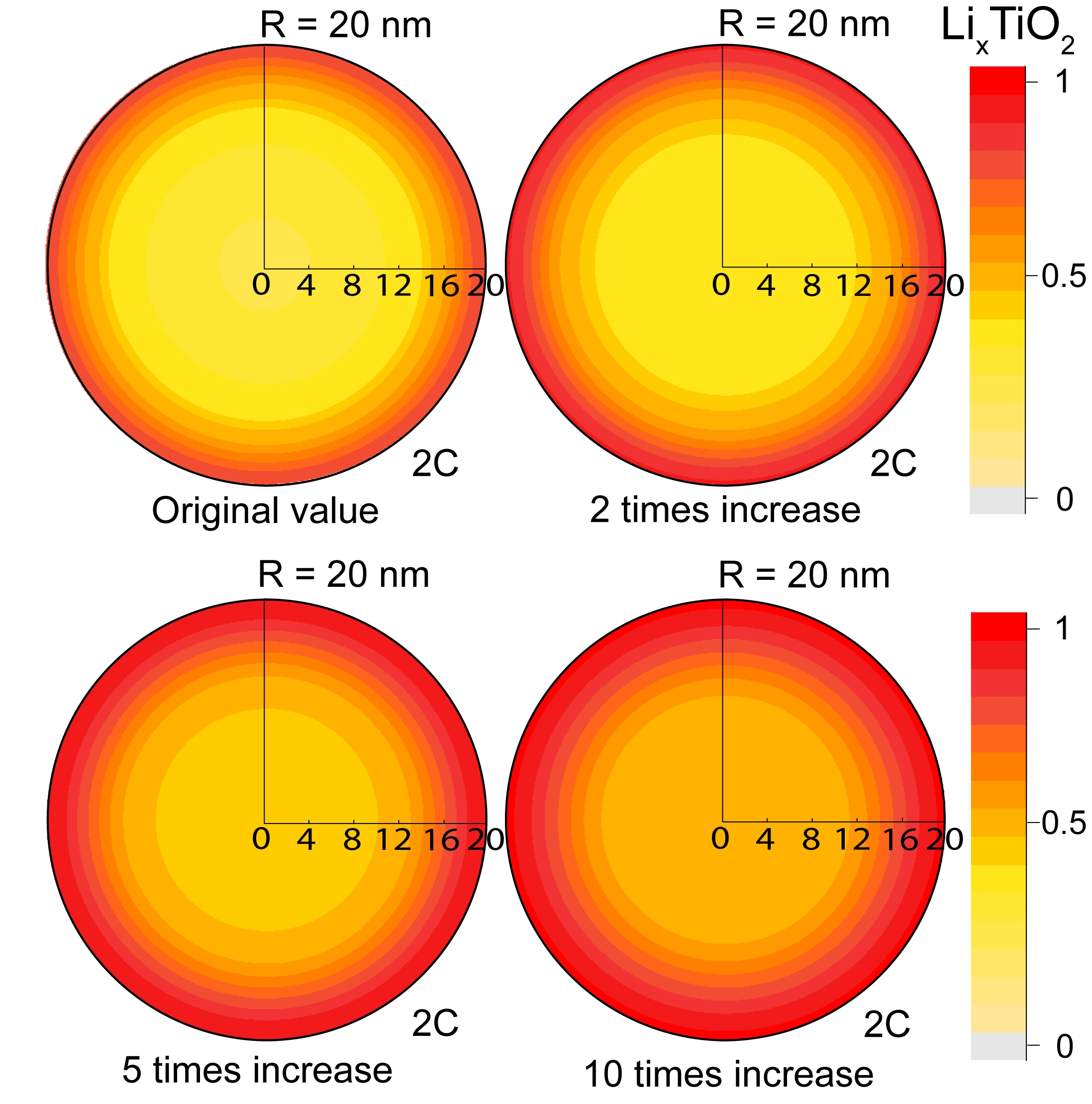} 
			\caption{~}
		\end{subfigure}						
		\caption{Voltage profiles versus average concentration ($\tilde{X}$) for different increases in diffusivity at (a) 0.5C, and (b) 2C, and final concentration profiles inside the particles at (c) 0.5C and (d) 2C.}
		\label{fig:better_D}	
	\end{center}
\end{figure}
To capture the effects of a higher Li-diffusivity simulations were performed at 0.5C and 2C, where the Li-diffusion in both lattices is increased by a factor of 2, 5 and 10 compared to the values given in Table \ref{tab:parameters}. \\
As should be anticipated this results in larger capacities and higher voltages with increasing diffusivity, as shown in Figure \ref{fig:better_D}, consistent with experimental observations \cite{Shin_2012, Wang_2011, Ren_2014}.
For the lithium concentrations in the particle the higher diffusivity results in an extension of the \ce{Li1TiO2} phase from the surface of the particle. At 0.5C the inside of the particle transforms completely to the \ce{Li_{0.5}TiO2} phase, even for the original diffusivity. While at 2C an increase of the diffusivity by a factor of 5 is necessary to transform the inside of the particle to the \ce{Li_{0.5}TiO2} phase.   
These results confirm that increasing the diffusivity is a promising way to increase the capacity of anatase electrodes, especially when aiming at high (dis)charge rates.

\subsection{Impact of surface area}
The simulations shown in Figure \ref{fig:c-rates} and \ref{fig:better_D} predict large charge-transfer overpotentials during the lithiation of anatase particles. These large overpotentials are caused by a high Li-concentration near the surface of anatase particles, making it hard for Li-ions to enter into the anatase particles. 
A reduction in the surface Li-concentration can be achieved by increasing the surface area, which will lead to smaller charge-transfer overpotentials. An additional advantage is that a larger surface areas also lowers the current density, which further decreases the charge-transfer overpotential. \\
This leads to larger capacities, as has been demonstrated by various experimental studies on high surface area anatase particles \cite{Shin_2011, Madej_2015, Madej_2014B}. 
To capture the surface area effect simulations were performed on a spherical and a cylindrical particle (infinitely long, i.e. neglecting the top and bottom surface of the cylinder). The spherical particle has a surface to volume ratio of $\frac{3}{R}$, whereas the cylindrical particle has a smaller surface to volume ratio of $\frac{2}{R}$. \\
\begin{figure}[htbp]
	\begin{center}
		\begin{subfigure}{0.45\textwidth}
			\includegraphics[width=\textwidth]{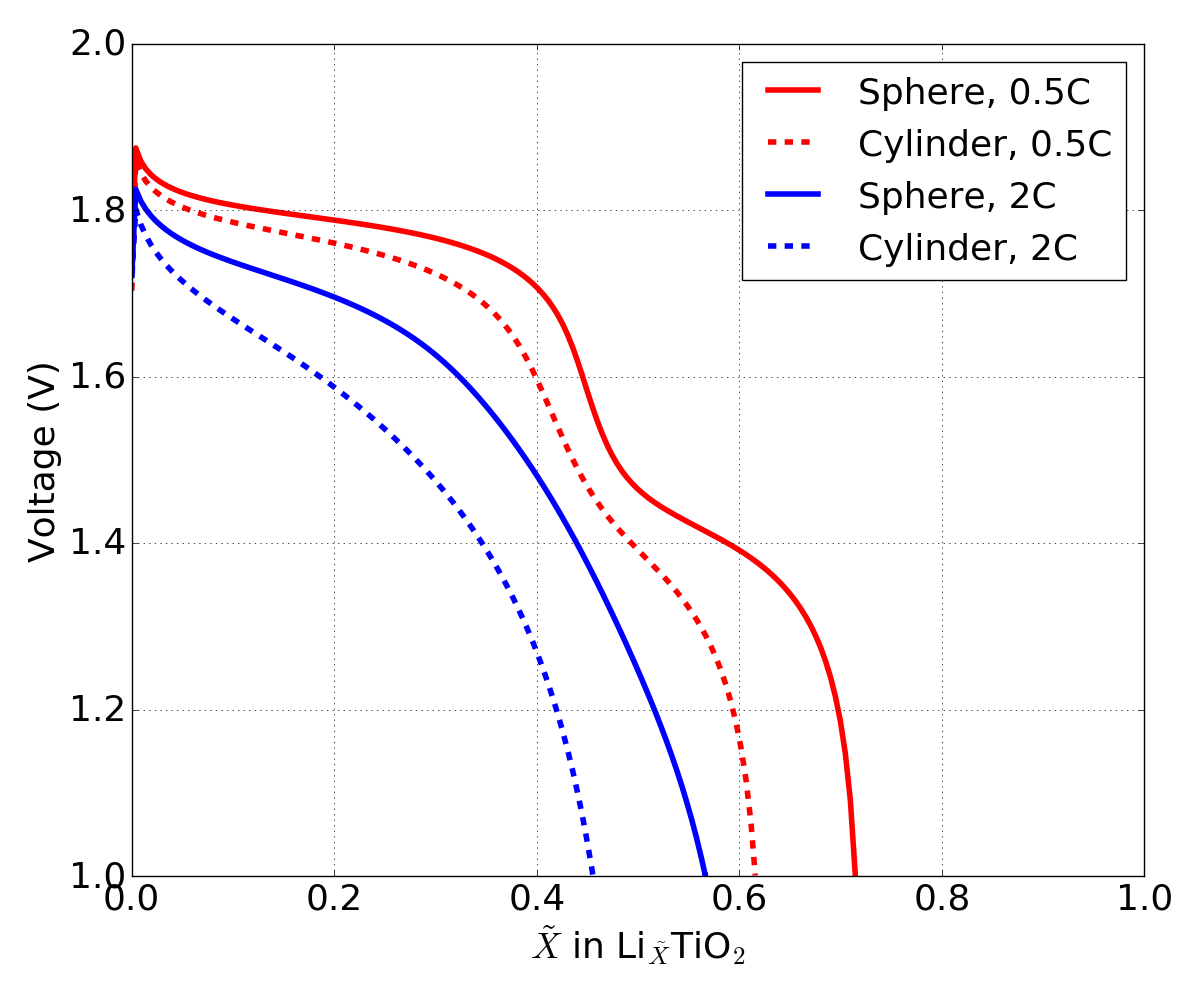}
			\caption{~} 
		\end{subfigure}	
		\begin{subfigure}{0.45\textwidth}
			\includegraphics[width=\textwidth]{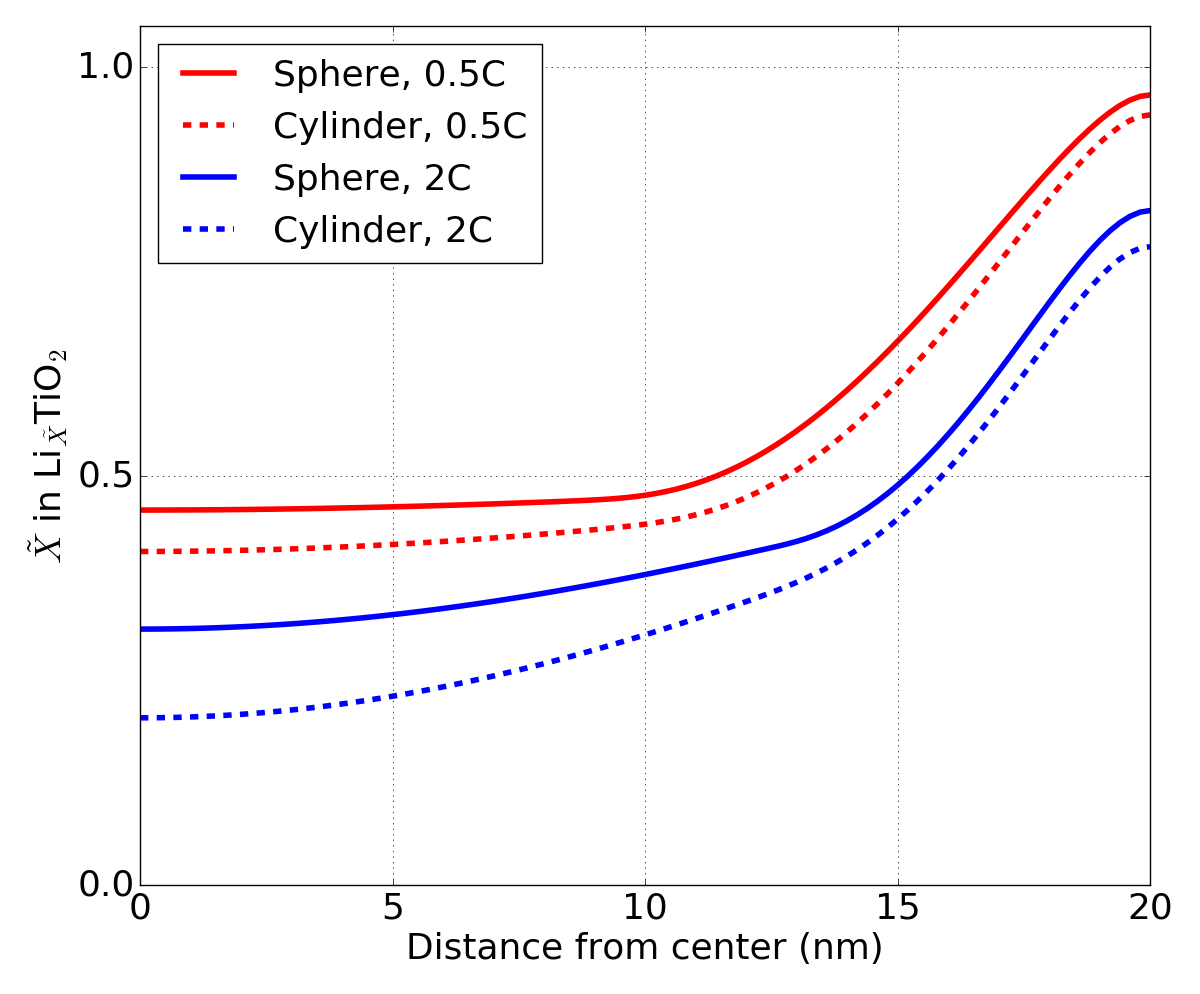} 
			\caption{~} 
		\end{subfigure}				
		\caption{(a) Voltage profiles versus average concentration ($\tilde{X}$) and (b) final concentration profiles of a spherical and a cylindrical particle at 0.5C and 2C.}
		\label{fig:sphere_vs_cylinder}	
	\end{center}
\end{figure}
The concentration profiles in Figure \ref{fig:sphere_vs_cylinder}b show that when the cut-off voltage is reached the Li-concentration near the surface is comparable in the spherical and cylindrical particle. However, the larger surface area of the spherical particle allows for a larger Li-ion flux into the anatase \ce{TiO2} particle, resulting in a final Li-fraction approximately 20\% larger compared to the cylindrical particle.  
As shown in Figure \ref{fig:sphere_vs_cylinder}a, at 0.5C relatively small differences for the first voltage plateau are predicted, the difference in capacity being primarily caused by the second lattice. This is because the slower diffusion in the second phase leads to high Li-concentrations near the surface more quickly, and thus high charge-transfer overpotentials at an earlier stage. 
Increasing the C-rate to 2C significantly augments this effect, raising the difference in charge-transfer overpotential between the spherical and cylindrical particle, although the decrease in final Li-composition is similar when compared to 0.5C. \\
Experimentally similar observations are reported upon increasing the surface area \cite{Madej_2015, Madej_2014B, Shin_2011}. 
However, higher surface areas are usually achieved by reducing the particle size \cite{Shin_2011, Madej_2015}, which has a similar impact on the voltage profiles, as shown in Figure \ref{fig:particle_sizes} and discussed below, making it hard to distinguish between the effects of nano-sizing and particle shape. 

\subsection{Impact of particle size}
\begin{figure}[htbp]
	\begin{center}
		\begin{subfigure}{0.45\textwidth}
			\includegraphics[width=\textwidth]{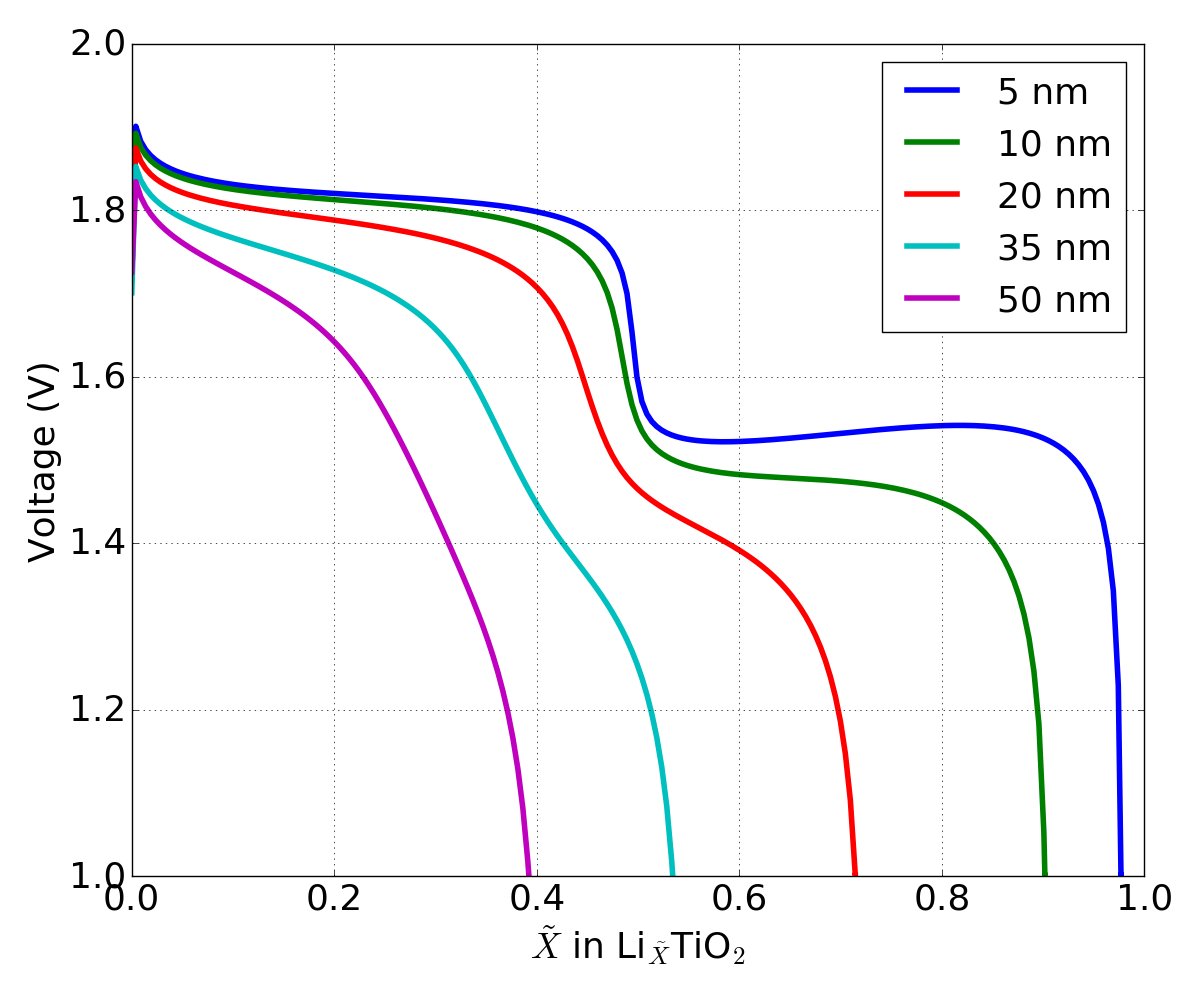} 
			\caption{~}
		\end{subfigure}
		\begin{subfigure}{0.45\textwidth}
			\includegraphics[width=\textwidth]{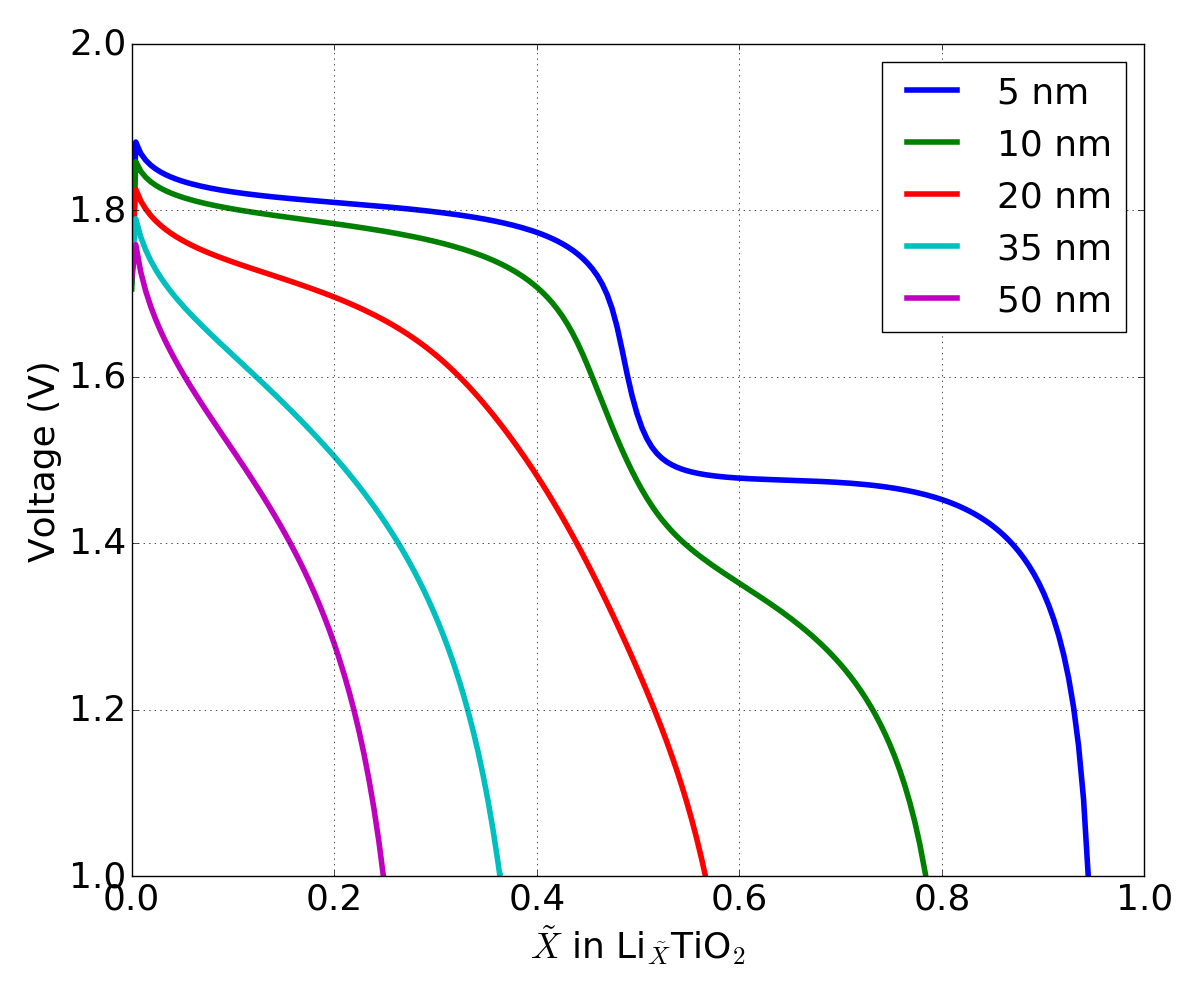} 
			\caption{~}
		\end{subfigure}
		\caption{Voltage profiles versus average concentration ($\tilde{X}$) for particles with different radii at (a) 0.5C and (b) 2C.}
		\label{fig:particle_sizes}
	\end{center}
\end{figure}

\begin{figure}[htbp]
	\begin{center}
		\includegraphics[width=0.5\textwidth]{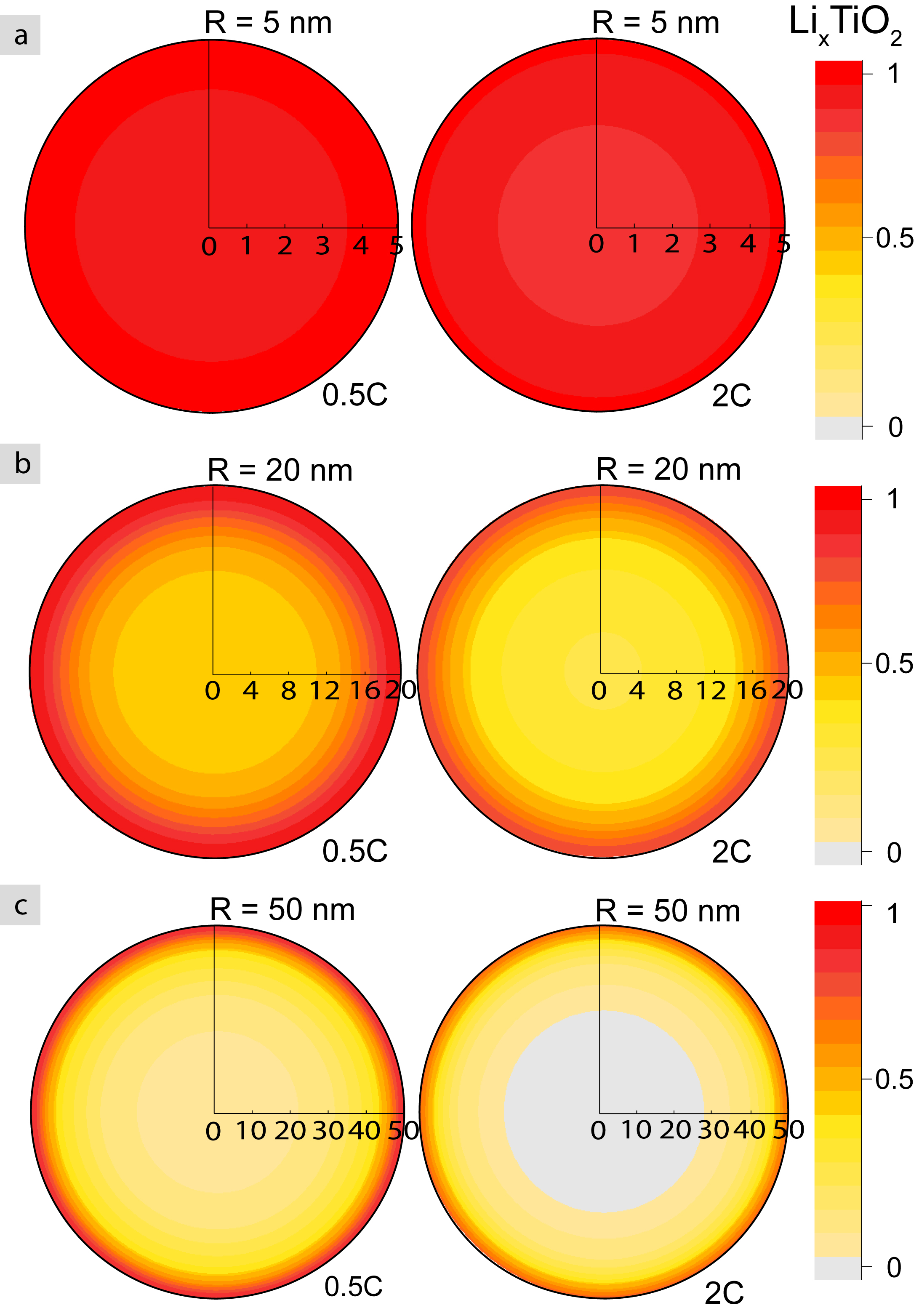} 					
	\caption{Final concentration profiles at the end of a simulation at 0.5C and 2C for particles with a radius of (a) 5 nm, (b) 20 nm, and (c) 50 nm.}	
	\label{fig:conc_profiles}
	\end{center}
\end{figure}

\begin{figure}[htbp]
	\begin{center}
		\begin{subfigure}{0.45\textwidth}
			\includegraphics[width=\textwidth]{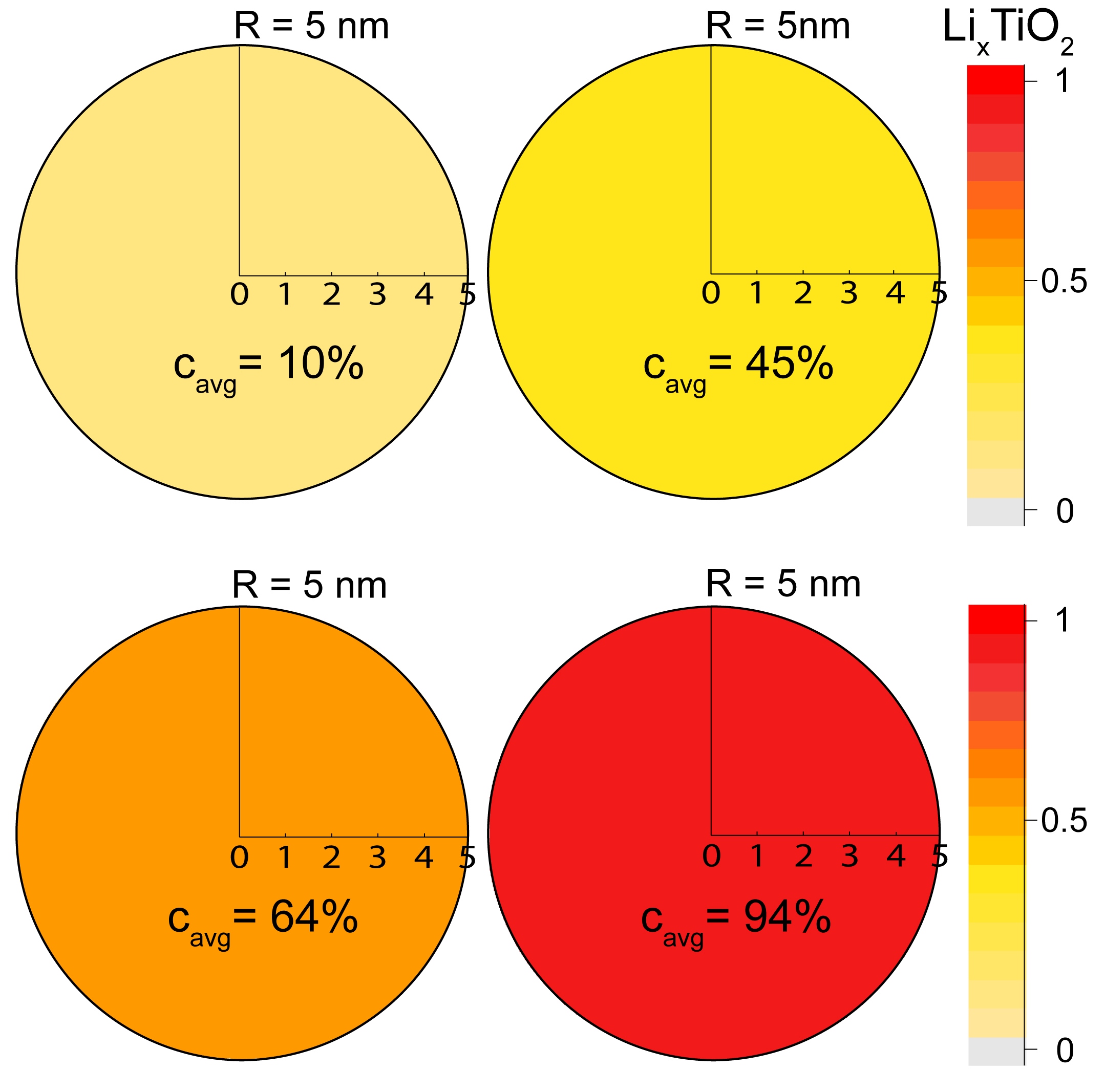} 
			\caption{~}
		\end{subfigure}
		\begin{subfigure}{0.45\textwidth}
			\includegraphics[width=\textwidth]{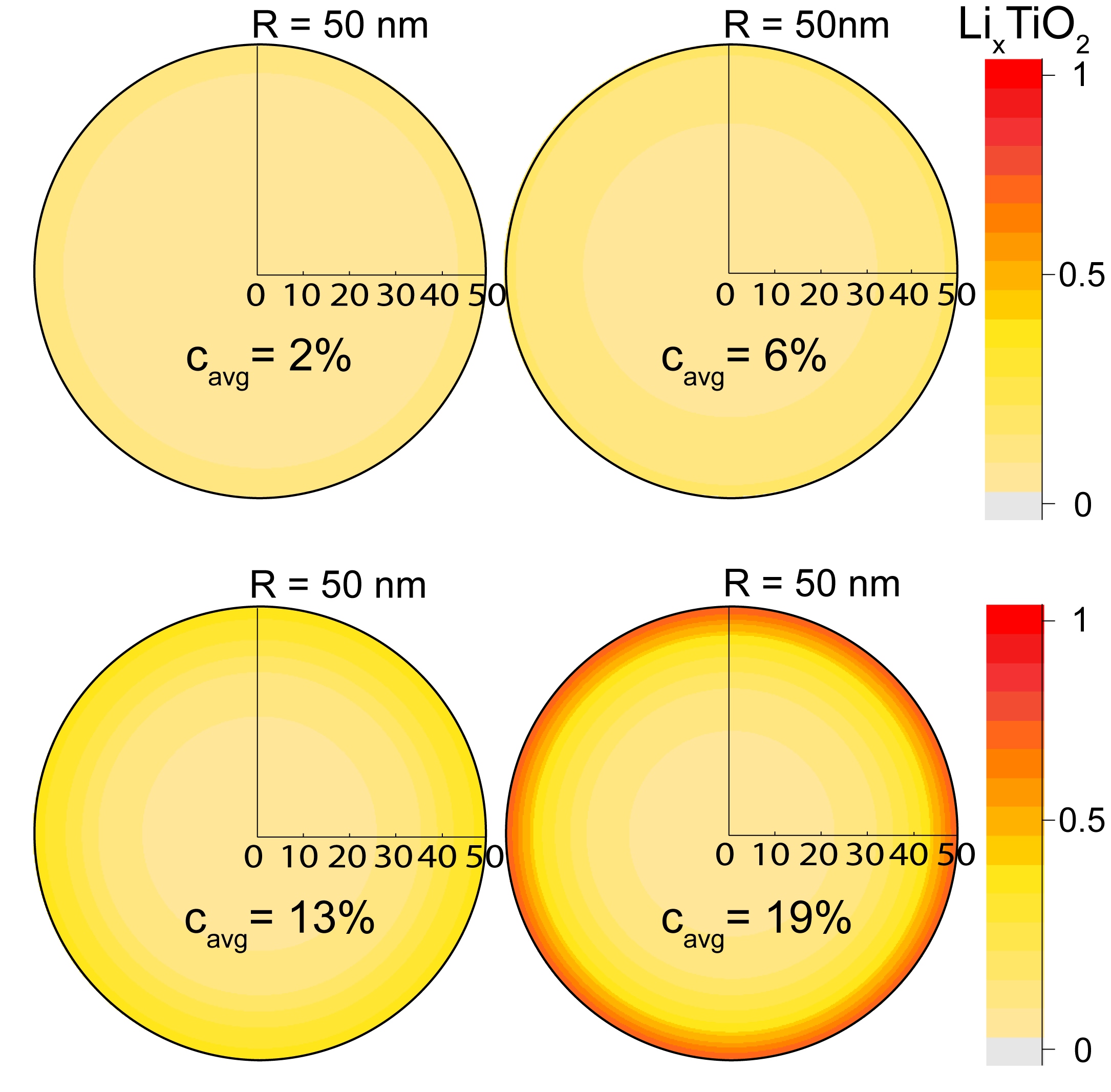} 
			\caption{~}
		\end{subfigure}
		\caption{Concentration profiles during a 0.5C simulation at different lithiation stages for particles with a radius of (a) 5 nm and (b) 50 nm.}
		\label{fig:profiles_sizes}
	\end{center}
\end{figure}

\begin{figure}[htbp]
	\begin{center}
		\begin{subfigure}{0.45\textwidth}
			\includegraphics[width=\textwidth]{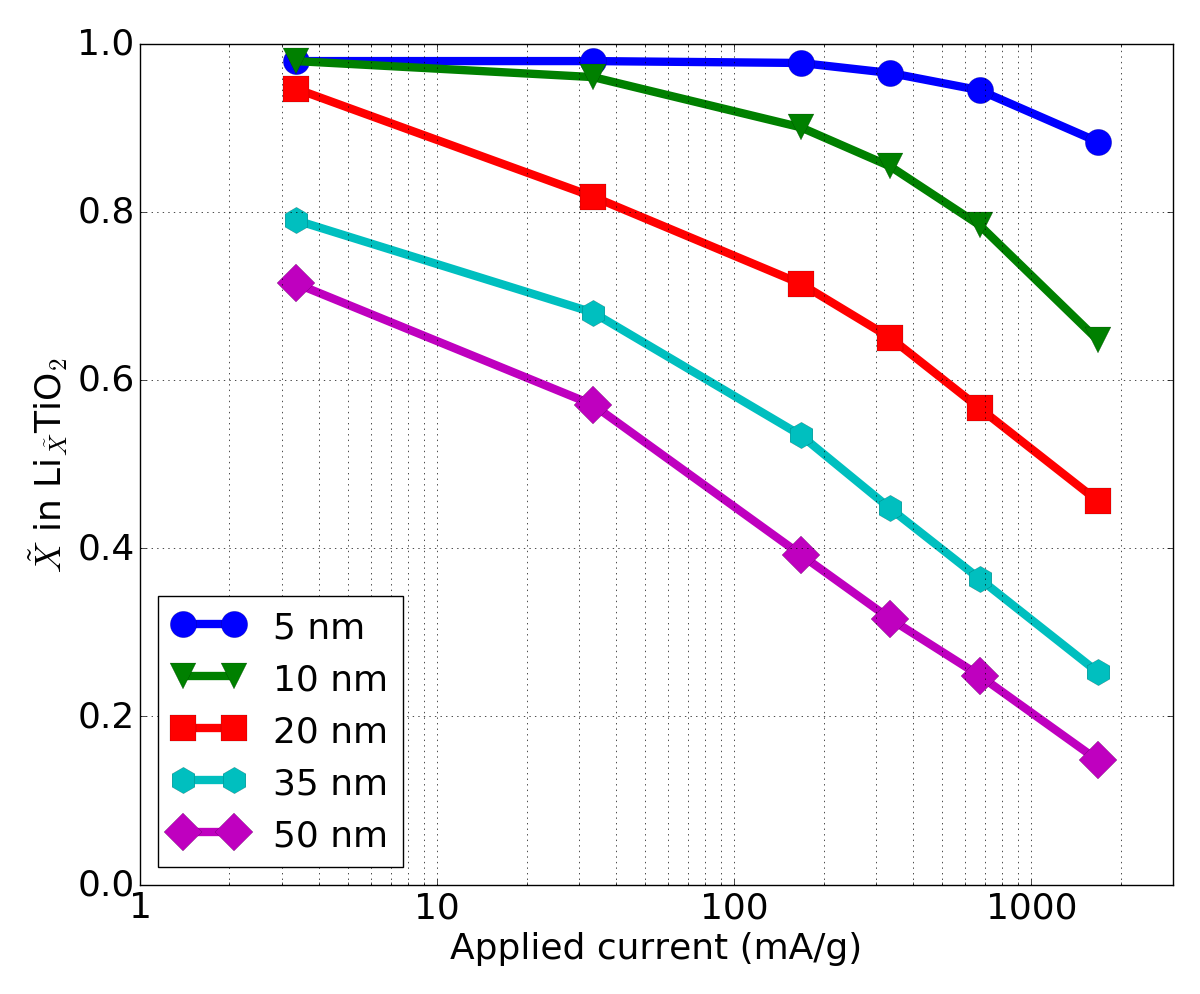}
			\caption{~} 
		\end{subfigure}
		\begin{subfigure}{0.45\textwidth}
			\includegraphics[width=\textwidth]{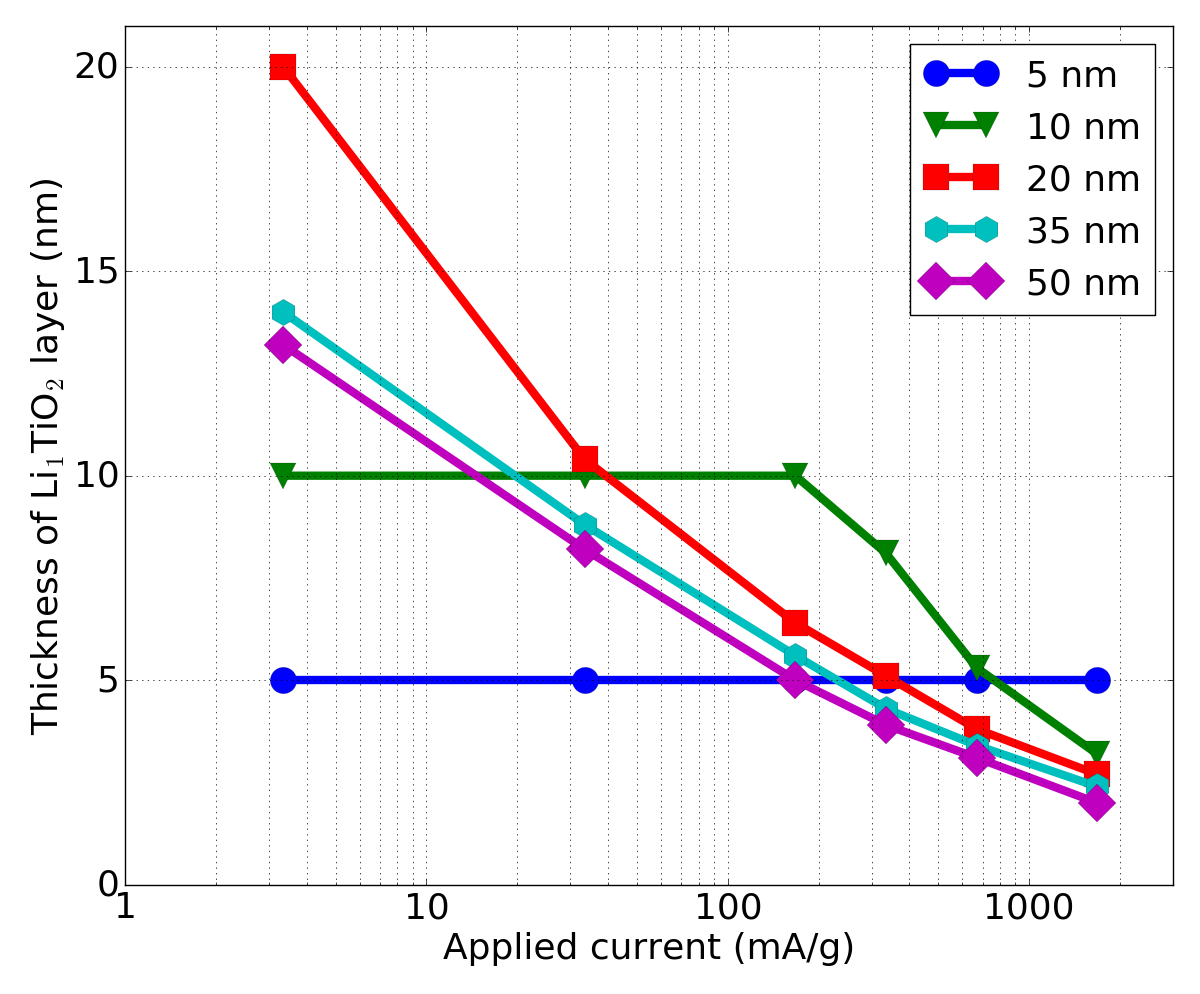} 
			\caption{~}
		\end{subfigure}
		\caption{For different particle sizes and C-rates the (a) final Li-concentrations and (b) final thickness of the \ce{Li1TiO2} layer (defined as the depth at which \ce{Li_{0.6}TiO2} occurs), the lines are a guide to the eye.}
		\label{fig:over_part_crate}
	\end{center}
\end{figure}

When Li-ion diffusion limits the electrode performance, decreasing the diffusion distance through the electrode material by particle size reduction is a well-established strategy to reach improved rate performance. 
Smaller particles increase the surface to volume ratio, which has been shown to be beneficial in the prior section.
Additionally, particle size reduction has been shown to change the thermodynamics by increasing the solubility limits, which in small particles can even lead to suppression of phase separation, as shown for anatase \ce{TiO2} \cite{Wagemaker_2007A, Wagemaker_2009} and \ce{LiFePO4} \cite{Burch_2009, Wagemaker_2011}. \\
A solid solution reaction can be expected to enhance Li-ion kinetics, in anatase specifically by suppressing the phase transition towards \ce{Li1TiO2} at the particle surface, thereby promoting Li-ion transport and resulting in higher voltages and larger capacities at high C-rates.   
Indeed for anatase \ce{TiO2} particle size reduction has been shown to improve performance drastically, for instance resulting in complete lithiation of 7 nm particles at C/20 \cite{Lafont_2010}, whereas large particles cannot practically be lithiated to compositions exceeding \ce{Li$_{0.6}$TiO2} \cite{Wagemaker_2007A, Gentili_2012}. 
Consistently, phase-field simulations for different particle sizes, shown in Figure \ref{fig:particle_sizes}, predict that decreasing the particle size diminishes kinetic limitations, resulting in nearly complete lithiation of 5 nm radius particles at 0.5C. \\  
Experimentally it has been reported that only the first few nanometres near the surface transform to the \ce{Li1TiO2} phase \cite{Wagemaker_2007A, Borghols_2009a, Wagemaker_2004}, explaining why only nano-particles are able to reach the theoretical \ce{Li1TiO2} composition \cite{Wagemaker_2007A, Lafont_2010}, and fully supported by the predicted concentration profiles in Figure \ref{fig:conc_profiles}. At 0.5C and 2C particles with a 5 nm radius transform completely to the \ce{Li1TiO2} phase, while in bigger particles only the first few nanometres at the surface transform to the \ce{Li1TiO2} phase. As a consequence decreasing the particle size increases the capacity, in particular because of the increasing utilization of the second phase transition. \\
Increasing the C-rate makes these effects more pronounced. For example, the final Li-composition of the 50 nm radius particle decreases from \ce{Li_{0.39}TiO2} to \ce{Li_{0.25}TiO2} when increasing the rate from 0.5C to 2C, while it only drops from \ce{Li_{0.98}TiO2} to \ce{Li_{0.94}TiO2} in the 5 nm particle. In the first place improved performance upon particle size reduction can be understood through the trivial effect of reducing the diffusion distance. The phase-field simulations also demonstrate that the non-trivial size effect, the destabilization of the first order phase transition, plays a crucial role in the enhanced performance of the smaller particles. As shown in Figure \ref{fig:over_part_crate} the thickness of the \ce{Li1TiO2}-layer depends on particle size, with smaller particles having thicker \ce{Li1TiO2}-layers. 
When the particle size approaches the width of the interface between the coexisting phases, phase separation is suppressed as previously shown for \ce{LiFePO4} \cite{Burch_2009, Wagemaker_2011}. \\
For the \ce{TiO2}-\ce{Li_{0.5}TiO2} and \ce{Li_{0.5}TiO2}-\ce{Li1TiO2} transitions the interface widths are approximately 40 nm and 5 nm, respectively \cite{Wagemaker_2007A}, directly related to the gradient penalty ($\kappa$). This explains the evolution of the Li-ion concentration throughout the 5 and 50 nm particles shown for 0.5C in Figure \ref{fig:profiles_sizes}.
The 5 nm particle completely lithiates through a solid solution reaction in which both phase transitions are suppressed due to the small particle size. In the 50 nm particle the first transformation from \ce{TiO2} to \ce{Li_{0.5}TiO2} is largely suppressed because it approaches the interface width of approximately 40 nm. However, at this particle size the second transition from \ce{Li_{0.5}TiO2} to \ce{Li1TiO2} is not suppressed. \\
The consequence of the particle size induced solid solution behaviour is that the formation of the blocking \ce{Li1TiO2} phase at the particle surface is prevented, as observed in Figure \ref{fig:conc_profiles}, resulting in higher voltages and larger capacities for small particles.
In larger particles phase separation will occur, and Li-concentrations near the surface will approach \ce{Li1TiO2}. The poor Li-diffusivity in the \ce{Li1TiO2} phase prevents moving of the phase boundary away from the surface, and the \ce{Li1TiO2} surface layer will block further lithiation of the particle. \\
To show how this affects battery capacity the final capacity vs. C-rate and final thickness of the \ce{Li1TiO2}-layer vs. C-rate are shown for different particle sizes in Figure \ref{fig:over_part_crate}. Figure \ref{fig:over_part_crate} shows that the thickness of the blocking \ce{Li1TiO2} layer strongly depends on the particle size and C-rate, and that this is strongly correlated with the final capacity. 
As shown in Figure \ref{fig:over_part_crate} to obtain high capacities small particles or low C-rates are required, otherwise the layer of \ce{Li1TiO2} forming at the surface will block Li-intercalation. 
To obtain a general prediction of the maximum obtainable capacity as a function of applied current the concept of Sand's time \cite{Bai_2016} could be used. However, to incorporate the non-linear Li-flux through the anatase particles, which also depends on the particle size, modifications to the formulation of Sand's time are necessary.
For small particles the simulations show good agreement with experiments, but for large particles (\textgreater 35 nm.) the simulated maximum particle composition is significantly smaller than \ce{Li$_{0.6}$TiO2} at moderate C-rates (\textgreater 0.5C), underestimating experimental observations \cite{Gentili_2012, Wagemaker_2007A}. 
We propose two arguments for this. First, in experiments the surface area is larger compared to that of spherical particles assumed in the simulations. As a consequence the present simulations underestimate the amount of \ce{Li1TiO2} formed, explaining the smaller capacities predicted. 
Second, although the phase-field calculations predict the extended solubility limits, the consequential enhanced Li-ion diffusivity is not implemented. Implementing this would require a diffusion term which depends on the local Li-concentration and the gradient of the Li-concentration, which requires further research outside of the scope of this study. 
Therefore the Li-ion diffusivity in particles where the solubility limits are significantly affected will be underestimated. 
This will be relevant for particle sizes comparable to the interface width, where in particular the larger interface width of the first phase transition enhances the Li-ion solubility limits. Thus explaining why for particles larger than 35 nm the present phase-field calculations underestimate the capacities as compared to experiments.  

\section{Discussion}
\begin{figure}[htbp]
	\begin{center}
	\includegraphics[width=\textwidth]{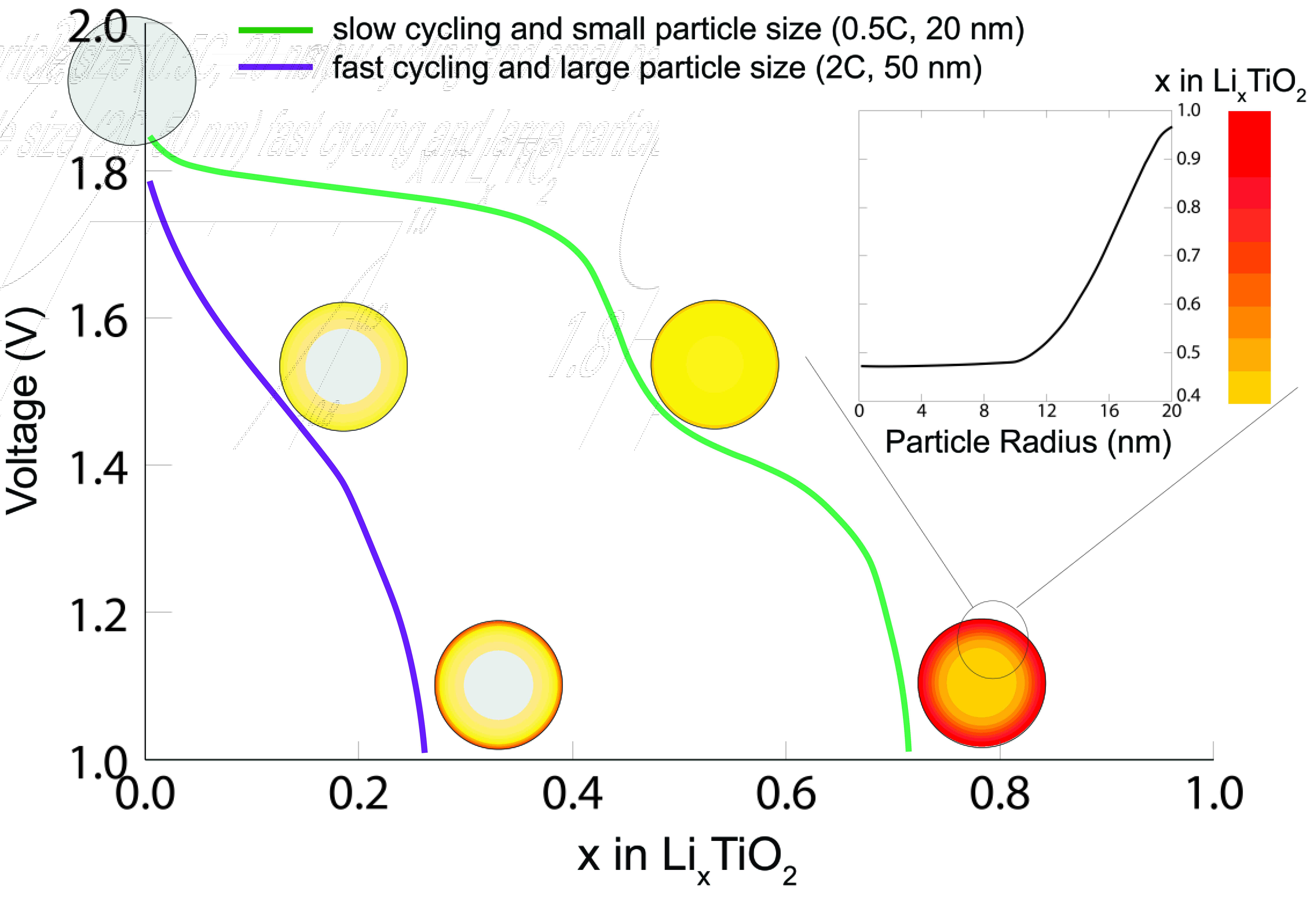}
	\caption{Overview of Li-intercalation in anatase.}
	\label{fig:overview}	
	\end{center}
\end{figure}

As summarised in Figure \ref{fig:overview}, the second phase transformation towards the \ce{Li1TiO2} phase at the surface of anatase \ce{TiO2} particles is primarily responsible for the performance of anatase as Li-ion battery electrode. 
During lithiation \ce{TiO2}-anatase is transformed to \ce{Li_{0.5}TiO2}, and this is followed by the formation of a thin layer of \ce{Li1TiO2} at the particle surface. Because the Li-diffusion is poor in the \ce{Li1TiO2}-layer, the thickness of this layer primarily depends on the applied C-rate, as shown in Figure \ref{fig:over_part_crate}. \\
When applying small C-rates the entire anatase particle is able to transform to the \ce{Li1TiO2}-phase, electrochemically represented by a second voltage plateau. At moderate C-rates only the onset of the second plateau will occur, as consistently predicted by the present phase-field simulations.  
For large C-rates and/or large anatase particles the lower voltage will drive the formation of a thin layer of the \ce{Li1TiO2} phase. The slow Li transport through this layer prevents transformation of the complete particle towards the \ce{Li1TiO2} phase. This is the origin of the observed large charge-transfer overpotentials, causing a shortening of the first voltage plateau, followed by a sloped tail in the voltage profile, thus rationalising the decreasing capacities with increasing lithiation rates and increasing particle size. \\
The strongly concentration dependent Li-diffusion provides an explanation of the memory effect observed for anatase electrodes \cite{Madej_2014B}, which presents itself through an decrease in the voltage in the charge-cycle (lithiation) following upon an incomplete charge/discharge cycle. The memory effect is caused by particles that are not completely delithiated before they are lithiating again, and thus more regions with poor diffusivity will remain in comparison to a complete charge/discharge cycle. In the memory-cycle the regions with poor diffusion will increase the charge-transfer overpotential in comparison to completely delithiated particles, explaining the observed memory effect. The phase-field simulations show that the charge-transfer overpotential strongly depends on the applied C-rate and the surface to volume ratio, explaining the experimental observations \cite{Madej_2014B} of a larger memory effect at high C-rates and with smaller surface to volume ratios. \\
The blocking mechanism through the \ce{Li1TiO2} phase formation at the surface of anatase particles, as predicted by the present simulations, indicates that the single grain lithiation limits the rate performance of complete electrodes. This is explicitly demonstrated by the multi-particle phase-field simulation, where all particles transform concurrently. All grains are active, implying that the electrode is unable to provide a higher current due to the single particle limitations. This is unlike other electrode materials, such as \ce{LiFePO4}, \ce{Li4Ti5O_{12}}, and \ce{LiCoO2}, in which ionic and electronic wiring dominate the internal resistance \cite{Liu_2016, Strobridge_2015, Li_2015, Singh_2013, Kim_2013}. 
The phase-field model predicts that the applied overpotential is unable to induce solid solution behaviour through widening of the phase interface regions, which has been shown for \ce{LiFePO4} \cite{Zhang_2014}, because at high overpotentials a layer of \ce{Li1TiO2} phase will form at the surface, after which the poor Li-ion diffusivity in this phase consumes the applied overpotential. 

\section{Conclusions}
The present phase-field model with parameters from literature is able to qualitatively explain practically all experimental phenomena observed during lithiation of anatase electrodes. This includes the impact of particle size, C-rate, Li-ion diffusivity, surface area, and the observed memory effect. 
The kinetic restrictions of the \ce{Li1TiO2} phase forming at the surface of anatase particles is the origin of the performance restrictions, and prevents fast transformation of anatase particles, even under high current. The simulations reveal that the transformation of single anatase particles limits the performance of the complete electrode, rather than the ionic and electronic wiring that appears to be rate limiting for other electrode materials. 
This demonstrates that the limitation of anatase can only be addressed if the formation of a blocking \ce{Li1TiO2} layer can be suppressed. This can be achieved by particle size reduction, where the reduction of the Li-ion diffusion distance as well as the suppression of the phase transitions enhances the Li-ion transport, increasing the capacity at high C-rates.
A second strategy is to improve the Li-ion diffusivity in the anatase \ce{TiO2} lattice by doping. The simulations give detailed insight on the impact of both approaches, providing a rational strategy towards improved performance. 
Thus, a comprehensive model for lithiation of anatase \ce{TiO2} has been presented. The good agreement with literature provides deep insight into the lithiation mechanism of anatase, and validation of the physical description of the phase-field model. This further motivates developing phase-field models for electrode materials, which will provide fundamental understanding of the limitations of battery materials, necessary for the formulation of strategies towards improved battery materials, design, and performance. 

\section{Acknowledgements}
The research leading to these results has received funding from the European Research Council under the European Union's Seventh Framework Programme (FP/2007-2013)/ERC Grant Agreement nr. [307161] of M.W. Financial support from the Advanced Dutch Energy Materials (ADEM) program of the Dutch Ministry of Economic Affairs, Agriculture and Innovation is gratefully acknowledged. R.B.S. and M.Z.B. acknowledge partial support from Samsung-MIT Alliance and the D3BATT project of the Toyota Research Institute.

\section{Supplementary Information}
See the Supplemental Material for a description of the many particle effects at 0.01C, and a comparison between the minimal and maximum charge transfer coefficients obtained from experiments.

\bibliography{bib_anatase_phase_field}

\begin{thebibliography}{78}%
\makeatletter
\providecommand \@ifxundefined [1]{%
 \@ifx{#1\undefined}
}%
\providecommand \@ifnum [1]{%
 \ifnum #1\expandafter \@firstoftwo
 \else \expandafter \@secondoftwo
 \fi
}%
\providecommand \@ifx [1]{%
 \ifx #1\expandafter \@firstoftwo
 \else \expandafter \@secondoftwo
 \fi
}%
\providecommand \natexlab [1]{#1}%
\providecommand \enquote  [1]{``#1''}%
\providecommand \bibnamefont  [1]{#1}%
\providecommand \bibfnamefont [1]{#1}%
\providecommand \citenamefont [1]{#1}%
\providecommand \href@noop [0]{\@secondoftwo}%
\providecommand \href [0]{\begingroup \@sanitize@url \@href}%
\providecommand \@href[1]{\@@startlink{#1}\@@href}%
\providecommand \@@href[1]{\endgroup#1\@@endlink}%
\providecommand \@sanitize@url [0]{\catcode `\\12\catcode `\$12\catcode
  `\&12\catcode `\#12\catcode `\^12\catcode `\_12\catcode `\%12\relax}%
\providecommand \@@startlink[1]{}%
\providecommand \@@endlink[0]{}%
\providecommand \url  [0]{\begingroup\@sanitize@url \@url }%
\providecommand \@url [1]{\endgroup\@href {#1}{\urlprefix }}%
\providecommand \urlprefix  [0]{URL }%
\providecommand \Eprint [0]{\href }%
\providecommand \doibase [0]{http://dx.doi.org/}%
\providecommand \selectlanguage [0]{\@gobble}%
\providecommand \bibinfo  [0]{\@secondoftwo}%
\providecommand \bibfield  [0]{\@secondoftwo}%
\providecommand \translation [1]{[#1]}%
\providecommand \BibitemOpen [0]{}%
\providecommand \bibitemStop [0]{}%
\providecommand \bibitemNoStop [0]{.\EOS\space}%
\providecommand \EOS [0]{\spacefactor3000\relax}%
\providecommand \BibitemShut  [1]{\csname bibitem#1\endcsname}%
\let\auto@bib@innerbib\@empty
\bibitem [{\citenamefont {Deng}(2015)}]{Deng_2015}%
  \BibitemOpen
  \bibfield  {author} {\bibinfo {author} {\bibfnamefont {D.}~\bibnamefont
  {Deng}},\ }\href {\doibase 10.1002/ese3.95} {\bibfield  {journal} {\bibinfo
  {journal} {Energy Science and Engineering}\ }\textbf {\bibinfo {volume}
  {3}},\ \bibinfo {pages} {385} (\bibinfo {year} {2015})}\BibitemShut {NoStop}%
\bibitem [{\citenamefont {Franco}(2013)}]{Franco_2013}%
  \BibitemOpen
  \bibfield  {author} {\bibinfo {author} {\bibfnamefont {A.~A.}\ \bibnamefont
  {Franco}},\ }\href {\doibase 10.1039/c3ra23502e} {\bibfield  {journal}
  {\bibinfo  {journal} {Rsc Advances}\ }\textbf {\bibinfo {volume} {3}},\
  \bibinfo {pages} {13027} (\bibinfo {year} {2013})}\BibitemShut {NoStop}%
\bibitem [{\citenamefont {Zhang}\ \emph
  {et~al.}(2015{\natexlab{a}})\citenamefont {Zhang}, \citenamefont {Verhallen},
  \citenamefont {Labohm},\ and\ \citenamefont {Wagemaker}}]{Zhang_2015a}%
  \BibitemOpen
  \bibfield  {author} {\bibinfo {author} {\bibfnamefont {X.}~\bibnamefont
  {Zhang}}, \bibinfo {author} {\bibfnamefont {T.~W.}\ \bibnamefont
  {Verhallen}}, \bibinfo {author} {\bibfnamefont {F.}~\bibnamefont {Labohm}}, \
  and\ \bibinfo {author} {\bibfnamefont {M.}~\bibnamefont {Wagemaker}},\ }\href
  {\doibase 10.1002/aenm.201500498} {\bibfield  {journal} {\bibinfo  {journal}
  {Advanced Energy Materials}\ }\textbf {\bibinfo {volume} {5}} (\bibinfo
  {year} {2015}{\natexlab{a}}),\ 10.1002/aenm.201500498}\BibitemShut {NoStop}%
\bibitem [{\citenamefont {Notten}\ and\ \citenamefont
  {Danilov}(2014)}]{Notten_2014}%
  \BibitemOpen
  \bibfield  {author} {\bibinfo {author} {\bibfnamefont {P.~H.~L.}\
  \bibnamefont {Notten}}\ and\ \bibinfo {author} {\bibfnamefont {D.~L.}\
  \bibnamefont {Danilov}},\ }\href {\doibase 10.4236/aces.2014.41009}
  {\bibfield  {journal} {\bibinfo  {journal} {Advances in Chemical Engineering
  and Science}\ }\textbf {\bibinfo {volume} {04}},\ \bibinfo {pages} {62}
  (\bibinfo {year} {2014})}\BibitemShut {NoStop}%
\bibitem [{\citenamefont {Vo}\ \emph {et~al.}(2015)\citenamefont {Vo},
  \citenamefont {Chen}, \citenamefont {Shen},\ and\ \citenamefont
  {Kapoor}}]{Vo_2015}%
  \BibitemOpen
  \bibfield  {author} {\bibinfo {author} {\bibfnamefont {T.~T.}\ \bibnamefont
  {Vo}}, \bibinfo {author} {\bibfnamefont {X.}~\bibnamefont {Chen}}, \bibinfo
  {author} {\bibfnamefont {W.}~\bibnamefont {Shen}}, \ and\ \bibinfo {author}
  {\bibfnamefont {A.}~\bibnamefont {Kapoor}},\ }\href {\doibase
  10.1016/j.jpowsour.2014.09.108} {\bibfield  {journal} {\bibinfo  {journal}
  {Journal of Power Sources}\ }\textbf {\bibinfo {volume} {273}},\ \bibinfo
  {pages} {413} (\bibinfo {year} {2015})}\BibitemShut {NoStop}%
\bibitem [{\citenamefont {Sethuraman}\ \emph {et~al.}(2012)\citenamefont
  {Sethuraman}, \citenamefont {Srinivasan},\ and\ \citenamefont
  {Newman}}]{Sethuraman_2012}%
  \BibitemOpen
  \bibfield  {author} {\bibinfo {author} {\bibfnamefont {V.~A.}\ \bibnamefont
  {Sethuraman}}, \bibinfo {author} {\bibfnamefont {V.}~\bibnamefont
  {Srinivasan}}, \ and\ \bibinfo {author} {\bibfnamefont {J.}~\bibnamefont
  {Newman}},\ }\href {\doibase 10.1149/2.008303jes} {\bibfield  {journal}
  {\bibinfo  {journal} {Journal of the Electrochemical Society}\ }\textbf
  {\bibinfo {volume} {160}},\ \bibinfo {pages} {A394} (\bibinfo {year}
  {2012})}\BibitemShut {NoStop}%
\bibitem [{\citenamefont {Landstorfer}\ \emph {et~al.}(2011)\citenamefont
  {Landstorfer}, \citenamefont {Funken},\ and\ \citenamefont
  {Jacob}}]{Landstorfer_2011}%
  \BibitemOpen
  \bibfield  {author} {\bibinfo {author} {\bibfnamefont {M.}~\bibnamefont
  {Landstorfer}}, \bibinfo {author} {\bibfnamefont {S.}~\bibnamefont {Funken}},
  \ and\ \bibinfo {author} {\bibfnamefont {T.}~\bibnamefont {Jacob}},\ }\href
  {\doibase 10.1039/c0cp02473b} {\bibfield  {journal} {\bibinfo  {journal}
  {Physical chemistry chemical physics : PCCP}\ }\textbf {\bibinfo {volume}
  {13}},\ \bibinfo {pages} {12817} (\bibinfo {year} {2011})}\BibitemShut
  {NoStop}%
\bibitem [{\citenamefont {Dargaville}\ and\ \citenamefont
  {Farrell}(2013)}]{Dargaville_2013}%
  \BibitemOpen
  \bibfield  {author} {\bibinfo {author} {\bibfnamefont {S.}~\bibnamefont
  {Dargaville}}\ and\ \bibinfo {author} {\bibfnamefont {T.~W.}\ \bibnamefont
  {Farrell}},\ }\href {\doibase 10.1016/j.electacta.2013.08.014} {\bibfield
  {journal} {\bibinfo  {journal} {Electrochimica Acta}\ }\textbf {\bibinfo
  {volume} {111}},\ \bibinfo {pages} {474} (\bibinfo {year}
  {2013})}\BibitemShut {NoStop}%
\bibitem [{\citenamefont {Bazant}(2013)}]{Bazant_2013}%
  \BibitemOpen
  \bibfield  {author} {\bibinfo {author} {\bibfnamefont {M.~Z.}\ \bibnamefont
  {Bazant}},\ }\href {\doibase 10.1021/ar300145c} {\bibfield  {journal}
  {\bibinfo  {journal} {Accounts of Chemical Research}\ }\textbf {\bibinfo
  {volume} {46}},\ \bibinfo {pages} {1144} (\bibinfo {year}
  {2013})}\BibitemShut {NoStop}%
\bibitem [{\citenamefont {Salvadori}\ \emph {et~al.}(2015)\citenamefont
  {Salvadori}, \citenamefont {Grazioli},\ and\ \citenamefont
  {Geers}}]{Salvadori_2015}%
  \BibitemOpen
  \bibfield  {author} {\bibinfo {author} {\bibfnamefont {A.}~\bibnamefont
  {Salvadori}}, \bibinfo {author} {\bibfnamefont {D.}~\bibnamefont {Grazioli}},
  \ and\ \bibinfo {author} {\bibfnamefont {M.~G.~D.}\ \bibnamefont {Geers}},\
  }\href {\doibase 10.1016/j.ijsolstr.2015.01.014} {\bibfield  {journal}
  {\bibinfo  {journal} {International Journal of Solids and Structures}\
  }\textbf {\bibinfo {volume} {59}},\ \bibinfo {pages} {90} (\bibinfo {year}
  {2015})}\BibitemShut {NoStop}%
\bibitem [{\citenamefont {Li}\ \emph {et~al.}(2014)\citenamefont {Li},
  \citenamefont {El~Gabaly}, \citenamefont {Ferguson}, \citenamefont {Smith},
  \citenamefont {Bartelt}, \citenamefont {Sugar}, \citenamefont {Fenton},
  \citenamefont {Cogswell}, \citenamefont {Kilcoyne}, \citenamefont
  {Tyliszczak}, \citenamefont {Bazant},\ and\ \citenamefont {Chueh}}]{Li_2014}%
  \BibitemOpen
  \bibfield  {author} {\bibinfo {author} {\bibfnamefont {Y.}~\bibnamefont
  {Li}}, \bibinfo {author} {\bibfnamefont {F.}~\bibnamefont {El~Gabaly}},
  \bibinfo {author} {\bibfnamefont {T.~R.}\ \bibnamefont {Ferguson}}, \bibinfo
  {author} {\bibfnamefont {R.~B.}\ \bibnamefont {Smith}}, \bibinfo {author}
  {\bibfnamefont {N.~C.}\ \bibnamefont {Bartelt}}, \bibinfo {author}
  {\bibfnamefont {J.~D.}\ \bibnamefont {Sugar}}, \bibinfo {author}
  {\bibfnamefont {K.~R.}\ \bibnamefont {Fenton}}, \bibinfo {author}
  {\bibfnamefont {D.~A.}\ \bibnamefont {Cogswell}}, \bibinfo {author}
  {\bibfnamefont {A.~L.~D.}\ \bibnamefont {Kilcoyne}}, \bibinfo {author}
  {\bibfnamefont {T.}~\bibnamefont {Tyliszczak}}, \bibinfo {author}
  {\bibfnamefont {M.~Z.}\ \bibnamefont {Bazant}}, \ and\ \bibinfo {author}
  {\bibfnamefont {W.~C.}\ \bibnamefont {Chueh}},\ }\href {\doibase
  10.1038/nmat4084} {\bibfield  {journal} {\bibinfo  {journal} {Nature
  Materials}\ }\textbf {\bibinfo {volume} {13}},\ \bibinfo {pages} {1149}
  (\bibinfo {year} {2014})}\BibitemShut {NoStop}%
\bibitem [{\citenamefont {Singh}\ \emph {et~al.}(2008)\citenamefont {Singh},
  \citenamefont {Ceder},\ and\ \citenamefont {Bazant}}]{Singh_2008}%
  \BibitemOpen
  \bibfield  {author} {\bibinfo {author} {\bibfnamefont {G.~K.}\ \bibnamefont
  {Singh}}, \bibinfo {author} {\bibfnamefont {G.}~\bibnamefont {Ceder}}, \ and\
  \bibinfo {author} {\bibfnamefont {M.~Z.}\ \bibnamefont {Bazant}},\ }\href
  {\doibase 10.1016/j.electacta.2008.03.083} {\bibfield  {journal} {\bibinfo
  {journal} {Electrochimica Acta}\ }\textbf {\bibinfo {volume} {53}},\ \bibinfo
  {pages} {7599} (\bibinfo {year} {2008})}\BibitemShut {NoStop}%
\bibitem [{\citenamefont {Burch}\ and\ \citenamefont
  {Bazant}(2009)}]{Burch_2009}%
  \BibitemOpen
  \bibfield  {author} {\bibinfo {author} {\bibfnamefont {D.}~\bibnamefont
  {Burch}}\ and\ \bibinfo {author} {\bibfnamefont {M.~Z.}\ \bibnamefont
  {Bazant}},\ }\href {\doibase 10.1021/nl9019787} {\bibfield  {journal}
  {\bibinfo  {journal} {Nano Letters}\ }\textbf {\bibinfo {volume} {9}},\
  \bibinfo {pages} {3795} (\bibinfo {year} {2009})}\BibitemShut {NoStop}%
\bibitem [{\citenamefont {Ferguson}\ and\ \citenamefont
  {Bazant}(2012)}]{Ferguson_2012}%
  \BibitemOpen
  \bibfield  {author} {\bibinfo {author} {\bibfnamefont {T.~R.}\ \bibnamefont
  {Ferguson}}\ and\ \bibinfo {author} {\bibfnamefont {M.~Z.}\ \bibnamefont
  {Bazant}},\ }\href {\doibase 10.1149/2.048212jes} {\bibfield  {journal}
  {\bibinfo  {journal} {Journal of the Electrochemical Society}\ }\textbf
  {\bibinfo {volume} {159}},\ \bibinfo {pages} {A1967} (\bibinfo {year}
  {2012})}\BibitemShut {NoStop}%
\bibitem [{\citenamefont {Wagemaker}\ \emph {et~al.}(2011)\citenamefont
  {Wagemaker}, \citenamefont {Singh}, \citenamefont {Borghols}, \citenamefont
  {Lafont}, \citenamefont {Haverkate}, \citenamefont {Peterson},\ and\
  \citenamefont {Mulder}}]{Wagemaker_2011}%
  \BibitemOpen
  \bibfield  {author} {\bibinfo {author} {\bibfnamefont {M.}~\bibnamefont
  {Wagemaker}}, \bibinfo {author} {\bibfnamefont {D.~P.}\ \bibnamefont
  {Singh}}, \bibinfo {author} {\bibfnamefont {W.~J.~H.}\ \bibnamefont
  {Borghols}}, \bibinfo {author} {\bibfnamefont {U.}~\bibnamefont {Lafont}},
  \bibinfo {author} {\bibfnamefont {L.}~\bibnamefont {Haverkate}}, \bibinfo
  {author} {\bibfnamefont {V.~K.}\ \bibnamefont {Peterson}}, \ and\ \bibinfo
  {author} {\bibfnamefont {F.~M.}\ \bibnamefont {Mulder}},\ }\href {\doibase
  10.1021/ja2026213} {\bibfield  {journal} {\bibinfo  {journal} {Journal of the
  American Chemical Society}\ }\textbf {\bibinfo {volume} {133}},\ \bibinfo
  {pages} {10222} (\bibinfo {year} {2011})}\BibitemShut {NoStop}%
\bibitem [{\citenamefont {Welland}\ \emph {et~al.}(2015)\citenamefont
  {Welland}, \citenamefont {Karpeyev}, \citenamefont {O'Connor},\ and\
  \citenamefont {Heinonen}}]{Welland_2015}%
  \BibitemOpen
  \bibfield  {author} {\bibinfo {author} {\bibfnamefont {M.~J.}\ \bibnamefont
  {Welland}}, \bibinfo {author} {\bibfnamefont {D.}~\bibnamefont {Karpeyev}},
  \bibinfo {author} {\bibfnamefont {D.~T.}\ \bibnamefont {O'Connor}}, \ and\
  \bibinfo {author} {\bibfnamefont {O.}~\bibnamefont {Heinonen}},\ }\href
  {\doibase 10.1021/acsnano.5b02555} {\bibfield  {journal} {\bibinfo  {journal}
  {ACS Nano}\ }\textbf {\bibinfo {volume} {9}},\ \bibinfo {pages} {9757}
  (\bibinfo {year} {2015})}\BibitemShut {NoStop}%
\bibitem [{\citenamefont {Zhang}\ \emph {et~al.}(2014)\citenamefont {Zhang},
  \citenamefont {van Hulzen}, \citenamefont {Singh}, \citenamefont {Brownrigg},
  \citenamefont {Wright}, \citenamefont {van Dijk},\ and\ \citenamefont
  {Wagemaker}}]{Zhang_2014}%
  \BibitemOpen
  \bibfield  {author} {\bibinfo {author} {\bibfnamefont {X.}~\bibnamefont
  {Zhang}}, \bibinfo {author} {\bibfnamefont {M.}~\bibnamefont {van Hulzen}},
  \bibinfo {author} {\bibfnamefont {D.~P.}\ \bibnamefont {Singh}}, \bibinfo
  {author} {\bibfnamefont {A.}~\bibnamefont {Brownrigg}}, \bibinfo {author}
  {\bibfnamefont {J.~P.}\ \bibnamefont {Wright}}, \bibinfo {author}
  {\bibfnamefont {N.~H.}\ \bibnamefont {van Dijk}}, \ and\ \bibinfo {author}
  {\bibfnamefont {M.}~\bibnamefont {Wagemaker}},\ }\href {\doibase
  10.1021/nl404285y} {\bibfield  {journal} {\bibinfo  {journal} {Nano Lett.}\
  }\textbf {\bibinfo {volume} {14}},\ \bibinfo {pages} {2279} (\bibinfo {year}
  {2014})}\BibitemShut {NoStop}%
\bibitem [{\citenamefont {Liu}\ \emph {et~al.}(2014)\citenamefont {Liu},
  \citenamefont {Strobridge}, \citenamefont {Borkiewicz}, \citenamefont
  {Wiaderek}, \citenamefont {Chapman}, \citenamefont {Chupas},\ and\
  \citenamefont {Grey}}]{Liu_2014}%
  \BibitemOpen
  \bibfield  {author} {\bibinfo {author} {\bibfnamefont {H.}~\bibnamefont
  {Liu}}, \bibinfo {author} {\bibfnamefont {F.~C.}\ \bibnamefont {Strobridge}},
  \bibinfo {author} {\bibfnamefont {O.~J.}\ \bibnamefont {Borkiewicz}},
  \bibinfo {author} {\bibfnamefont {K.~M.}\ \bibnamefont {Wiaderek}}, \bibinfo
  {author} {\bibfnamefont {K.~W.}\ \bibnamefont {Chapman}}, \bibinfo {author}
  {\bibfnamefont {P.~J.}\ \bibnamefont {Chupas}}, \ and\ \bibinfo {author}
  {\bibfnamefont {C.~P.}\ \bibnamefont {Grey}},\ }\href {\doibase
  10.1126/science.1252817} {\bibfield  {journal} {\bibinfo  {journal}
  {Science}\ }\textbf {\bibinfo {volume} {344}},\ \bibinfo {pages} {1252817}
  (\bibinfo {year} {2014})}\BibitemShut {NoStop}%
\bibitem [{\citenamefont {Zhang}\ \emph
  {et~al.}(2015{\natexlab{b}})\citenamefont {Zhang}, \citenamefont {van
  Hulzen}, \citenamefont {Singh}, \citenamefont {Brownrigg}, \citenamefont
  {Wright}, \citenamefont {van Dijk},\ and\ \citenamefont
  {Wagemaker}}]{Zhang_2015b}%
  \BibitemOpen
  \bibfield  {author} {\bibinfo {author} {\bibfnamefont {X.}~\bibnamefont
  {Zhang}}, \bibinfo {author} {\bibfnamefont {M.}~\bibnamefont {van Hulzen}},
  \bibinfo {author} {\bibfnamefont {D.~P.}\ \bibnamefont {Singh}}, \bibinfo
  {author} {\bibfnamefont {A.}~\bibnamefont {Brownrigg}}, \bibinfo {author}
  {\bibfnamefont {J.~P.}\ \bibnamefont {Wright}}, \bibinfo {author}
  {\bibfnamefont {N.~H.}\ \bibnamefont {van Dijk}}, \ and\ \bibinfo {author}
  {\bibfnamefont {M.}~\bibnamefont {Wagemaker}},\ }\href {\doibase
  10.1038/ncomms9333} {\bibfield  {journal} {\bibinfo  {journal} {Nat Commun}\
  }\textbf {\bibinfo {volume} {6}},\ \bibinfo {pages} {8333} (\bibinfo {year}
  {2015}{\natexlab{b}})}\BibitemShut {NoStop}%
\bibitem [{\citenamefont {O’Connor}\ \emph {et~al.}(2016)\citenamefont
  {O’Connor}, \citenamefont {Welland}, \citenamefont {Kam~Liu},\ and\
  \citenamefont {Voorhees}}]{Oconnor_2016}%
  \BibitemOpen
  \bibfield  {author} {\bibinfo {author} {\bibfnamefont {D.~T.}\ \bibnamefont
  {O’Connor}}, \bibinfo {author} {\bibfnamefont {M.~J.}\ \bibnamefont
  {Welland}}, \bibinfo {author} {\bibfnamefont {W.}~\bibnamefont {Kam~Liu}}, \
  and\ \bibinfo {author} {\bibfnamefont {P.~W.}\ \bibnamefont {Voorhees}},\
  }\href {\doibase 10.1088/0965-0393/24/3/035020} {\bibfield  {journal}
  {\bibinfo  {journal} {Modell. Simul. Mater. Sci. Eng.}\ }\textbf {\bibinfo
  {volume} {24}},\ \bibinfo {pages} {035020} (\bibinfo {year}
  {2016})}\BibitemShut {NoStop}%
\bibitem [{\citenamefont {Ferguson}\ and\ \citenamefont
  {Bazant}(2014)}]{Ferguson_2014}%
  \BibitemOpen
  \bibfield  {author} {\bibinfo {author} {\bibfnamefont {T.~R.}\ \bibnamefont
  {Ferguson}}\ and\ \bibinfo {author} {\bibfnamefont {M.~Z.}\ \bibnamefont
  {Bazant}},\ }\href {\doibase 10.1016/j.electacta.2014.08.083} {\bibfield
  {journal} {\bibinfo  {journal} {Electrochimica Acta}\ }\textbf {\bibinfo
  {volume} {146}},\ \bibinfo {pages} {89} (\bibinfo {year} {2014})}\BibitemShut
  {NoStop}%
\bibitem [{\citenamefont {Guo}\ \emph {et~al.}(2016)\citenamefont {Guo},
  \citenamefont {Smith}, \citenamefont {Yu}, \citenamefont {Efetov},
  \citenamefont {Wang}, \citenamefont {Kim}, \citenamefont {Bazant},\ and\
  \citenamefont {Brus}}]{Guo_2016}%
  \BibitemOpen
  \bibfield  {author} {\bibinfo {author} {\bibfnamefont {Y.}~\bibnamefont
  {Guo}}, \bibinfo {author} {\bibfnamefont {R.~B.}\ \bibnamefont {Smith}},
  \bibinfo {author} {\bibfnamefont {Z.}~\bibnamefont {Yu}}, \bibinfo {author}
  {\bibfnamefont {D.~K.}\ \bibnamefont {Efetov}}, \bibinfo {author}
  {\bibfnamefont {J.}~\bibnamefont {Wang}}, \bibinfo {author} {\bibfnamefont
  {P.}~\bibnamefont {Kim}}, \bibinfo {author} {\bibfnamefont {M.~Z.}\
  \bibnamefont {Bazant}}, \ and\ \bibinfo {author} {\bibfnamefont {L.~E.}\
  \bibnamefont {Brus}},\ }\href {\doibase 10.1021/acs.jpclett.6b00625}
  {\bibfield  {journal} {\bibinfo  {journal} {J Phys Chem Lett}\ }\textbf
  {\bibinfo {volume} {7}},\ \bibinfo {pages} {2151} (\bibinfo {year}
  {2016})}\BibitemShut {NoStop}%
\bibitem [{\citenamefont {Lafont}\ \emph {et~al.}(2010)\citenamefont {Lafont},
  \citenamefont {Carta}, \citenamefont {Mountjoy}, \citenamefont {Chadwick},\
  and\ \citenamefont {Kelder}}]{Lafont_2010}%
  \BibitemOpen
  \bibfield  {author} {\bibinfo {author} {\bibfnamefont {U.}~\bibnamefont
  {Lafont}}, \bibinfo {author} {\bibfnamefont {D.}~\bibnamefont {Carta}},
  \bibinfo {author} {\bibfnamefont {G.}~\bibnamefont {Mountjoy}}, \bibinfo
  {author} {\bibfnamefont {A.~V.}\ \bibnamefont {Chadwick}}, \ and\ \bibinfo
  {author} {\bibfnamefont {E.~M.}\ \bibnamefont {Kelder}},\ }\href {\doibase
  10.1021/jp908786t} {\bibfield  {journal} {\bibinfo  {journal} {Journal of
  Physical Chemistry C}\ }\textbf {\bibinfo {volume} {114}},\ \bibinfo {pages}
  {1372} (\bibinfo {year} {2010})}\BibitemShut {NoStop}%
\bibitem [{\citenamefont {Singh}\ \emph
  {et~al.}(2013{\natexlab{a}})\citenamefont {Singh}, \citenamefont {George},
  \citenamefont {Kumar}, \citenamefont {ten Elshof},\ and\ \citenamefont
  {Wagemaker}}]{Singh_JPC}%
  \BibitemOpen
  \bibfield  {author} {\bibinfo {author} {\bibfnamefont {D.~P.}\ \bibnamefont
  {Singh}}, \bibinfo {author} {\bibfnamefont {A.}~\bibnamefont {George}},
  \bibinfo {author} {\bibfnamefont {R.~V.}\ \bibnamefont {Kumar}}, \bibinfo
  {author} {\bibfnamefont {J.~E.}\ \bibnamefont {ten Elshof}}, \ and\ \bibinfo
  {author} {\bibfnamefont {M.}~\bibnamefont {Wagemaker}},\ }\href {\doibase
  10.1021/jp3118659} {\bibfield  {journal} {\bibinfo  {journal} {Journal of
  Physical Chemistry C}\ }\textbf {\bibinfo {volume} {117}},\ \bibinfo {pages}
  {19809} (\bibinfo {year} {2013}{\natexlab{a}})}\BibitemShut {NoStop}%
\bibitem [{\citenamefont {Shen}\ \emph {et~al.}(2014)\citenamefont {Shen},
  \citenamefont {Chen}, \citenamefont {Klaver}, \citenamefont {Mulder},\ and\
  \citenamefont {Wagemaker}}]{Shen_2014}%
  \BibitemOpen
  \bibfield  {author} {\bibinfo {author} {\bibfnamefont {K.}~\bibnamefont
  {Shen}}, \bibinfo {author} {\bibfnamefont {H.}~\bibnamefont {Chen}}, \bibinfo
  {author} {\bibfnamefont {F.}~\bibnamefont {Klaver}}, \bibinfo {author}
  {\bibfnamefont {F.~M.}\ \bibnamefont {Mulder}}, \ and\ \bibinfo {author}
  {\bibfnamefont {M.}~\bibnamefont {Wagemaker}},\ }\href {\doibase
  10.1021/cm4037346} {\bibfield  {journal} {\bibinfo  {journal} {Chemistry of
  Materials}\ }\textbf {\bibinfo {volume} {26}},\ \bibinfo {pages} {1608}
  (\bibinfo {year} {2014})}\BibitemShut {NoStop}%
\bibitem [{\citenamefont {Wagemaker}\ \emph {et~al.}(2007)\citenamefont
  {Wagemaker}, \citenamefont {Borghols},\ and\ \citenamefont
  {Mulder}}]{Wagemaker_2007A}%
  \BibitemOpen
  \bibfield  {author} {\bibinfo {author} {\bibfnamefont {M.}~\bibnamefont
  {Wagemaker}}, \bibinfo {author} {\bibfnamefont {W.~J.~H.}\ \bibnamefont
  {Borghols}}, \ and\ \bibinfo {author} {\bibfnamefont {F.~M.}\ \bibnamefont
  {Mulder}},\ }\href {\doibase 10.1021/ja067733p} {\bibfield  {journal}
  {\bibinfo  {journal} {Journal of the American Chemical Society}\ }\textbf
  {\bibinfo {volume} {129}},\ \bibinfo {pages} {4323} (\bibinfo {year}
  {2007})}\BibitemShut {NoStop}%
\bibitem [{\citenamefont {Sudant}\ \emph {et~al.}(2005)\citenamefont {Sudant},
  \citenamefont {Baudrin}, \citenamefont {Larcher},\ and\ \citenamefont
  {Tarascon}}]{Sudant_2005}%
  \BibitemOpen
  \bibfield  {author} {\bibinfo {author} {\bibfnamefont {G.}~\bibnamefont
  {Sudant}}, \bibinfo {author} {\bibfnamefont {E.}~\bibnamefont {Baudrin}},
  \bibinfo {author} {\bibfnamefont {D.}~\bibnamefont {Larcher}}, \ and\
  \bibinfo {author} {\bibfnamefont {J.-M.}\ \bibnamefont {Tarascon}},\ }\href
  {\doibase 10.1039/b416176a} {\bibfield  {journal} {\bibinfo  {journal}
  {Journal of Materials Chemistry}\ } (\bibinfo {year} {2005}),\
  10.1039/b416176a}\BibitemShut {NoStop}%
\bibitem [{\citenamefont {Morgan}\ and\ \citenamefont
  {Watson}(2011)}]{Morgan_2011}%
  \BibitemOpen
  \bibfield  {author} {\bibinfo {author} {\bibfnamefont {B.~J.}\ \bibnamefont
  {Morgan}}\ and\ \bibinfo {author} {\bibfnamefont {G.~W.}\ \bibnamefont
  {Watson}},\ }\href {\doibase 10.1021/jz200718e} {\bibfield  {journal}
  {\bibinfo  {journal} {Journal of Physical Chemistry Letters}\ }\textbf
  {\bibinfo {volume} {2}},\ \bibinfo {pages} {1657} (\bibinfo {year}
  {2011})}\BibitemShut {NoStop}%
\bibitem [{\citenamefont {S{\o}ndergaard}\ \emph {et~al.}(2015)\citenamefont
  {S{\o}ndergaard}, \citenamefont {Shen}, \citenamefont {Mamakhel},
  \citenamefont {Marinaro}, \citenamefont {Wohlfahrt-Mehrens}, \citenamefont
  {Wonsyld}, \citenamefont {Dahl},\ and\ \citenamefont
  {Iversen}}]{Sondergaard_2015}%
  \BibitemOpen
  \bibfield  {author} {\bibinfo {author} {\bibfnamefont {M.}~\bibnamefont
  {S{\o}ndergaard}}, \bibinfo {author} {\bibfnamefont {Y.}~\bibnamefont
  {Shen}}, \bibinfo {author} {\bibfnamefont {A.}~\bibnamefont {Mamakhel}},
  \bibinfo {author} {\bibfnamefont {M.}~\bibnamefont {Marinaro}}, \bibinfo
  {author} {\bibfnamefont {M.}~\bibnamefont {Wohlfahrt-Mehrens}}, \bibinfo
  {author} {\bibfnamefont {K.}~\bibnamefont {Wonsyld}}, \bibinfo {author}
  {\bibfnamefont {S.}~\bibnamefont {Dahl}}, \ and\ \bibinfo {author}
  {\bibfnamefont {B.~B.}\ \bibnamefont {Iversen}},\ }\href {\doibase
  10.1021/cm503479h} {\bibfield  {journal} {\bibinfo  {journal} {Chemistry of
  Materials}\ }\textbf {\bibinfo {volume} {27}},\ \bibinfo {pages} {119}
  (\bibinfo {year} {2015})}\BibitemShut {NoStop}%
\bibitem [{\citenamefont {Madej}\ \emph
  {et~al.}(2014{\natexlab{a}})\citenamefont {Madej}, \citenamefont {La~Mantia},
  \citenamefont {Mei}, \citenamefont {Klink}, \citenamefont {Muhler},
  \citenamefont {Schuhmann},\ and\ \citenamefont {Ventosa}}]{Madej_2014A}%
  \BibitemOpen
  \bibfield  {author} {\bibinfo {author} {\bibfnamefont {E.}~\bibnamefont
  {Madej}}, \bibinfo {author} {\bibfnamefont {F.}~\bibnamefont {La~Mantia}},
  \bibinfo {author} {\bibfnamefont {B.}~\bibnamefont {Mei}}, \bibinfo {author}
  {\bibfnamefont {S.}~\bibnamefont {Klink}}, \bibinfo {author} {\bibfnamefont
  {M.}~\bibnamefont {Muhler}}, \bibinfo {author} {\bibfnamefont
  {W.}~\bibnamefont {Schuhmann}}, \ and\ \bibinfo {author} {\bibfnamefont
  {E.}~\bibnamefont {Ventosa}},\ }\href {\doibase
  10.1016/j.jpowsour.2014.05.018} {\bibfield  {journal} {\bibinfo  {journal}
  {Journal of Power Sources}\ }\textbf {\bibinfo {volume} {266}},\ \bibinfo
  {pages} {155} (\bibinfo {year} {2014}{\natexlab{a}})}\BibitemShut {NoStop}%
\bibitem [{\citenamefont {Wang}\ \emph {et~al.}(2011)\citenamefont {Wang},
  \citenamefont {Liu}, \citenamefont {Huang}, \citenamefont {Fang},\ and\
  \citenamefont {Zhuang}}]{Wang_2011}%
  \BibitemOpen
  \bibfield  {author} {\bibinfo {author} {\bibfnamefont {Y.}~\bibnamefont
  {Wang}}, \bibinfo {author} {\bibfnamefont {S.}~\bibnamefont {Liu}}, \bibinfo
  {author} {\bibfnamefont {K.}~\bibnamefont {Huang}}, \bibinfo {author}
  {\bibfnamefont {D.}~\bibnamefont {Fang}}, \ and\ \bibinfo {author}
  {\bibfnamefont {S.}~\bibnamefont {Zhuang}},\ }\href {\doibase
  10.1007/s10008-011-1417-5} {\bibfield  {journal} {\bibinfo  {journal}
  {Journal of Solid State Electrochemistry}\ }\textbf {\bibinfo {volume}
  {16}},\ \bibinfo {pages} {723} (\bibinfo {year} {2011})}\BibitemShut
  {NoStop}%
\bibitem [{\citenamefont {Sun}\ \emph {et~al.}(2010)\citenamefont {Sun},
  \citenamefont {Yang}, \citenamefont {Chen}, \citenamefont {Li}, \citenamefont
  {Lou}, \citenamefont {Li}, \citenamefont {Smith}, \citenamefont {Lu},\ and\
  \citenamefont {Yang}}]{Sun_2010}%
  \BibitemOpen
  \bibfield  {author} {\bibinfo {author} {\bibfnamefont {C.~H.}\ \bibnamefont
  {Sun}}, \bibinfo {author} {\bibfnamefont {X.~H.}\ \bibnamefont {Yang}},
  \bibinfo {author} {\bibfnamefont {J.~S.}\ \bibnamefont {Chen}}, \bibinfo
  {author} {\bibfnamefont {Z.}~\bibnamefont {Li}}, \bibinfo {author}
  {\bibfnamefont {X.~W.}\ \bibnamefont {Lou}}, \bibinfo {author} {\bibfnamefont
  {C.}~\bibnamefont {Li}}, \bibinfo {author} {\bibfnamefont {S.~C.}\
  \bibnamefont {Smith}}, \bibinfo {author} {\bibfnamefont {G.~Q.}\ \bibnamefont
  {Lu}}, \ and\ \bibinfo {author} {\bibfnamefont {H.~G.}\ \bibnamefont
  {Yang}},\ }\href {\doibase 10.1039/c0cc00832j} {\bibfield  {journal}
  {\bibinfo  {journal} {Chemical Communications}\ }\textbf {\bibinfo {volume}
  {46}},\ \bibinfo {pages} {6129} (\bibinfo {year} {2010})}\BibitemShut
  {NoStop}%
\bibitem [{\citenamefont {Madej}\ \emph
  {et~al.}(2014{\natexlab{b}})\citenamefont {Madej}, \citenamefont {Ventosa},
  \citenamefont {Klink}, \citenamefont {Schuhmann},\ and\ \citenamefont
  {La~Mantia}}]{Madej_2014C}%
  \BibitemOpen
  \bibfield  {author} {\bibinfo {author} {\bibfnamefont {E.}~\bibnamefont
  {Madej}}, \bibinfo {author} {\bibfnamefont {E.}~\bibnamefont {Ventosa}},
  \bibinfo {author} {\bibfnamefont {S.}~\bibnamefont {Klink}}, \bibinfo
  {author} {\bibfnamefont {W.}~\bibnamefont {Schuhmann}}, \ and\ \bibinfo
  {author} {\bibfnamefont {F.}~\bibnamefont {La~Mantia}},\ }\href {\doibase
  10.1039/c4cp00630e} {\bibfield  {journal} {\bibinfo  {journal} {Phys Chem
  Chem Phys}\ }\textbf {\bibinfo {volume} {16}},\ \bibinfo {pages} {7939}
  (\bibinfo {year} {2014}{\natexlab{b}})}\BibitemShut {NoStop}%
\bibitem [{\citenamefont {Gentili}\ \emph {et~al.}(2012)\citenamefont
  {Gentili}, \citenamefont {Brutti}, \citenamefont {Hardwick}, \citenamefont
  {Armstrong}, \citenamefont {Panero},\ and\ \citenamefont
  {Bruce}}]{Gentili_2012}%
  \BibitemOpen
  \bibfield  {author} {\bibinfo {author} {\bibfnamefont {V.}~\bibnamefont
  {Gentili}}, \bibinfo {author} {\bibfnamefont {S.}~\bibnamefont {Brutti}},
  \bibinfo {author} {\bibfnamefont {L.~J.}\ \bibnamefont {Hardwick}}, \bibinfo
  {author} {\bibfnamefont {A.~R.}\ \bibnamefont {Armstrong}}, \bibinfo {author}
  {\bibfnamefont {S.}~\bibnamefont {Panero}}, \ and\ \bibinfo {author}
  {\bibfnamefont {P.~G.}\ \bibnamefont {Bruce}},\ }\href {\doibase
  10.1021/cm302912f} {\bibfield  {journal} {\bibinfo  {journal} {Chemistry of
  Materials}\ }\textbf {\bibinfo {volume} {24}},\ \bibinfo {pages} {4468}
  (\bibinfo {year} {2012})}\BibitemShut {NoStop}%
\bibitem [{\citenamefont {Rai}\ \emph {et~al.}(2013)\citenamefont {Rai},
  \citenamefont {Anh}, \citenamefont {Gim}, \citenamefont {Mathew},
  \citenamefont {Kang}, \citenamefont {Paul}, \citenamefont {Song},\ and\
  \citenamefont {Kim}}]{Rai_2013}%
  \BibitemOpen
  \bibfield  {author} {\bibinfo {author} {\bibfnamefont {A.~K.}\ \bibnamefont
  {Rai}}, \bibinfo {author} {\bibfnamefont {L.~T.}\ \bibnamefont {Anh}},
  \bibinfo {author} {\bibfnamefont {J.}~\bibnamefont {Gim}}, \bibinfo {author}
  {\bibfnamefont {V.}~\bibnamefont {Mathew}}, \bibinfo {author} {\bibfnamefont
  {J.}~\bibnamefont {Kang}}, \bibinfo {author} {\bibfnamefont {B.~J.}\
  \bibnamefont {Paul}}, \bibinfo {author} {\bibfnamefont {J.}~\bibnamefont
  {Song}}, \ and\ \bibinfo {author} {\bibfnamefont {J.}~\bibnamefont {Kim}},\
  }\href {\doibase 10.1016/j.electacta.2012.11.104} {\bibfield  {journal}
  {\bibinfo  {journal} {Electrochimica Acta}\ }\textbf {\bibinfo {volume}
  {90}},\ \bibinfo {pages} {112} (\bibinfo {year} {2013})}\BibitemShut
  {NoStop}%
\bibitem [{\citenamefont {Zachau-Christiansen}\ \emph
  {et~al.}(1988)\citenamefont {Zachau-Christiansen}, \citenamefont {West},
  \citenamefont {Jacobsen},\ and\ \citenamefont {Atlung}}]{Zachau_1988}%
  \BibitemOpen
  \bibfield  {author} {\bibinfo {author} {\bibfnamefont {B.}~\bibnamefont
  {Zachau-Christiansen}}, \bibinfo {author} {\bibfnamefont {K.}~\bibnamefont
  {West}}, \bibinfo {author} {\bibfnamefont {T.}~\bibnamefont {Jacobsen}}, \
  and\ \bibinfo {author} {\bibfnamefont {S.}~\bibnamefont {Atlung}},\ }\href
  {\doibase http://dx.doi.org/10.1016/0167-2738(88)90352-9} {\bibfield
  {journal} {\bibinfo  {journal} {Solid State Ionics}\ }\textbf {\bibinfo
  {volume} {28}},\ \bibinfo {pages} {1176} (\bibinfo {year}
  {1988})}\BibitemShut {NoStop}%
\bibitem [{\citenamefont {Macklin}\ and\ \citenamefont
  {Neat}(1992)}]{Macklin_1992}%
  \BibitemOpen
  \bibfield  {author} {\bibinfo {author} {\bibfnamefont {W.~J.}\ \bibnamefont
  {Macklin}}\ and\ \bibinfo {author} {\bibfnamefont {R.~J.}\ \bibnamefont
  {Neat}},\ }\href {\doibase 10.1016/0167-2738(92)90449-y} {\bibfield
  {journal} {\bibinfo  {journal} {Solid State Ionics}\ }\textbf {\bibinfo
  {volume} {53}},\ \bibinfo {pages} {694} (\bibinfo {year} {1992})}\BibitemShut
  {NoStop}%
\bibitem [{\citenamefont {Belak}\ \emph {et~al.}(2012)\citenamefont {Belak},
  \citenamefont {Wang},\ and\ \citenamefont {Van~der Ven}}]{Belak_2012}%
  \BibitemOpen
  \bibfield  {author} {\bibinfo {author} {\bibfnamefont {A.~A.}\ \bibnamefont
  {Belak}}, \bibinfo {author} {\bibfnamefont {Y.}~\bibnamefont {Wang}}, \ and\
  \bibinfo {author} {\bibfnamefont {A.}~\bibnamefont {Van~der Ven}},\ }\href
  {\doibase 10.1021/cm300881t} {\bibfield  {journal} {\bibinfo  {journal}
  {Chemistry of Materials}\ }\textbf {\bibinfo {volume} {24}},\ \bibinfo
  {pages} {2894} (\bibinfo {year} {2012})}\BibitemShut {NoStop}%
\bibitem [{\citenamefont {Yildirim}\ \emph {et~al.}(2011)\citenamefont
  {Yildirim}, \citenamefont {Greeley},\ and\ \citenamefont
  {Sankaranarayanan}}]{Yildirim_2011}%
  \BibitemOpen
  \bibfield  {author} {\bibinfo {author} {\bibfnamefont {H.}~\bibnamefont
  {Yildirim}}, \bibinfo {author} {\bibfnamefont {J.}~\bibnamefont {Greeley}}, \
  and\ \bibinfo {author} {\bibfnamefont {S.~K. R.~S.}\ \bibnamefont
  {Sankaranarayanan}},\ }\href {\doibase 10.1021/jp202514j} {\bibfield
  {journal} {\bibinfo  {journal} {The Journal of Physical Chemistry C}\
  }\textbf {\bibinfo {volume} {115}},\ \bibinfo {pages} {15661} (\bibinfo
  {year} {2011})}\BibitemShut {NoStop}%
\bibitem [{\citenamefont {Borghols}\ \emph
  {et~al.}(2009{\natexlab{a}})\citenamefont {Borghols}, \citenamefont
  {L{\"u}tzenkirchen-Hecht}, \citenamefont {Haake}, \citenamefont {van Eck},
  \citenamefont {Mulder},\ and\ \citenamefont {Wagemaker}}]{Borghols_2009a}%
  \BibitemOpen
  \bibfield  {author} {\bibinfo {author} {\bibfnamefont {W.~J.~H.}\
  \bibnamefont {Borghols}}, \bibinfo {author} {\bibfnamefont {D.}~\bibnamefont
  {L{\"u}tzenkirchen-Hecht}}, \bibinfo {author} {\bibfnamefont
  {U.}~\bibnamefont {Haake}}, \bibinfo {author} {\bibfnamefont {E.~R.~H.}\
  \bibnamefont {van Eck}}, \bibinfo {author} {\bibfnamefont {F.~M.}\
  \bibnamefont {Mulder}}, \ and\ \bibinfo {author} {\bibfnamefont
  {M.}~\bibnamefont {Wagemaker}},\ }\href {\doibase 10.1039/b823142g}
  {\bibfield  {journal} {\bibinfo  {journal} {Physical Chemistry Chemical
  Physics}\ }\textbf {\bibinfo {volume} {11}},\ \bibinfo {pages} {5742}
  (\bibinfo {year} {2009}{\natexlab{a}})}\BibitemShut {NoStop}%
\bibitem [{\citenamefont {Smith}\ and\ \citenamefont
  {Bazant}(2017)}]{Smith_MPET}%
  \BibitemOpen
  \bibfield  {author} {\bibinfo {author} {\bibfnamefont {R.~B.}\ \bibnamefont
  {Smith}}\ and\ \bibinfo {author} {\bibfnamefont {M.~Z.}\ \bibnamefont
  {Bazant}},\ }\href {\doibase 10.1149/2.0171711jes} {\bibfield  {journal}
  {\bibinfo  {journal} {J. Electrochem. Soc.}\ }\textbf {\bibinfo {volume}
  {164}},\ \bibinfo {pages} {E3291} (\bibinfo {year} {2017})}\BibitemShut
  {NoStop}%
\bibitem [{\citenamefont {Munichandraiah}\ \emph {et~al.}(1994)\citenamefont
  {Munichandraiah}, \citenamefont {Scanlon}, \citenamefont {Marsh},
  \citenamefont {Kumar},\ and\ \citenamefont {Sircar}}]{Munich_1994}%
  \BibitemOpen
  \bibfield  {author} {\bibinfo {author} {\bibfnamefont {N.}~\bibnamefont
  {Munichandraiah}}, \bibinfo {author} {\bibfnamefont {L.~G.}\ \bibnamefont
  {Scanlon}}, \bibinfo {author} {\bibfnamefont {R.~A.}\ \bibnamefont {Marsh}},
  \bibinfo {author} {\bibfnamefont {B.}~\bibnamefont {Kumar}}, \ and\ \bibinfo
  {author} {\bibfnamefont {A.~K.}\ \bibnamefont {Sircar}},\ }\href {\doibase
  10.1016/0022-0728(94)87174-4} {\bibfield  {journal} {\bibinfo  {journal}
  {Journal of Electroanalytical Chemistry}\ }\textbf {\bibinfo {volume}
  {379}},\ \bibinfo {pages} {495} (\bibinfo {year} {1994})}\BibitemShut
  {NoStop}%
\bibitem [{\citenamefont {Smith}\ \emph {et~al.}(2017)\citenamefont {Smith},
  \citenamefont {Khoo},\ and\ \citenamefont {Bazant}}]{Smith_2017}%
  \BibitemOpen
  \bibfield  {author} {\bibinfo {author} {\bibfnamefont {R.~B.}\ \bibnamefont
  {Smith}}, \bibinfo {author} {\bibfnamefont {E.}~\bibnamefont {Khoo}}, \ and\
  \bibinfo {author} {\bibfnamefont {M.~Z.}\ \bibnamefont {Bazant}},\ }\href
  {\doibase 10.1021/acs.jpcc.7b00185} {\bibfield  {journal} {\bibinfo
  {journal} {J. Phys. Chem. C}\ }\textbf {\bibinfo {volume} {121}},\ \bibinfo
  {pages} {12505} (\bibinfo {year} {2017})}\BibitemShut {NoStop}%
\bibitem [{\citenamefont {Bai}\ \emph {et~al.}(2011)\citenamefont {Bai},
  \citenamefont {Cogswell},\ and\ \citenamefont {Bazant}}]{Bai_2011}%
  \BibitemOpen
  \bibfield  {author} {\bibinfo {author} {\bibfnamefont {P.}~\bibnamefont
  {Bai}}, \bibinfo {author} {\bibfnamefont {D.~A.}\ \bibnamefont {Cogswell}}, \
  and\ \bibinfo {author} {\bibfnamefont {M.~Z.}\ \bibnamefont {Bazant}},\
  }\href {\doibase 10.1021/nl202764f} {\bibfield  {journal} {\bibinfo
  {journal} {Nano Letters}\ }\textbf {\bibinfo {volume} {11}},\ \bibinfo
  {pages} {4890} (\bibinfo {year} {2011})}\BibitemShut {NoStop}%
\bibitem [{\citenamefont {Borghols}\ \emph
  {et~al.}(2009{\natexlab{b}})\citenamefont {Borghols}, \citenamefont
  {Wagemaker}, \citenamefont {Lafont}, \citenamefont {Kelder},\ and\
  \citenamefont {Mulder}}]{Borghols_2009b}%
  \BibitemOpen
  \bibfield  {author} {\bibinfo {author} {\bibfnamefont {W.~J.}\ \bibnamefont
  {Borghols}}, \bibinfo {author} {\bibfnamefont {M.}~\bibnamefont {Wagemaker}},
  \bibinfo {author} {\bibfnamefont {U.}~\bibnamefont {Lafont}}, \bibinfo
  {author} {\bibfnamefont {E.~M.}\ \bibnamefont {Kelder}}, \ and\ \bibinfo
  {author} {\bibfnamefont {F.~M.}\ \bibnamefont {Mulder}},\ }\href {\doibase
  10.1021/ja902423e} {\bibfield  {journal} {\bibinfo  {journal} {J Am Chem
  Soc}\ }\textbf {\bibinfo {volume} {131}},\ \bibinfo {pages} {17786} (\bibinfo
  {year} {2009}{\natexlab{b}})}\BibitemShut {NoStop}%
\bibitem [{\citenamefont {Lunell}\ \emph {et~al.}(1997)\citenamefont {Lunell},
  \citenamefont {Stashans}, \citenamefont {Ojamäe}, \citenamefont
  {Lindström},\ and\ \citenamefont {Hagfeldt}}]{Lunell_1997}%
  \BibitemOpen
  \bibfield  {author} {\bibinfo {author} {\bibfnamefont {S.}~\bibnamefont
  {Lunell}}, \bibinfo {author} {\bibfnamefont {A.}~\bibnamefont {Stashans}},
  \bibinfo {author} {\bibfnamefont {L.}~\bibnamefont {Ojamäe}}, \bibinfo
  {author} {\bibfnamefont {H.}~\bibnamefont {Lindström}}, \ and\ \bibinfo
  {author} {\bibfnamefont {A.}~\bibnamefont {Hagfeldt}},\ }\href {\doibase
  10.1021/ja9708629} {\bibfield  {journal} {\bibinfo  {journal} {Journal of the
  American Chemical Society}\ }\textbf {\bibinfo {volume} {119}},\ \bibinfo
  {pages} {7374} (\bibinfo {year} {1997})}\BibitemShut {NoStop}%
\bibitem [{\citenamefont {Tielens}\ \emph {et~al.}(2005)\citenamefont
  {Tielens}, \citenamefont {Calatayud}, \citenamefont {Beltran}, \citenamefont
  {Minot},\ and\ \citenamefont {Andres}}]{Tielens_2005}%
  \BibitemOpen
  \bibfield  {author} {\bibinfo {author} {\bibfnamefont {F.}~\bibnamefont
  {Tielens}}, \bibinfo {author} {\bibfnamefont {M.}~\bibnamefont {Calatayud}},
  \bibinfo {author} {\bibfnamefont {A.}~\bibnamefont {Beltran}}, \bibinfo
  {author} {\bibfnamefont {C.}~\bibnamefont {Minot}}, \ and\ \bibinfo {author}
  {\bibfnamefont {J.}~\bibnamefont {Andres}},\ }\href {\doibase
  10.1016/j.jelechem.2005.04.009} {\bibfield  {journal} {\bibinfo  {journal}
  {Journal of Electroanalytical Chemistry}\ }\textbf {\bibinfo {volume}
  {581}},\ \bibinfo {pages} {216} (\bibinfo {year} {2005})}\BibitemShut
  {NoStop}%
\bibitem [{\citenamefont {Sussman}\ \emph {et~al.}(2014)\citenamefont
  {Sussman}, \citenamefont {Yasin},\ and\ \citenamefont
  {Demopoulos}}]{Sussman_2014}%
  \BibitemOpen
  \bibfield  {author} {\bibinfo {author} {\bibfnamefont {M.~J.}\ \bibnamefont
  {Sussman}}, \bibinfo {author} {\bibfnamefont {A.}~\bibnamefont {Yasin}}, \
  and\ \bibinfo {author} {\bibfnamefont {G.~P.}\ \bibnamefont {Demopoulos}},\
  }\href {\doibase 10.1016/j.jpowsour.2014.08.050} {\bibfield  {journal}
  {\bibinfo  {journal} {Journal of Power Sources}\ }\textbf {\bibinfo {volume}
  {272}},\ \bibinfo {pages} {58} (\bibinfo {year} {2014})}\BibitemShut
  {NoStop}%
\bibitem [{\citenamefont {Nauman}\ and\ \citenamefont
  {He}(2001)}]{Nauman_2001}%
  \BibitemOpen
  \bibfield  {author} {\bibinfo {author} {\bibfnamefont {E.~B.}\ \bibnamefont
  {Nauman}}\ and\ \bibinfo {author} {\bibfnamefont {D.~Q.}\ \bibnamefont
  {He}},\ }\href {\doibase 10.1016/s0009-2509(01)00005-7} {\bibfield  {journal}
  {\bibinfo  {journal} {Chem. Eng. Sci.}\ }\textbf {\bibinfo {volume} {56}},\
  \bibinfo {pages} {1999} (\bibinfo {year} {2001})}\BibitemShut {NoStop}%
\bibitem [{\citenamefont {Zeng}\ and\ \citenamefont
  {Bazant}(2014)}]{Zeng_2014}%
  \BibitemOpen
  \bibfield  {author} {\bibinfo {author} {\bibfnamefont {Y.}~\bibnamefont
  {Zeng}}\ and\ \bibinfo {author} {\bibfnamefont {M.~Z.}\ \bibnamefont
  {Bazant}},\ }\href {\doibase 10.1137/130937548} {\bibfield  {journal}
  {\bibinfo  {journal} {SIAM Journal on Applied Mathematics}\ }\textbf
  {\bibinfo {volume} {74}},\ \bibinfo {pages} {980} (\bibinfo {year}
  {2014})}\BibitemShut {NoStop}%
\bibitem [{\citenamefont {Wang}\ \emph {et~al.}(2007)\citenamefont {Wang},
  \citenamefont {Polleux}, \citenamefont {Lim},\ and\ \citenamefont
  {Dunn}}]{Wang_2007}%
  \BibitemOpen
  \bibfield  {author} {\bibinfo {author} {\bibfnamefont {J.}~\bibnamefont
  {Wang}}, \bibinfo {author} {\bibfnamefont {J.}~\bibnamefont {Polleux}},
  \bibinfo {author} {\bibfnamefont {J.}~\bibnamefont {Lim}}, \ and\ \bibinfo
  {author} {\bibfnamefont {B.}~\bibnamefont {Dunn}},\ }\href {\doibase
  10.1021/jp074464w} {\bibfield  {journal} {\bibinfo  {journal} {Journal of
  Physical Chemistry C}\ }\textbf {\bibinfo {volume} {111}},\ \bibinfo {pages}
  {14925} (\bibinfo {year} {2007})}\BibitemShut {NoStop}%
\bibitem [{\citenamefont {Kim}\ and\ \citenamefont {Choi}(2015)}]{Hwan_2015}%
  \BibitemOpen
  \bibfield  {author} {\bibinfo {author} {\bibfnamefont {S.-H.}\ \bibnamefont
  {Kim}}\ and\ \bibinfo {author} {\bibfnamefont {S.-Y.}\ \bibnamefont {Choi}},\
  }\href {\doibase 10.1016/j.jelechem.2015.03.007} {\bibfield  {journal}
  {\bibinfo  {journal} {Journal of Electroanalytical Chemistry}\ }\textbf
  {\bibinfo {volume} {744}},\ \bibinfo {pages} {45} (\bibinfo {year}
  {2015})}\BibitemShut {NoStop}%
\bibitem [{\citenamefont {Lindstrom}\ \emph {et~al.}(1997)\citenamefont
  {Lindstrom}, \citenamefont {Sodergren}, \citenamefont {Solbrand},
  \citenamefont {Rensmo}, \citenamefont {Hjelm}, \citenamefont {Hagfeldt},\
  and\ \citenamefont {Lindquist}}]{Lindstrom_1997}%
  \BibitemOpen
  \bibfield  {author} {\bibinfo {author} {\bibfnamefont {H.}~\bibnamefont
  {Lindstrom}}, \bibinfo {author} {\bibfnamefont {S.}~\bibnamefont
  {Sodergren}}, \bibinfo {author} {\bibfnamefont {A.}~\bibnamefont {Solbrand}},
  \bibinfo {author} {\bibfnamefont {H.}~\bibnamefont {Rensmo}}, \bibinfo
  {author} {\bibfnamefont {J.}~\bibnamefont {Hjelm}}, \bibinfo {author}
  {\bibfnamefont {A.}~\bibnamefont {Hagfeldt}}, \ and\ \bibinfo {author}
  {\bibfnamefont {S.~E.}\ \bibnamefont {Lindquist}},\ }\href {\doibase
  10.1021/jp970490q} {\bibfield  {journal} {\bibinfo  {journal} {Journal of
  Physical Chemistry B}\ }\textbf {\bibinfo {volume} {101}},\ \bibinfo {pages}
  {7717} (\bibinfo {year} {1997})}\BibitemShut {NoStop}%
\bibitem [{\citenamefont {Wagemaker}\ \emph {et~al.}(2003)\citenamefont
  {Wagemaker}, \citenamefont {Kearley}, \citenamefont {van Well}, \citenamefont
  {Mutka},\ and\ \citenamefont {Mulder}}]{Wagemaker_2003}%
  \BibitemOpen
  \bibfield  {author} {\bibinfo {author} {\bibfnamefont {M.}~\bibnamefont
  {Wagemaker}}, \bibinfo {author} {\bibfnamefont {G.~J.}\ \bibnamefont
  {Kearley}}, \bibinfo {author} {\bibfnamefont {A.~A.}\ \bibnamefont {van
  Well}}, \bibinfo {author} {\bibfnamefont {H.}~\bibnamefont {Mutka}}, \ and\
  \bibinfo {author} {\bibfnamefont {F.~M.}\ \bibnamefont {Mulder}},\ }\href
  {\doibase 10.1021/ja028165q} {\bibfield  {journal} {\bibinfo  {journal}
  {Journal of the American Chemical Society}\ }\textbf {\bibinfo {volume}
  {125}},\ \bibinfo {pages} {840} (\bibinfo {year} {2003})}\BibitemShut
  {NoStop}%
\bibitem [{\citenamefont {Ganapathy}\ \emph {et~al.}(2011)\citenamefont
  {Ganapathy}, \citenamefont {van Eck}, \citenamefont {Kentgens}, \citenamefont
  {Mulder},\ and\ \citenamefont {Wagemaker}}]{Ganapathy_2009}%
  \BibitemOpen
  \bibfield  {author} {\bibinfo {author} {\bibfnamefont {S.}~\bibnamefont
  {Ganapathy}}, \bibinfo {author} {\bibfnamefont {E.~R.}\ \bibnamefont {van
  Eck}}, \bibinfo {author} {\bibfnamefont {A.~P.}\ \bibnamefont {Kentgens}},
  \bibinfo {author} {\bibfnamefont {F.~M.}\ \bibnamefont {Mulder}}, \ and\
  \bibinfo {author} {\bibfnamefont {M.}~\bibnamefont {Wagemaker}},\ }\href
  {\doibase 10.1002/chem.201101431} {\bibfield  {journal} {\bibinfo  {journal}
  {Chemistry}\ }\textbf {\bibinfo {volume} {17}},\ \bibinfo {pages} {14811}
  (\bibinfo {year} {2011})}\BibitemShut {NoStop}%
\bibitem [{\citenamefont {Shin}\ \emph {et~al.}(2012)\citenamefont {Shin},
  \citenamefont {Joo}, \citenamefont {Samuelis},\ and\ \citenamefont
  {Maier}}]{Shin_2012}%
  \BibitemOpen
  \bibfield  {author} {\bibinfo {author} {\bibfnamefont {J.-Y.}\ \bibnamefont
  {Shin}}, \bibinfo {author} {\bibfnamefont {J.~H.}\ \bibnamefont {Joo}},
  \bibinfo {author} {\bibfnamefont {D.}~\bibnamefont {Samuelis}}, \ and\
  \bibinfo {author} {\bibfnamefont {J.}~\bibnamefont {Maier}},\ }\href
  {\doibase 10.1021/cm2031009} {\bibfield  {journal} {\bibinfo  {journal}
  {Chemistry of Materials}\ }\textbf {\bibinfo {volume} {24}},\ \bibinfo
  {pages} {543} (\bibinfo {year} {2012})}\BibitemShut {NoStop}%
\bibitem [{\citenamefont {Kavan}(2014)}]{Kavan_2014}%
  \BibitemOpen
  \bibfield  {author} {\bibinfo {author} {\bibfnamefont {L.}~\bibnamefont
  {Kavan}},\ }\href {\doibase 10.1007/s10008-014-2435-x} {\bibfield  {journal}
  {\bibinfo  {journal} {Journal of Solid State Electrochemistry}\ }\textbf
  {\bibinfo {volume} {18}},\ \bibinfo {pages} {2297} (\bibinfo {year}
  {2014})}\BibitemShut {NoStop}%
\bibitem [{\citenamefont {Wang}\ \emph {et~al.}(2014)\citenamefont {Wang},
  \citenamefont {Sha}, \citenamefont {Liu}, \citenamefont {He}, \citenamefont
  {Shi}, \citenamefont {Li},\ and\ \citenamefont {Zhao}}]{Zhiyuan_2014}%
  \BibitemOpen
  \bibfield  {author} {\bibinfo {author} {\bibfnamefont {Z.}~\bibnamefont
  {Wang}}, \bibinfo {author} {\bibfnamefont {J.}~\bibnamefont {Sha}}, \bibinfo
  {author} {\bibfnamefont {E.}~\bibnamefont {Liu}}, \bibinfo {author}
  {\bibfnamefont {C.}~\bibnamefont {He}}, \bibinfo {author} {\bibfnamefont
  {C.}~\bibnamefont {Shi}}, \bibinfo {author} {\bibfnamefont {J.}~\bibnamefont
  {Li}}, \ and\ \bibinfo {author} {\bibfnamefont {N.}~\bibnamefont {Zhao}},\
  }\href {\doibase 10.1039/c4ta00574k} {\bibfield  {journal} {\bibinfo
  {journal} {Journal of Materials Chemistry A}\ }\textbf {\bibinfo {volume}
  {2}},\ \bibinfo {pages} {8893} (\bibinfo {year} {2014})}\BibitemShut
  {NoStop}%
\bibitem [{\citenamefont {Yang}\ \emph {et~al.}(2015)\citenamefont {Yang},
  \citenamefont {Yang}, \citenamefont {Hou}, \citenamefont {Zhang},
  \citenamefont {Fang}, \citenamefont {Chen},\ and\ \citenamefont
  {Ji}}]{Yang_2015}%
  \BibitemOpen
  \bibfield  {author} {\bibinfo {author} {\bibfnamefont {X.}~\bibnamefont
  {Yang}}, \bibinfo {author} {\bibfnamefont {Y.}~\bibnamefont {Yang}}, \bibinfo
  {author} {\bibfnamefont {H.}~\bibnamefont {Hou}}, \bibinfo {author}
  {\bibfnamefont {Y.}~\bibnamefont {Zhang}}, \bibinfo {author} {\bibfnamefont
  {L.}~\bibnamefont {Fang}}, \bibinfo {author} {\bibfnamefont {J.}~\bibnamefont
  {Chen}}, \ and\ \bibinfo {author} {\bibfnamefont {X.}~\bibnamefont {Ji}},\
  }\href {\doibase 10.1021/jp512289g} {\bibfield  {journal} {\bibinfo
  {journal} {Journal of Physical Chemistry C}\ }\textbf {\bibinfo {volume}
  {119}},\ \bibinfo {pages} {3923} (\bibinfo {year} {2015})}\BibitemShut
  {NoStop}%
\bibitem [{\citenamefont {Wagemaker}\ \emph {et~al.}(2001)\citenamefont
  {Wagemaker}, \citenamefont {van~de Krol}, \citenamefont {Kentgens},
  \citenamefont {van Well},\ and\ \citenamefont {Mulder}}]{Wagemaker_2001}%
  \BibitemOpen
  \bibfield  {author} {\bibinfo {author} {\bibfnamefont {M.}~\bibnamefont
  {Wagemaker}}, \bibinfo {author} {\bibfnamefont {R.}~\bibnamefont {van~de
  Krol}}, \bibinfo {author} {\bibfnamefont {A.~P.~M.}\ \bibnamefont
  {Kentgens}}, \bibinfo {author} {\bibfnamefont {A.~A.}\ \bibnamefont {van
  Well}}, \ and\ \bibinfo {author} {\bibfnamefont {F.~M.}\ \bibnamefont
  {Mulder}},\ }\href {\doibase 10.1021/ja0161148} {\bibfield  {journal}
  {\bibinfo  {journal} {Journal of the American Chemical Society}\ }\textbf
  {\bibinfo {volume} {123}},\ \bibinfo {pages} {11454} (\bibinfo {year}
  {2001})}\BibitemShut {NoStop}%
\bibitem [{\citenamefont {Wagemaker}\ \emph {et~al.}(2002)\citenamefont
  {Wagemaker}, \citenamefont {Kentgens},\ and\ \citenamefont
  {Mulder}}]{Wagemaker_2002}%
  \BibitemOpen
  \bibfield  {author} {\bibinfo {author} {\bibfnamefont {M.}~\bibnamefont
  {Wagemaker}}, \bibinfo {author} {\bibfnamefont {A.~P.~M.}\ \bibnamefont
  {Kentgens}}, \ and\ \bibinfo {author} {\bibfnamefont {F.~M.}\ \bibnamefont
  {Mulder}},\ }\href {\doibase 10.1038/nature00901} {\bibfield  {journal}
  {\bibinfo  {journal} {Nature}\ }\textbf {\bibinfo {volume} {418}},\ \bibinfo
  {pages} {397} (\bibinfo {year} {2002})}\BibitemShut {NoStop}%
\bibitem [{\citenamefont {Smith}(2017)}]{MPET_code}%
  \BibitemOpen
  \bibfield  {author} {\bibinfo {author} {\bibfnamefont {R.~B.}\ \bibnamefont
  {Smith}},\ }\href@noop {} {\bibfield  {journal} {\bibinfo  {journal}
  {https://bitbucket.org/bazantgroup/mpet}\ } (\bibinfo {year}
  {2017})}\BibitemShut {NoStop}%
\bibitem [{\citenamefont {Nikoli\'{c}}(2016)}]{Nikolic_2016}%
  \BibitemOpen
  \bibfield  {author} {\bibinfo {author} {\bibfnamefont {D.~D.}\ \bibnamefont
  {Nikoli\'{c}}},\ }\href {\doibase 10.7717/peerj-cs.54} {\bibfield  {journal}
  {\bibinfo  {journal} {PeerJ Computer Science}\ }\textbf {\bibinfo {volume}
  {2}},\ \bibinfo {pages} {e54} (\bibinfo {year} {2016})}\BibitemShut {NoStop}%
\bibitem [{\citenamefont {Kim}\ \emph {et~al.}(2010)\citenamefont {Kim},
  \citenamefont {Kim},\ and\ \citenamefont {Cho}}]{Kim_2010}%
  \BibitemOpen
  \bibfield  {author} {\bibinfo {author} {\bibfnamefont {M.~G.}\ \bibnamefont
  {Kim}}, \bibinfo {author} {\bibfnamefont {H.}~\bibnamefont {Kim}}, \ and\
  \bibinfo {author} {\bibfnamefont {J.}~\bibnamefont {Cho}},\ }\href {\doibase
  10.1149/1.3425619} {\bibfield  {journal} {\bibinfo  {journal} {Journal of The
  Electrochemical Society}\ }\textbf {\bibinfo {volume} {157}},\ \bibinfo
  {pages} {A802} (\bibinfo {year} {2010})}\BibitemShut {NoStop}%
\bibitem [{\citenamefont {Bruce}\ \emph {et~al.}(2008)\citenamefont {Bruce},
  \citenamefont {Scrosati},\ and\ \citenamefont {Tarascon}}]{Bruce_2008}%
  \BibitemOpen
  \bibfield  {author} {\bibinfo {author} {\bibfnamefont {P.~G.}\ \bibnamefont
  {Bruce}}, \bibinfo {author} {\bibfnamefont {B.}~\bibnamefont {Scrosati}}, \
  and\ \bibinfo {author} {\bibfnamefont {J.~M.}\ \bibnamefont {Tarascon}},\
  }\href {\doibase 10.1002/anie.200702505} {\bibfield  {journal} {\bibinfo
  {journal} {Angew Chem Int Ed Engl}\ }\textbf {\bibinfo {volume} {47}},\
  \bibinfo {pages} {2930} (\bibinfo {year} {2008})}\BibitemShut {NoStop}%
\bibitem [{\citenamefont {Orvananos}\ \emph {et~al.}(2015)\citenamefont
  {Orvananos}, \citenamefont {Yu}, \citenamefont {Abdellahi}, \citenamefont
  {Malik}, \citenamefont {Grey}, \citenamefont {Ceder},\ and\ \citenamefont
  {Thornton}}]{Orvananos_2015}%
  \BibitemOpen
  \bibfield  {author} {\bibinfo {author} {\bibfnamefont {B.}~\bibnamefont
  {Orvananos}}, \bibinfo {author} {\bibfnamefont {H.~C.}\ \bibnamefont {Yu}},
  \bibinfo {author} {\bibfnamefont {A.}~\bibnamefont {Abdellahi}}, \bibinfo
  {author} {\bibfnamefont {R.}~\bibnamefont {Malik}}, \bibinfo {author}
  {\bibfnamefont {C.~P.}\ \bibnamefont {Grey}}, \bibinfo {author}
  {\bibfnamefont {G.}~\bibnamefont {Ceder}}, \ and\ \bibinfo {author}
  {\bibfnamefont {K.}~\bibnamefont {Thornton}},\ }\href {\doibase
  10.1149/2.0481506jes} {\bibfield  {journal} {\bibinfo  {journal} {Journal of
  the Electrochemical Society}\ }\textbf {\bibinfo {volume} {162}},\ \bibinfo
  {pages} {A965} (\bibinfo {year} {2015})}\BibitemShut {NoStop}%
\bibitem [{\citenamefont {Liu}\ \emph {et~al.}(2016)\citenamefont {Liu},
  \citenamefont {Verhallen}, \citenamefont {Singh}, \citenamefont {Wang},
  \citenamefont {Wagemaker},\ and\ \citenamefont {Barnett}}]{Liu_2016}%
  \BibitemOpen
  \bibfield  {author} {\bibinfo {author} {\bibfnamefont {Z.}~\bibnamefont
  {Liu}}, \bibinfo {author} {\bibfnamefont {T.~W.}\ \bibnamefont {Verhallen}},
  \bibinfo {author} {\bibfnamefont {D.~P.}\ \bibnamefont {Singh}}, \bibinfo
  {author} {\bibfnamefont {H.}~\bibnamefont {Wang}}, \bibinfo {author}
  {\bibfnamefont {M.}~\bibnamefont {Wagemaker}}, \ and\ \bibinfo {author}
  {\bibfnamefont {S.}~\bibnamefont {Barnett}},\ }\href {\doibase
  10.1016/j.jpowsour.2016.05.097} {\bibfield  {journal} {\bibinfo  {journal}
  {Journal of Power Sources}\ }\textbf {\bibinfo {volume} {324}},\ \bibinfo
  {pages} {358} (\bibinfo {year} {2016})}\BibitemShut {NoStop}%
\bibitem [{\citenamefont {Strobridge}\ \emph {et~al.}(2015)\citenamefont
  {Strobridge}, \citenamefont {Orvananos}, \citenamefont {Croft}, \citenamefont
  {Yu}, \citenamefont {Robert}, \citenamefont {Liu}, \citenamefont {Zhong},
  \citenamefont {Connolley}, \citenamefont {Drakopoulos}, \citenamefont
  {Thornton},\ and\ \citenamefont {Grey}}]{Strobridge_2015}%
  \BibitemOpen
  \bibfield  {author} {\bibinfo {author} {\bibfnamefont {F.~C.}\ \bibnamefont
  {Strobridge}}, \bibinfo {author} {\bibfnamefont {B.}~\bibnamefont
  {Orvananos}}, \bibinfo {author} {\bibfnamefont {M.}~\bibnamefont {Croft}},
  \bibinfo {author} {\bibfnamefont {H.-C.}\ \bibnamefont {Yu}}, \bibinfo
  {author} {\bibfnamefont {R.}~\bibnamefont {Robert}}, \bibinfo {author}
  {\bibfnamefont {H.}~\bibnamefont {Liu}}, \bibinfo {author} {\bibfnamefont
  {Z.}~\bibnamefont {Zhong}}, \bibinfo {author} {\bibfnamefont
  {T.}~\bibnamefont {Connolley}}, \bibinfo {author} {\bibfnamefont
  {M.}~\bibnamefont {Drakopoulos}}, \bibinfo {author} {\bibfnamefont
  {K.}~\bibnamefont {Thornton}}, \ and\ \bibinfo {author} {\bibfnamefont
  {C.~P.}\ \bibnamefont {Grey}},\ }\href {\doibase 10.1021/cm504317a}
  {\bibfield  {journal} {\bibinfo  {journal} {Chemistry of Materials}\ }\textbf
  {\bibinfo {volume} {27}},\ \bibinfo {pages} {2374} (\bibinfo {year}
  {2015})}\BibitemShut {NoStop}%
\bibitem [{\citenamefont {Li}\ \emph {et~al.}(2015)\citenamefont {Li},
  \citenamefont {Meyer}, \citenamefont {Lim}, \citenamefont {Lee},
  \citenamefont {Gent}, \citenamefont {Marchesini}, \citenamefont {Krishnan},
  \citenamefont {Tyliszczak}, \citenamefont {Shapiro}, \citenamefont
  {Kilcoyne},\ and\ \citenamefont {Chueh}}]{Li_2015}%
  \BibitemOpen
  \bibfield  {author} {\bibinfo {author} {\bibfnamefont {Y.}~\bibnamefont
  {Li}}, \bibinfo {author} {\bibfnamefont {S.}~\bibnamefont {Meyer}}, \bibinfo
  {author} {\bibfnamefont {J.}~\bibnamefont {Lim}}, \bibinfo {author}
  {\bibfnamefont {S.~C.}\ \bibnamefont {Lee}}, \bibinfo {author} {\bibfnamefont
  {W.~E.}\ \bibnamefont {Gent}}, \bibinfo {author} {\bibfnamefont
  {S.}~\bibnamefont {Marchesini}}, \bibinfo {author} {\bibfnamefont
  {H.}~\bibnamefont {Krishnan}}, \bibinfo {author} {\bibfnamefont
  {T.}~\bibnamefont {Tyliszczak}}, \bibinfo {author} {\bibfnamefont
  {D.}~\bibnamefont {Shapiro}}, \bibinfo {author} {\bibfnamefont {A.~L.}\
  \bibnamefont {Kilcoyne}}, \ and\ \bibinfo {author} {\bibfnamefont {W.~C.}\
  \bibnamefont {Chueh}},\ }\href {\doibase 10.1002/adma.201502276} {\bibfield
  {journal} {\bibinfo  {journal} {Adv. Mater.}\ }\textbf {\bibinfo {volume}
  {27}},\ \bibinfo {pages} {6591} (\bibinfo {year} {2015})}\BibitemShut
  {NoStop}%
\bibitem [{\citenamefont {Singh}\ \emph
  {et~al.}(2013{\natexlab{b}})\citenamefont {Singh}, \citenamefont {Mulder},\
  and\ \citenamefont {Wagemaker}}]{Singh_2013}%
  \BibitemOpen
  \bibfield  {author} {\bibinfo {author} {\bibfnamefont {D.~P.}\ \bibnamefont
  {Singh}}, \bibinfo {author} {\bibfnamefont {F.~M.}\ \bibnamefont {Mulder}}, \
  and\ \bibinfo {author} {\bibfnamefont {M.}~\bibnamefont {Wagemaker}},\ }\href
  {\doibase 10.1016/j.elecom.2013.08.014} {\bibfield  {journal} {\bibinfo
  {journal} {Electrochemistry Communications}\ }\textbf {\bibinfo {volume}
  {35}},\ \bibinfo {pages} {124} (\bibinfo {year}
  {2013}{\natexlab{b}})}\BibitemShut {NoStop}%
\bibitem [{\citenamefont {Kim}\ \emph {et~al.}(2013)\citenamefont {Kim},
  \citenamefont {Norberg}, \citenamefont {Alexander}, \citenamefont
  {Kostecki},\ and\ \citenamefont {Cabana}}]{Kim_2013}%
  \BibitemOpen
  \bibfield  {author} {\bibinfo {author} {\bibfnamefont {C.}~\bibnamefont
  {Kim}}, \bibinfo {author} {\bibfnamefont {N.~S.}\ \bibnamefont {Norberg}},
  \bibinfo {author} {\bibfnamefont {C.~T.}\ \bibnamefont {Alexander}}, \bibinfo
  {author} {\bibfnamefont {R.}~\bibnamefont {Kostecki}}, \ and\ \bibinfo
  {author} {\bibfnamefont {J.}~\bibnamefont {Cabana}},\ }\href {\doibase
  10.1002/adfm.201201684} {\bibfield  {journal} {\bibinfo  {journal} {Advanced
  Functional Materials}\ }\textbf {\bibinfo {volume} {23}},\ \bibinfo {pages}
  {1214} (\bibinfo {year} {2013})}\BibitemShut {NoStop}%
\bibitem [{\citenamefont {Ren}\ \emph {et~al.}(2014)\citenamefont {Ren},
  \citenamefont {Li},\ and\ \citenamefont {Yu}}]{Ren_2014}%
  \BibitemOpen
  \bibfield  {author} {\bibinfo {author} {\bibfnamefont {Y.}~\bibnamefont
  {Ren}}, \bibinfo {author} {\bibfnamefont {J.}~\bibnamefont {Li}}, \ and\
  \bibinfo {author} {\bibfnamefont {J.}~\bibnamefont {Yu}},\ }\href {\doibase
  10.1016/j.electacta.2014.06.068} {\bibfield  {journal} {\bibinfo  {journal}
  {Electrochimica Acta}\ }\textbf {\bibinfo {volume} {138}},\ \bibinfo {pages}
  {41} (\bibinfo {year} {2014})}\BibitemShut {NoStop}%
\bibitem [{\citenamefont {Shin}\ \emph {et~al.}(2011)\citenamefont {Shin},
  \citenamefont {Samuelis},\ and\ \citenamefont {Maier}}]{Shin_2011}%
  \BibitemOpen
  \bibfield  {author} {\bibinfo {author} {\bibfnamefont {J.-Y.}\ \bibnamefont
  {Shin}}, \bibinfo {author} {\bibfnamefont {D.}~\bibnamefont {Samuelis}}, \
  and\ \bibinfo {author} {\bibfnamefont {J.}~\bibnamefont {Maier}},\ }\href
  {\doibase 10.1002/adfm.201002527} {\bibfield  {journal} {\bibinfo  {journal}
  {Advanced Functional Materials}\ }\textbf {\bibinfo {volume} {21}},\ \bibinfo
  {pages} {3464} (\bibinfo {year} {2011})}\BibitemShut {NoStop}%
\bibitem [{\citenamefont {Madej}\ \emph {et~al.}(2015)\citenamefont {Madej},
  \citenamefont {Klink}, \citenamefont {Schuhmann}, \citenamefont {Ventosa},\
  and\ \citenamefont {La~Mantia}}]{Madej_2015}%
  \BibitemOpen
  \bibfield  {author} {\bibinfo {author} {\bibfnamefont {E.}~\bibnamefont
  {Madej}}, \bibinfo {author} {\bibfnamefont {S.}~\bibnamefont {Klink}},
  \bibinfo {author} {\bibfnamefont {W.}~\bibnamefont {Schuhmann}}, \bibinfo
  {author} {\bibfnamefont {E.}~\bibnamefont {Ventosa}}, \ and\ \bibinfo
  {author} {\bibfnamefont {F.}~\bibnamefont {La~Mantia}},\ }\href {\doibase
  10.1016/j.jpowsour.2015.07.079} {\bibfield  {journal} {\bibinfo  {journal}
  {Journal of Power Sources}\ }\textbf {\bibinfo {volume} {297}},\ \bibinfo
  {pages} {140} (\bibinfo {year} {2015})}\BibitemShut {NoStop}%
\bibitem [{\citenamefont {Madej}\ \emph
  {et~al.}(2014{\natexlab{c}})\citenamefont {Madej}, \citenamefont {La~Mantia},
  \citenamefont {Schuhmann},\ and\ \citenamefont {Ventosa}}]{Madej_2014B}%
  \BibitemOpen
  \bibfield  {author} {\bibinfo {author} {\bibfnamefont {E.}~\bibnamefont
  {Madej}}, \bibinfo {author} {\bibfnamefont {F.}~\bibnamefont {La~Mantia}},
  \bibinfo {author} {\bibfnamefont {W.}~\bibnamefont {Schuhmann}}, \ and\
  \bibinfo {author} {\bibfnamefont {E.}~\bibnamefont {Ventosa}},\ }\href
  {\doibase 10.1002/aenm.201400829} {\bibfield  {journal} {\bibinfo  {journal}
  {Advanced Energy Materials}\ }\textbf {\bibinfo {volume} {4}},\ \bibinfo
  {pages} {n/a} (\bibinfo {year} {2014}{\natexlab{c}})}\BibitemShut {NoStop}%
\bibitem [{\citenamefont {Wagemaker}\ \emph {et~al.}(2009)\citenamefont
  {Wagemaker}, \citenamefont {Mulder},\ and\ \citenamefont {Van~der
  Ven}}]{Wagemaker_2009}%
  \BibitemOpen
  \bibfield  {author} {\bibinfo {author} {\bibfnamefont {M.}~\bibnamefont
  {Wagemaker}}, \bibinfo {author} {\bibfnamefont {F.~M.}\ \bibnamefont
  {Mulder}}, \ and\ \bibinfo {author} {\bibfnamefont {A.}~\bibnamefont {Van~der
  Ven}},\ }\href {\doibase 10.1002/adma.200803038} {\bibfield  {journal}
  {\bibinfo  {journal} {Advanced Materials}\ }\textbf {\bibinfo {volume}
  {21}},\ \bibinfo {pages} {2703} (\bibinfo {year} {2009})}\BibitemShut
  {NoStop}%
\bibitem [{\citenamefont {Wagemaker}\ \emph {et~al.}(2004)\citenamefont
  {Wagemaker}, \citenamefont {L{\"u}enkirchen-Hecht}, \citenamefont {van
  Well},\ and\ \citenamefont {Frahm}}]{Wagemaker_2004}%
  \BibitemOpen
  \bibfield  {author} {\bibinfo {author} {\bibfnamefont {M.}~\bibnamefont
  {Wagemaker}}, \bibinfo {author} {\bibfnamefont {D.}~\bibnamefont
  {L{\"u}enkirchen-Hecht}}, \bibinfo {author} {\bibfnamefont {A.~A.}\
  \bibnamefont {van Well}}, \ and\ \bibinfo {author} {\bibfnamefont
  {R.}~\bibnamefont {Frahm}},\ }\href {\doibase 10.1021/jp048567f} {\bibfield
  {journal} {\bibinfo  {journal} {The Journal of Physical Chemistry B}\
  }\textbf {\bibinfo {volume} {108}},\ \bibinfo {pages} {12456} (\bibinfo
  {year} {2004})}\BibitemShut {NoStop}%
\bibitem [{\citenamefont {Bai}\ \emph {et~al.}(2016)\citenamefont {Bai},
  \citenamefont {Li}, \citenamefont {Brushett},\ and\ \citenamefont
  {Bazant}}]{Bai_2016}%
  \BibitemOpen
  \bibfield  {author} {\bibinfo {author} {\bibfnamefont {P.}~\bibnamefont
  {Bai}}, \bibinfo {author} {\bibfnamefont {J.}~\bibnamefont {Li}}, \bibinfo
  {author} {\bibfnamefont {F.~R.}\ \bibnamefont {Brushett}}, \ and\ \bibinfo
  {author} {\bibfnamefont {M.~Z.}\ \bibnamefont {Bazant}},\ }\href {\doibase
  10.1039/c6ee01674j} {\bibfield  {journal} {\bibinfo  {journal} {Energy
  Environ. Sci.}\ }\textbf {\bibinfo {volume} {9}},\ \bibinfo {pages} {3221}
  (\bibinfo {year} {2016})}\BibitemShut {NoStop}%
\end{thebibliography}%


\begin{thebibliography}{3}%
\makeatletter
\providecommand \@ifxundefined [1]{%
 \@ifx{#1\undefined}
}%
\providecommand \@ifnum [1]{%
 \ifnum #1\expandafter \@firstoftwo
 \else \expandafter \@secondoftwo
 \fi
}%
\providecommand \@ifx [1]{%
 \ifx #1\expandafter \@firstoftwo
 \else \expandafter \@secondoftwo
 \fi
}%
\providecommand \natexlab [1]{#1}%
\providecommand \enquote  [1]{``#1''}%
\providecommand \bibnamefont  [1]{#1}%
\providecommand \bibfnamefont [1]{#1}%
\providecommand \citenamefont [1]{#1}%
\providecommand \href@noop [0]{\@secondoftwo}%
\providecommand \href [0]{\begingroup \@sanitize@url \@href}%
\providecommand \@href[1]{\@@startlink{#1}\@@href}%
\providecommand \@@href[1]{\endgroup#1\@@endlink}%
\providecommand \@sanitize@url [0]{\catcode `\\12\catcode `\$12\catcode
  `\&12\catcode `\#12\catcode `\^12\catcode `\_12\catcode `\%12\relax}%
\providecommand \@@startlink[1]{}%
\providecommand \@@endlink[0]{}%
\providecommand \url  [0]{\begingroup\@sanitize@url \@url }%
\providecommand \@url [1]{\endgroup\@href {#1}{\urlprefix }}%
\providecommand \urlprefix  [0]{URL }%
\providecommand \Eprint [0]{\href }%
\providecommand \doibase [0]{http://dx.doi.org/}%
\providecommand \selectlanguage [0]{\@gobble}%
\providecommand \bibinfo  [0]{\@secondoftwo}%
\providecommand \bibfield  [0]{\@secondoftwo}%
\providecommand \translation [1]{[#1]}%
\providecommand \BibitemOpen [0]{}%
\providecommand \bibitemStop [0]{}%
\providecommand \bibitemNoStop [0]{.\EOS\space}%
\providecommand \EOS [0]{\spacefactor3000\relax}%
\providecommand \BibitemShut  [1]{\csname bibitem#1\endcsname}%
\let\auto@bib@innerbib\@empty
\bibitem [{\citenamefont {Bai}\ \emph {et~al.}(2011)\citenamefont {Bai},
  \citenamefont {Cogswell},\ and\ \citenamefont {Bazant}}]{Bai_2011}%
  \BibitemOpen
  \bibfield  {author} {\bibinfo {author} {\bibfnamefont {P.}~\bibnamefont
  {Bai}}, \bibinfo {author} {\bibfnamefont {D.~A.}\ \bibnamefont {Cogswell}}, \
  and\ \bibinfo {author} {\bibfnamefont {M.~Z.}\ \bibnamefont {Bazant}},\
  }\href {\doibase 10.1021/nl202764f} {\bibfield  {journal} {\bibinfo
  {journal} {Nano Letters}\ }\textbf {\bibinfo {volume} {11}},\ \bibinfo
  {pages} {4890} (\bibinfo {year} {2011})}\BibitemShut {NoStop}%
\bibitem [{\citenamefont {Ganapathy}\ \emph {et~al.}(2011)\citenamefont
  {Ganapathy}, \citenamefont {van Eck}, \citenamefont {Kentgens}, \citenamefont
  {Mulder},\ and\ \citenamefont {Wagemaker}}]{Ganapathy_2009}%
  \BibitemOpen
  \bibfield  {author} {\bibinfo {author} {\bibfnamefont {S.}~\bibnamefont
  {Ganapathy}}, \bibinfo {author} {\bibfnamefont {E.~R.}\ \bibnamefont {van
  Eck}}, \bibinfo {author} {\bibfnamefont {A.~P.}\ \bibnamefont {Kentgens}},
  \bibinfo {author} {\bibfnamefont {F.~M.}\ \bibnamefont {Mulder}}, \ and\
  \bibinfo {author} {\bibfnamefont {M.}~\bibnamefont {Wagemaker}},\ }\href
  {\doibase 10.1002/chem.201101431} {\bibfield  {journal} {\bibinfo  {journal}
  {Chemistry}\ }\textbf {\bibinfo {volume} {17}},\ \bibinfo {pages} {14811}
  (\bibinfo {year} {2011})}\BibitemShut {NoStop}%
\bibitem [{\citenamefont {Olson}\ \emph {et~al.}(2006)\citenamefont {Olson},
  \citenamefont {Nelson},\ and\ \citenamefont {Islam}}]{Olson_2006}%
  \BibitemOpen
  \bibfield  {author} {\bibinfo {author} {\bibfnamefont {C.~L.}\ \bibnamefont
  {Olson}}, \bibinfo {author} {\bibfnamefont {J.}~\bibnamefont {Nelson}}, \
  and\ \bibinfo {author} {\bibfnamefont {M.~S.}\ \bibnamefont {Islam}},\ }\href
  {\doibase 10.1021/jp057261l} {\bibfield  {journal} {\bibinfo  {journal}
  {Journal of Physical Chemistry B}\ }\textbf {\bibinfo {volume} {110}},\
  \bibinfo {pages} {9995} (\bibinfo {year} {2006})}\BibitemShut {NoStop}%
\end{thebibliography}%

\end{document}


\title{Supplemental Material: Explaining key properties of lithiation in \ce{TiO2}-anatase Li-ion battery electrodes using phase-field modelling}

\author{Niek J.J. de Klerk}
 \altaffiliation{Contributed equally to this work.}
\author{Alexandros Vasileiadis}%
 \altaffiliation{Contributed equally to this work.}
\affiliation{Department of Radiation Science and Technology, Delft University of Technology, Mekelweg 15, 2629JB Delft, The Netherlands}

\author{Raymond B. Smith}
\affiliation{Department of Chemical Engineering, Massachusetts Institute of Technology, Cambridge, Massachusetts 02139, USA}
\author{Martin Z. Bazant}
\affiliation{Department of Chemical Engineering, Massachusetts Institute of Technology, Cambridge, Massachusetts 02139, USA}
\affiliation{Department of Mathematics, Massachusetts Institute of Technology, Cambridge, Massachusetts 02139, USA}
\author{Marnix Wagemaker}
\email{m.wagemaker@tudelft.nl}
\affiliation{Department of Radiation Science and Technology, Delft University of Technology, Mekelweg 15, 2629JB Delft, The Netherlands}


\maketitle

\section{Many particle effect at 0.01C}
\begin{figure}[htbp]
	\begin{center}
	\includegraphics[width=0.75\textwidth]{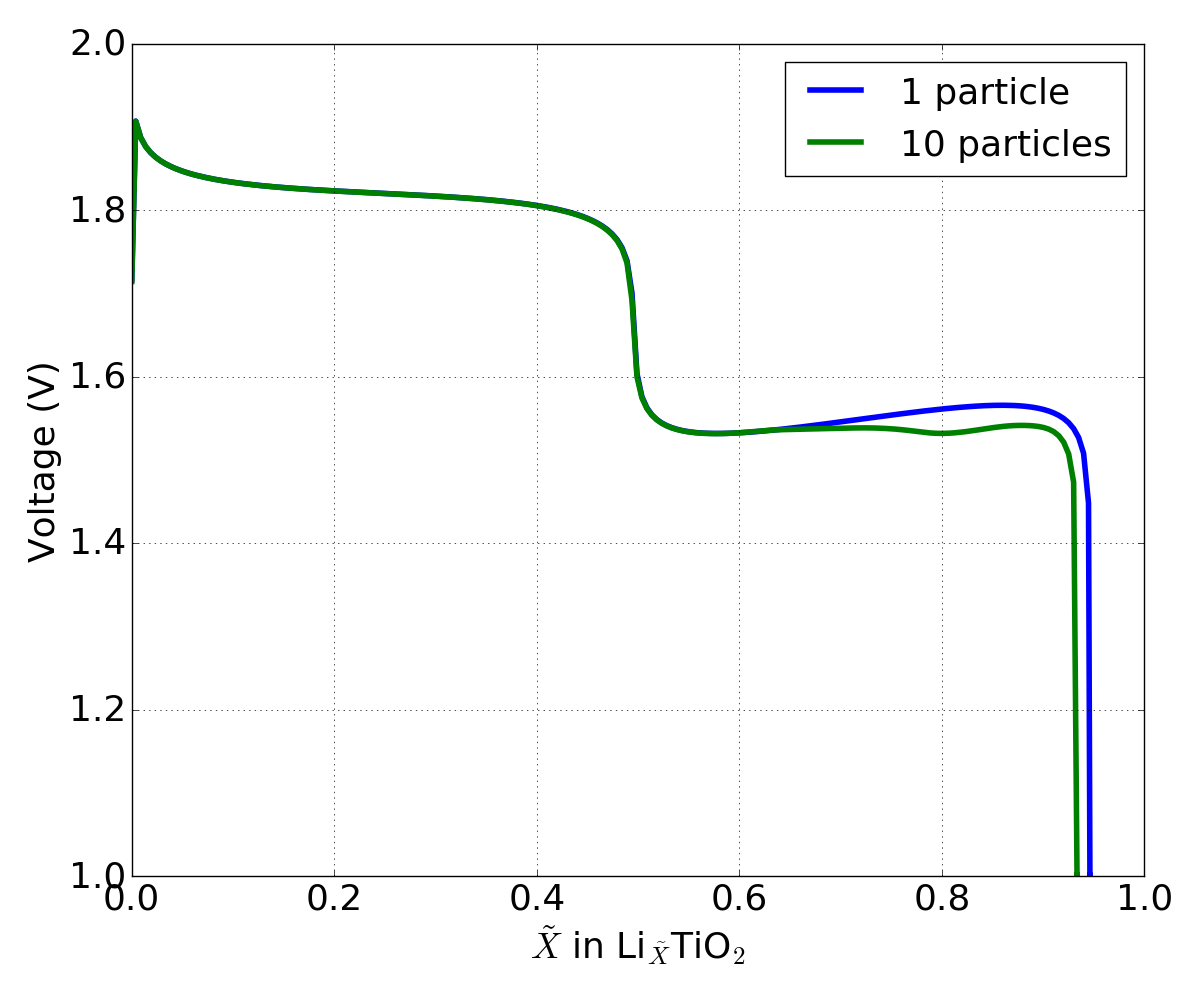} 
	\caption{Voltage profiles for simulations with 1 and 10 particles at 0.01C.
	\label{fig:10_particles_001C} }
	\end{center}
\end{figure}

In Figure 3 the voltage profile at C/100 tilts slightly upwards. This is caused by the simulation being performed on a single particle, repeating the same simulation with 10 particles gives a flat voltage profile, as shown in Figure \ref{fig:10_particles_001C}. The difference is caused by the fact that a single particle will always follow the spinodal line, which is inherently unstable \cite{Bai_2011}, but a single particle cannot depart from the spinodal line. A simulation with multiple particles does allow departure from the spinodal line, giving the flat voltage profile as shown in Figure \ref{fig:10_particles_001C}.

\section{Charge transfer coefficient}
\begin{figure}[htbp]
	\begin{center}
		\includegraphics[width=0.75\textwidth]{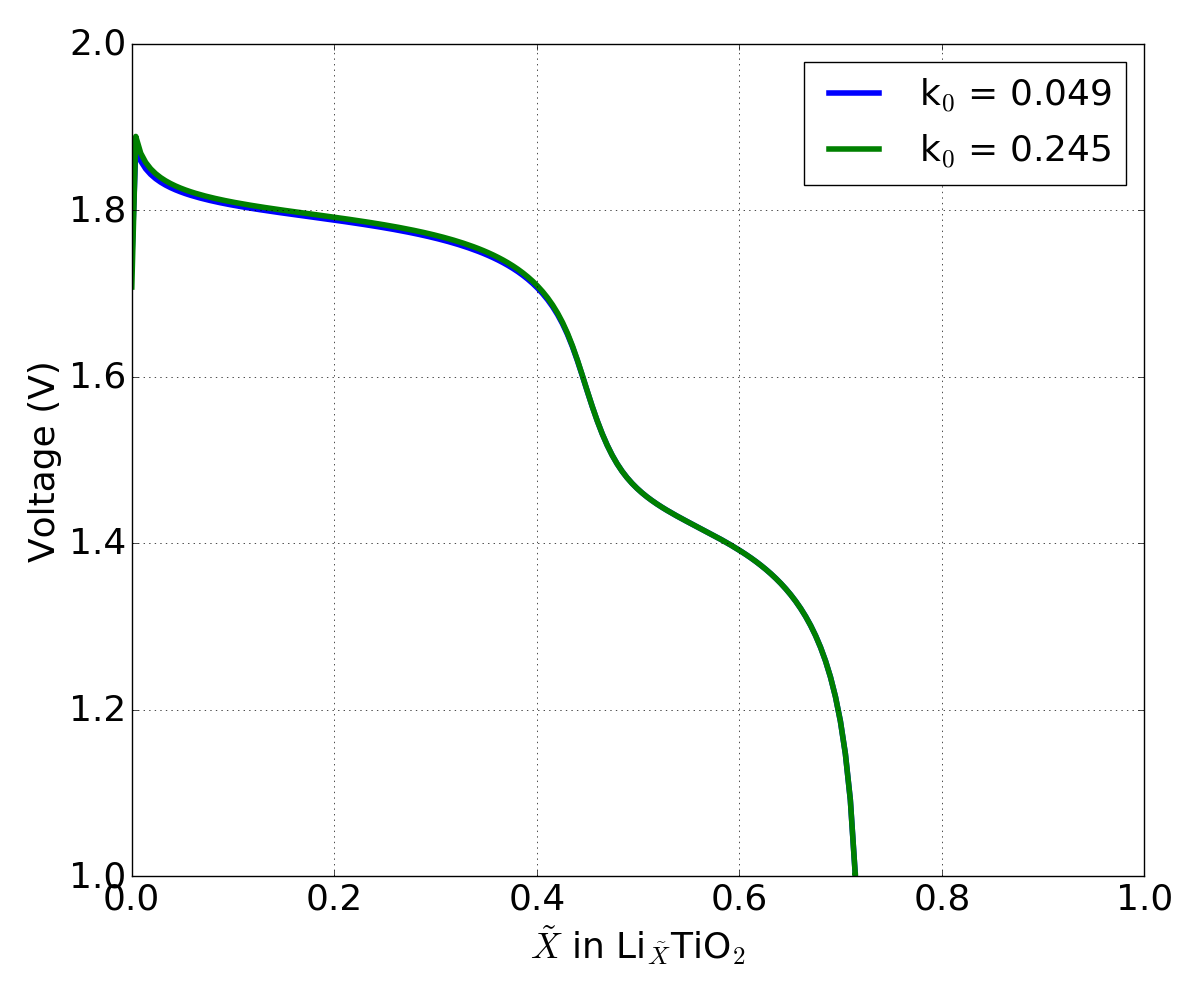} 
		\caption{Voltage profiles with high (0.245 A/m$^2$) and low (0.049 A/m$^2$) values for the charge transfer coefficient at 0.5C.}
		\label{fig:better_charge_transfer}	
	\end{center}
\end{figure}

The charge transfer coefficient ($k_0$) used for the simulations is based on the diffusion coefficient between anatase \ce{Li_{0.5}TiO2} and electrolyte as measured by NMR experiments \cite{Ganapathy_2009}.  
The measured diffusion coefficient is between 1*10$^{-12}$ and 5*10$^{-12}$ cm$^2$/sec at room temperature, depending on the electrolyte. 
The diffusivity can be related to jump rates using the Einstein-Smulochowski relation: $D = \frac{\tau a^2}{2d}$. Assuming a diffusion distance ($a$) equal to the particle size of 40 nm., and 1 dimensional diffusion ($d$) perpendicular to the surface, leads to a Li-jump frequency ($\tau$) between 0.125 and 0.625 sec$^{-1}$.
Combining this with the Li-density of 2.45*10$^{18}$ Li/m$^{2}$ for \ce{Li_{0.5}TiO2} at (101)-surfaces, which are the most stable facets of anatase crystals \cite{Olson_2006}, this gives a charge transfer between 0.31*10$^{18}$ and 1.53*10$^{18}$ Li sec$^{-1}$ m$^{-2}$, which is equal to 0.049 and 0.245 A/m$^2$, respectively. 
To be on the safe side the lower value of 0.049 A/m$^2$ is chosen for the simulations. However, as shown in Figure \ref{fig:better_charge_transfer} the difference between the voltage profiles for $k_0$-values of 0.049 and 0.245 is negligible. In the first plateau there is a tiny difference in voltage, smaller as 0.01 V, and during the second phase transformation no differences are visible.

\bibliography{bib_anatase_phase_field}